\documentclass[aps,prb,10pt,twocolumn,superscriptaddress]{revtex4-2}

\usepackage{amsmath}
\usepackage{amssymb}
\usepackage{graphicx}
\usepackage{dcolumn}
\usepackage{bm}
\usepackage[utf8]{inputenc}
\usepackage{verbatim}
\usepackage{color}
\renewcommand{\thesection}{\arabic{section}}
\renewcommand{\thesubsection}{\thesection.\arabic{subsection}}
\newcommand{\labelsubsec}[1]{
    \renewcommand\thesubsection{$\!$.\arabic{subsection}}
    \addtocounter{subsection}{-1}
    \refstepcounter{subsection}
    \label{#1}
    \renewcommand\thesubsection{\thesection.\arabic{subsection}}
}

\begin{document}
\title{An STM perspective on hexaborides: Surface states of the Kondo
insulator SmB$_6$}
\thanks{This article is dedicated to the memory of Stephan von Moln{\'a}r.}

\author{S. Wirth}
\affiliation{Max-Planck-Institute for Chemical Physics of Solids,
N\"othnitzer Str. 40, 01187 Dresden, Germany}
\author{P. Schlottmann}
\affiliation{Department of Physics, Florida State University, Tallahassee,
Florida 32306, USA}
\date{\today}
\begin{abstract}
Compounds within the hexaboride class of materials exhibit a wide variety of
interesting physical phenomena, including polaron formation and quadrupolar
order. In particular, SmB$_6$ has recently drawn attention as it is considered
a prototypical topological Kondo insulator. Evidence in favor of this concept,
however, has proven experimentally difficult and controversial, partly because
of the required temperatures and energy resolution. Here, a powerful tool
is Scanning Tunneling Microscopy (STM) with its unique ability to give local,
microscopic information that directly relates to the one-particle Green's
function. Yet, STM on hexaborides is met with its own set of challenges. This
article attempts to review the progress in STM investigations on hexaborides,
with emphasis on SmB$_6$ and its intriguing properties.
\end{abstract}
\maketitle

\section{Introduction}
Cubic hexaborides are known for more than a century \cite{moi1897} and their structure of type CaB$_6$ \cite{pau34} constitutes a class of compounds, the
different members of which exhibit a large variety of properties despite its
comparatively simple crystallographic structure. Perhaps best known is LaB$_6$
for its low work function, a material nowadays heavily used as a cathode
material in electron emission applications. Yet, the different hexaborides
exhibit intriguing physical phenomena. PrB$_6$, NdB$_6$ and GdB$_6$ are
antiferromagnets \cite{geb68}. EuB$_6$ is a semimetal, considered a prototype
material for polaron formation \cite{nyh97b,urb04} and colossal
magnetoresistance near its metal-semiconductor transition. Superconductivity
is observed in YB$_6$ below about 7.2 K, and in LaB$_6$ below 0.45 K
\cite{sch82}. CeB$_6$ features dense Kondo behaviour, a complex magnetic phase
diagram and quadrupolar order \cite{eff85}, and SmB$_6$ (see section
\ref{TKISmB6}), YbB$_6$ (under high pressure \cite{zho15}) and PuB$_6$
\cite{deng13} are topological semiconductors.

Renewed intense interest arose with the proposal of non-trivial topological
surface states in SmB$_6$ \cite{dze10}. This concept combined the rising field
of topological insulators \cite{has10} with that of Kondo insulators (KI)
\cite{aep92,ris00}. It provided a simple explanation for the long-standing
conundrum of the low-temperature resistance plateau observed in SmB$_6$
\cite{all79}. Yet, one issue became immediately obvious: Due to the small
energy scales and temperatures involved, any experimental verification of this
proposal would be challenging. The spectroscopic methods of choice here are
angle-resolved photoemission spectroscopy (ARPES) and scanning tunneling
microscopy and spectroscopy (STM/S) \cite{kir20}. Indeed, ARPES \cite{neu13,
jia13,nxu13,xu14b,den14} and in particular its spin-polarized version
\cite{nxu14,sug14} seemed to confirm this proposal, but were also challenged
\cite{hla18}. The severe difficulties involved in STS will be detailed below.

Along with the experimental efforts to confirm the topologically non-trivial
nature of the surface states there were also dedicated theoretical attempts
\cite{tak11,lu13,deng13,ale13,bar14,bar15,yu15}. Here, also specific surface
conditions were considered \cite{jkim14,kan15}. Despite the seemingly simple
crystallographic lattice of SmB$_6$, such calculations are particulary
complicated due to the intermediate Sm valence and the many-body nature of the
Kondo effect.

Albeit the focus of this review is clearly on STM of SmB$_6$, we will also
cross-compare to studies of other hexaborides. This will not only demonstrate
the diverse capabilities and issues of STM, but will also reinforce some of
the results, e.g.\ concerning the topography.

\section{Kondo insulators}
\label{Kondo}
KI, also known as heavy-fermion semiconductors, are stoichiometric compounds
with small-gap semiconducting properties \cite{aep92,ris00}. Most are
nonmagnetic, e.g., Ce$_3$Bi$_4$Pt$_3$ \cite{hun90}, CeFe$_4$P$_{12}$
\cite{mei85}, CeRu$_4$Sn$_6$ \cite{pas10}, YbB$_{12}$ \cite{iga88}, FeSi
\cite{dit97}, SmB$_6$ \cite{men69}, and SmS under pressure \cite{map71}, with
a van Vleck-like low-temperature susceptibility. The low-$T$ resistivity, and
in most cases the electronic specific heat, follow an exponential activation
law consistent with a gap in the density of states. Not all KI are perfect
semiconductors, e.g. CeNiSn \cite{tak92} and CeRhSb \cite{iza99}, because the
gap is frequently only a pseudogap and/or there are intrinsic or impurity
states in the band gap.

The gap in KI arises from the coherence of the heavy electron states at the
Fermi level $E_{\rm F}$ resulting from the hybridization between the rare earth
4$f$ and conduction bands. The hybridization gap at the Fermi level is strongly reduced by many-body interactions (Kondo effect) and can range from a few meV
to 50 meV. In contrast to usual band gaps, the gap of a KI is
temperature-dependent \cite{sch93}, as a consequence of the Kondo effect (or
many-body effects), and the materials become metallic at surprisingly low temperatures. The gap can also be gradually closed by large magnetic fields,
yielding a metallic state for fields larger than a critical value (for SmB$_6$
\cite{coo99}, for Ce$_3$Bi$_4$Pt$_3$ \cite{jai00}, and for YbB$_{12}$
\cite{sug88}). The band populations are changed by the Zeeman splitting and
the critical field corresponds to the beginning of the occupation of the bottom
of the conduction band. A discontinuous metal-insulator transition as a
function of pressure has been observed in SmB$_6$ \cite{coo95,coo95b}. With
pressure the valence of the rare earth ions increases, i.e. Ce and Sm ions
become less magnetic and Yb more magnetic, and the gap closing arises from
strong electronic correlations in analogy to a Mott-Hubbard insulator. The
high-pressure ground states of SmB$_6$ \cite{bar05} and SmS \cite{bar04} have
long-range ferromagnetic order.

There is evidence for in-gap states in SmB$_6$ \cite{mol82,nan93,gor99,nyh97,
ale93b,ale95}, FeSi \cite{sch93,yeo03,slu02} and YbB$_{12}$ \cite{bou98,nef99,
ale01} and they are thought to be a common feature in all Kondo insulators
\cite{ris03}. The nature of the in-gap bound states, whether intrinsic or
extrinsic, is still controversial. Extrinsic states arising from defects
(impurities or vacancies) are known as Kondo holes, and are usually donor or
acceptor bound states that even at low concentrations give rise to impurity
bands \cite{sol91,sch92}. Intrinsic in-gap states are for instance magnetic
excitons, i.e., bound states in the conduction and valence bands.
Experimentally, they have been observed as peaks in inelastic neutron
scattering in SmB$_6$ \cite{ale93b,ale95} and YbB$_{12}$ \cite{bou98,nef99,
ale01}, and models for these excitons have been proposed by Kasuya
\cite{kas96} and Riseborough \cite{ris00b}.

\subsection{Topological Kondo insulators}
\begin{figure}[t]
\includegraphics*[width=0.4\textwidth]{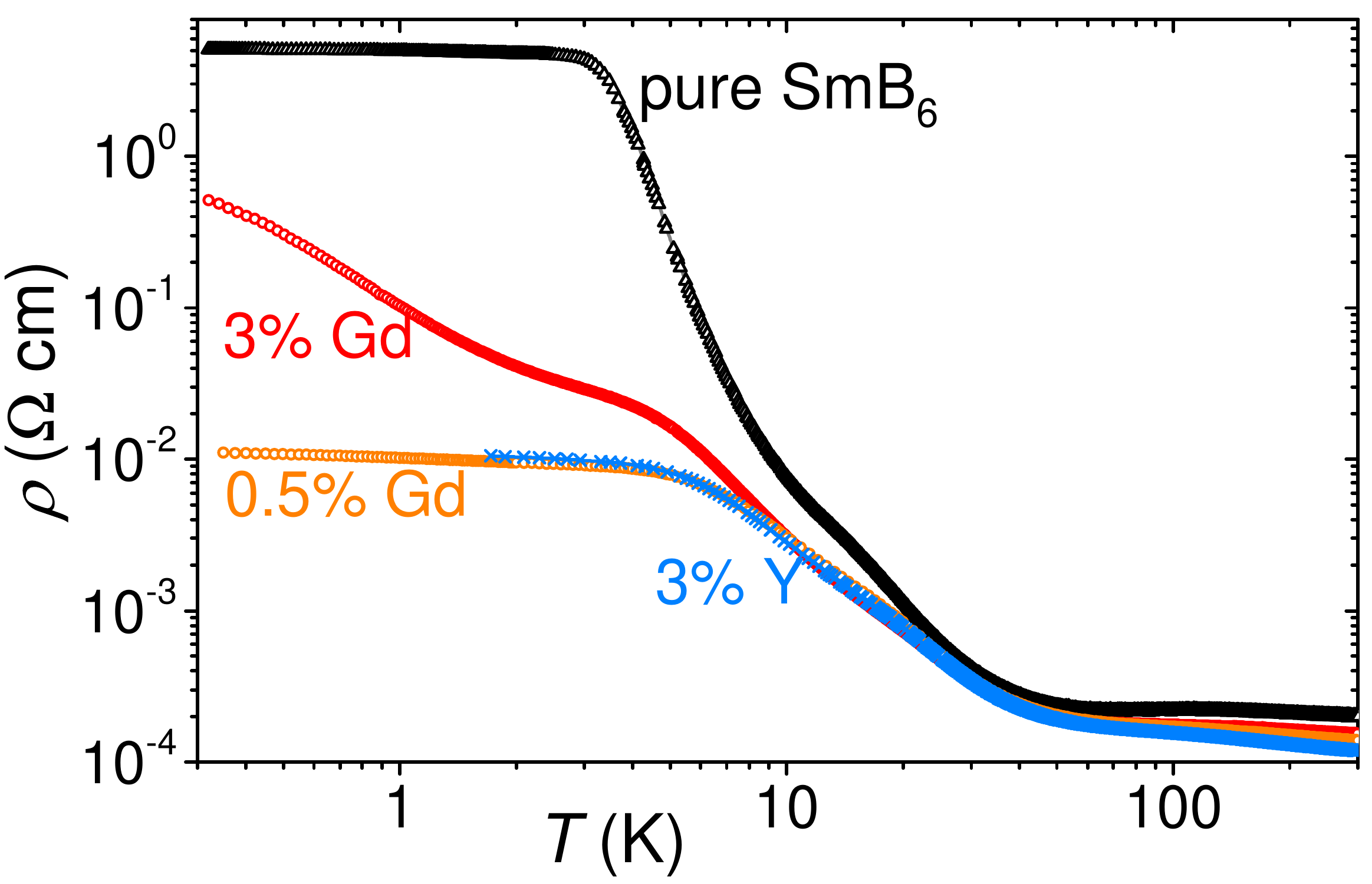}
\caption{Resistivity $\rho(T)$ of pure SmB$_6$ as well as substituted (0.5\%,
3\% Gd and 3\% Y) samples. The plateau of $\rho(T)$ at low temperature is
characteristic of the surface states. Adapted under the terms of a Creative
Commons Attribution 4.0 International License \cite{jiao18}. Copyright 2018
The Authors, published by AAAS.}  \label{resist}
\end{figure}

The remarkable discovery that three-dimensional insulators \cite{moo07,fuk07,
hsi08,roy09} can acquire a topological order through a spin-orbit driven band
inversion led to a new kind of insulators. The presence of topological order
is determined by a $\mathbb{Z}_2$-index, which is positive ($\mathbb{Z}_2=+1$)
in conventional insulators and negative ($\mathbb{Z}_2=-1$) in topological
insulators (TI). Dzero, Sun, Galitski and Coleman \cite{dze10} proposed to
combine topology with strong correlations and apply it to Kondo insulators. The
main requirements for Topological Kondo Insulators (TKI) are (i) the spin-orbit
coupling of the $4f$-electrons in a KI is much larger than the gap of the Kondo
insulator, which could drive the topological order, and (ii) the hybridization between the odd-parity $4f$-electron wave-functions with the even-parity
conduction band states (predominantly $5d$-electrons). Each time there is a
band-crossing between the two at a high symmetry point of the Brillouin zone,
the $\mathbb{Z}_2$-index changes sign, leading to a TI for an odd number of
sign changes \cite{dze16}. Each band crossing generates a Dirac cone of
protected spin-momentum locked surface states \cite{dze16}. These topological
surface states (TSS) give rise to a low-temperature resistivity that saturates
into a plateau, rather than to an exponential activation behavior, cf.\ {\bf
Figure} \ref{resist}. Prominent candidates for TKI are SmB$_6$,
Ce$_3$Bi$_4$Pt$_3$ \cite{wak16b}, and YbB$_{12}$ \cite{hag16}. Dilution with
nonmagnetic impurities (Y) generates an impurity band in the bulk, which
contributes to conduction. Gd is magnetic and breaks the time reversal symmetry
and hence, the protection of the TSS. In consequence, Gd impurities, on the
one hand, increase the conduction of the bulk due to the impurity band and, on
the other hand, blocks the surface conduction as a consequence of the loss of
protection of the TSS.

\subsection{The case of SmB$_6$}
\labelsubsec{TKISmB6}
The first intermediate valence semiconductor discovered was SmB$_6$ \cite{vai64,men69}. The ground state of Sm is a coherent superposition of two
ionic configurations, $4f^6$ and $4f^55d$, with approximate weight factors
0.4 and 0.6, respectively, yielding an effective $4f$ valence of $\sim$2.6 at
room temperature. The small indirect gap as measured from the electrical
resistivity is $E_g \approx 4.6$ meV \cite{aep92,coo95,slu99,all79} (in
transport, the activation energy is $E_g/2$) and the entropy (integrating the
specific heat \cite{mol82} over $T$) reaches $R\ln2$ at about 40~K. On the
other hand, a larger (direct) gap of 13.3--19 meV has been observed in optical
reflectivity data \cite{nan93,gor99}. Both gaps are properties of the bulk
material and are believed to be a consequence of the hybridization of the $4f$
and $5d$ states. Similar to other Kondo insulators, such as FeSi
\cite{slu02}, the behavior of SmB$_6$ can be classified for several
temperature ranges. Above 50 K, the electrical conductivity is almost
temperature-independent, the Hall coefficient is positive, and the material
behaves like a poor metal \cite{all79}. In this range, the magnetic
susceptibility is only weakly temperature-dependent \cite{coh70,tak81}. For
$6 \lesssim T \lesssim 50$ K, the conductivity $\sigma$ is activated, the Hall
coefficient is negative, and the susceptibility $\chi$ decreases slightly
before reaching a roughly constant value below 20 K \cite{tak81}. Below $\sim$6
K, the temperature dependence of $\sigma$ is smaller, and below 2--3 K,
$\sigma$ is almost constant \cite{coo95}, see {\bf Figure} \ref{resist}. The
scaling of the intensities of the low-temperature magnetic bulk properties
\cite{sch14}, i.e., the upturn of the susceptibility \cite{glu06}, the 16~meV
excitation measured by inelastic neutron scattering \cite{ale93b}, and the
Raman transition \cite{nyh97} suggest that they arise from the in-gap excitons
\cite{ris00b}. The $T$-dependence of the electrical conductivity below about
5~K, on the other hand, is believed to arise from the protected TSS.

The ground multiplet of the $4f$ electrons of Sm$^{3+}$ is $J=5/2$, which in
cubic symmetry splits into a $\Gamma_8$ quartet and a $\Gamma_7$ doublet, while
the ground state of Sm$^{2+}$ is a singlet, $J=0$. On the other hand, the $5d$ electrons in cubic symmetry in eightfold coordination belong to the $e_g$
ground doublet. Away from the $\Gamma$-point, the $e_g$ levels split and form
pockets at the three X-points of the lattice. This pocket is common to all the
hexaboride compounds, independently of the cation ion being trivalent (e.g.,
LaB$_6$, CeB$_6$, PrB$_6$, NdB$_6$, GdB$_6$) or divalent (e.g., CaB$_6$,
EuB$_6$, YbB$_6$). In the case of trivalent cations the pockets are half-full
\begin{figure}[t]
\includegraphics*[width=0.44\textwidth]{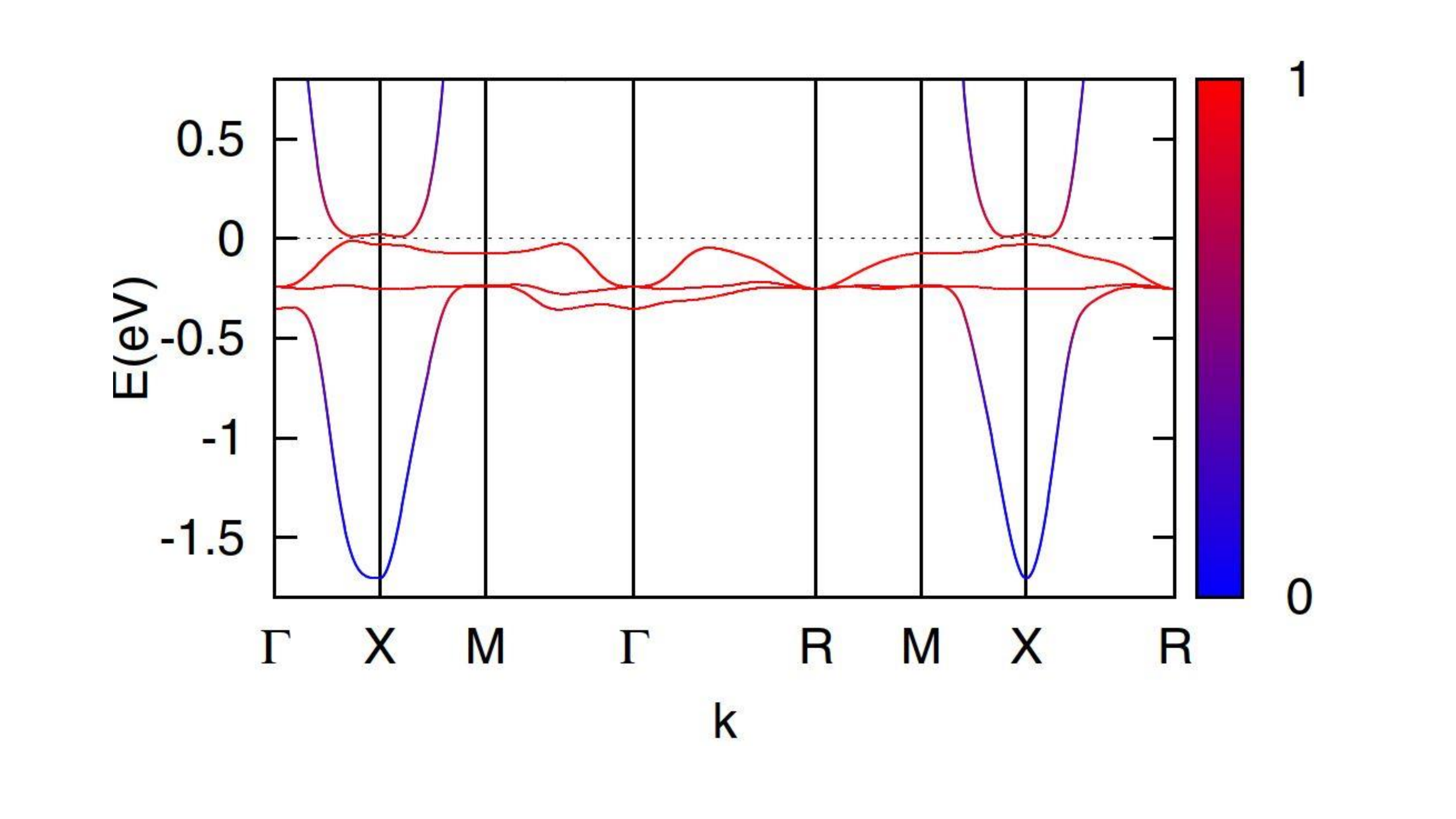}
\caption{Bulk dispersion of SmB$_6$ from a renormalized tight binding model
along a path in the 3D Brillouin zone.  The color code shows the $f$ weight.
Reproduced with permission \cite{bar14}. Copyright 2014, American Physical
Society.}  \label{bandstruc}
\end{figure}
and the system is metallic, while for CaB$_6$ and SrB$_6$ the pockets are empty
and the compound is an insulator; EuB$_6$ is a ferromagnetic semimetal, and the hybridization forces SmB$_6$ to be intermediate valent. For SmB$_6$ (and
YbB$_6$) the $5d$-band crosses through the $4f$-bands (they behave like three
Kramers doublets) at the three inequivalent X-points, resulting in a
topological band structure with $\mathbb{Z}_2 = (-1)^3 = -1$. The
band-inversion at the X-points leads to the formation of three Dirac cones on
the (100) surface, one at the surface ${\overline \Gamma}$ point and two at the
two surface ${\overline X}$ points \cite{tak11,deng13,bar14,lu13,dze16}. The
details of the quasi-particle band-structure of the bulk have been obtained by
a combination of LDA and Gutzwiller methods \cite{lu13} and LDA + DMFT
\cite{deng13} (the latter for the iso-electronic compound PuB$_6$). To study
the surface states on the (001) surface, a tight-binding Hamiltonian is
constructed by fitting to the bulk band-structure, which is then solved on a
25 to 60-layer slab constructed with this Hamiltonian. Results of such a
calculation \cite{bar14} are reproduced in {\bf Figure} \ref{bandstruc}.
Gapless edge states arise in the bulk gap around the ${\overline \Gamma}$ and
${\overline X}$ points. The Fermi surface at ${\overline X}$ is much larger
than at ${\overline \Gamma}$. Due to the strong spin-orbit interaction the
surface states are topologically protected and their spin and orbital momentum
are locked. A consequence of the spin-momentum locking is a
weak-antilocalization effect at the metallic surface \cite{tho16}.

The electronic structure of the surface state can be probed by ARPES. An
observation of an odd number of Dirac cones is a direct confirmation of the topological nature of the surface state \cite{dze16}. There are numerous
detailed ARPES studies of SmB$_6$ \cite{neu13,nxu13,jia13,zhu13,den13} that
show a general agreement with the theoretical predictions \cite{ale13} of two
large ${\overline X}$ pockets and a smaller one at the ${\overline \Gamma}$
point. Spin-resolved ARPES can in principle determine the spin texture of the
surface states, however, the results are still controversial due to the
limited energy resolution of ARPES as compared to the small gap of SmB$_6$ \cite{hsi09,nxu14,sug14,hla18}.

The transport properties of SmB$_6$ provide strong evidence for its TKI
character at low temperature. As already mentioned, at high $T$, SmB$_6$
behaves as a correlated metal, and below $\sim$50 K it becomes insulating with
the opening of the Kondo gap, i.e., the resistivity follows a diverging
activation law down to about 6 K. Below about 3 K the resistivity saturates at
a high value, which is attributed to the topologically protected metallic
surface state within the Kondo gap. Several experiments were carried out to
verify the surface conduction. Wolgast {\it et al.} \cite{wol13} used the
dependence of the bulk and surface contributions to the conduction on the size
and shape of the sample. They designed and fabricated a thin-plate shaped
device with electric current leads placed on top and bottom of the crystal.
Placing voltage leads at various places on the device, they measured three
paths as a function of temperature and could discern that the conduction at
low $T$ was dominated by the surface, whereas it was dictated by the bulk at
high temperatures. Surface and bulk conduction also differ sharply in the Hall
effect \cite{kim13}. For the bulk, the Hall voltage is inversely proportional
to the thickness of the Hall bar, but not so for the surface Hall voltage. Kim
{\it et al.} \cite{kim13} shaped the Hall bar in the form of a wedge and could
thereby vary the thickness of the sample. As a function of temperature the
surface and bulk Hall effects could be distinguished. Another method for
characterizing bulk and surface conductivities in SmB$_6$ is the {\it inverted
resistance measurement} \cite{eo18} by using Corbino disk geometries. This
method has also been employed to demonstrate that the transport gap in SmB$_6$
is robust and protected against disorder \cite{eo19}.

The coupled metallic surface and the bulk of SmB$_6$ have an equivalent
circuit consisting of two resistors, a capacitance and an induction. These
electronic components are strongly temperature dependent and vary with the
thickness of the sample. The bulk insulator is represented by a resistor $R_b$
and a capacitance $C_b$, while the surface states are described by a resistor
$R_s$ and the induction $L_s$. The voltage and current are then out of phase
and give rise to Lissajous curves which have been studied by Kim {\it et al.}
\cite{kim12}. Adding an external capacitor to the circuit drives oscillations
with a dc current source. Reducing the thickness of the SmB$_6$ plate requires
a smaller external capacitor until the device spontaneously starts oscillating
with a frequency of the order of $10^7$ Hz \cite{ste16}. The behaviors of the
oscillators agree well with a theoretical model describing the thermal Joule
heating and electronic dynamics of coupled surface and bulk states \cite{ste16,cas18}. An impedance study on SmB$_6$ single crystals, performed
at low $T$ and over a wide frequency range revealed the transition from
surface to bulk dominated electrical conduction between 2 and 10 K
\cite{sta21}. The device exhibits current-controlled negative differential
resistance at low $T$ brought by the Joule heating.

\section{The hexaboride family}
\begin{figure}[b]
\includegraphics*[width=0.36\textwidth]{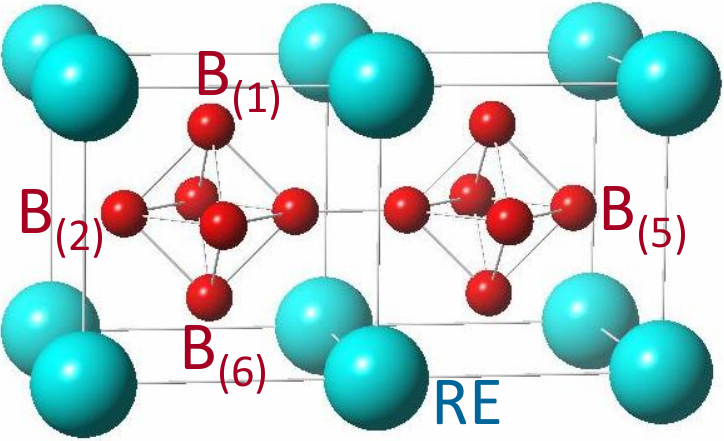}
\caption{Crystal structure of the rare earth (RE) hexaborides. Shown are two
unit cells.}  \label{structure}
\end{figure}

The hexaboride family is highly versatile as compounds can be formed with
alkaline earth and rare earth (RE) elements. They crystallize in the cubic
CaB$_6$ structure type, space group $Pm\bar{3}m$, {\bf Figure} \ref{structure}.
These ceramics typically exhibit high melting points, hardness and chemical
stability (see e.g.\ \cite{cah19}) and interesting thermoelectric properties
\cite{ber78}. The covalently-bonded B atoms form octahedra, which take up two
electrons from the metal element. In consequence, hexaborides with trivalent
REs are metals, while those with divalent elements (Eu, Yb) exhibit
semimetallic properties \cite{mer76,mas96}, see {\bf Table} \ref{tabprop}.
The likely best known hexaboride is LaB$_6$, which is an often-used cathode
material because of its low work function and low vapor pressure \cite{ahm72}.

At present, hexaboride single crystals are mostly prepared in two ways, either
by flux growth \cite{can92,rosa18} or by the floating-zone technique
\cite{tan80,pad81,hat13,pro18}. By means of the latter, large single crystals
can be prepared including samples suitable for neutron scattering experiments
\cite{ale95,fuh17}. Since there is no container or boat used, the samples are
expected to be clean of any contamination (except if a binder is applied).
But, because temperatures above the melting temperature of the polycrystalline
starting material are applied, there can be RE deficiencies with concomitant
changes of the lattice parameter \cite{phe16}. This may also give rise to
sample-to-sample differences \cite{phe16,ore17,eo20b}. Flux-grown samples are
typically smaller compared to floating-zone grown ones, and may contain
Al-inclusions stemming from the flux \cite{phe16,tho19}. However, since the
temperature during the growth is usually below the melting temperature,
decomposition can be prevented. SmB$_6$ samples tend to grow very close to
\begin{table}[t]
\caption{Lattice constant $a$ (averaged over different reports) and RE valence
$\nu$ of some hexaborides (low-$T$ value in case of SmB$_6$). Charge carriers
$n$ per RE from Hall measurements and results from x-ray chemical shift (XCS)
were taken from \cite{gru85}. AFM: antiferromagnet, AFQ: antiferroquadrupolar,
SC: superconductor.}
\label{tabprop}
\begin{ruledtabular}
\begin{tabular}{c|ccccl}
material & $a$ (\AA) & $\nu$ & $n$  & XCS  & remark\\ \hline
\rule{-3pt}{10pt}
LaB$_6$  & 4.154     & 3     & 0.93 & 0.86 & low work function, SC \\
CeB$_6$  & 4.141     & 3     & 1.06 & 1.00 & AFQ, AFM \\
NdB$_6$  & 4.125     & 3     & 0.88 & 1.01 & AFM \\
PrB$_6$  & 4.130     & 3     & 0.84 & 0.82 & AFM \\
GdB$_6$  & 4.112     & 3     & 0.94 & 1.00 & AFM \\
YB$_6$   & 4.102     & 3     & 0.96 & 1.00 & SC, $T_{\rm C}\approx$ 8.4 K\\
\hline \rule{-3pt}{8pt}
EuB$_6$  & 4.184     & 2     & 0.03 & $\leq$0.05 & magnetic semimetal \\
YbB$_6$  & 4.142     & 2     & 0.05 & $\leq$0.05 & high-pressure TKI\\
\hline \rule{-3pt}{8pt}
SmB$_6$  & 4.138     & $\sim$2.6 &  &      & proposed TKI \\
\end{tabular}
\end{ruledtabular}
\end{table}
stoichiometry \cite{eo19,ale21}. Apparently, the flux-grown samples can be
grown with a smaller number of RE vacancies compared to floating-zone grown
samples \cite{eo20b}. However, the exact details of the impact of defects for
the differently grown samples are not well understood, e.g., on the
low-temperature specific heat \cite{tho19} or thermal conductivity
\cite{har18}.

Just as intricate as the sample growth can be the surface preparation. Clean
surfaces can be obtained by cleaving but, as discussed in section \ref{STM},
there is no preferred cleaving plane, typically giving rise to atomically
rough surfaces. Polishing sample surfaces requires care because so-called
subsurface cracks may be formed \cite{eo20,ale21b}. Importantly, polishing or
etching may alter the surface, e.g. by disrupting the crystal structure at the
surface, introducing defects and/or, in case of SmB$_6$, changing the Sm
valence. Indeed, a Sm valence close to 3$+$ was observed near the surface
\cite{lut16,uts17}, which may indicate the presence of Sm$_2$O$_3$ at the
surface. In general, oxidation of the RE at the surface can be an issue (see
e.g. \cite{tre12,bel19}) and hence, storage under UHV conditions at and after
any SmB$_6$ surface treatment is advisable if clean surfaces are strived for.

\section{STM on hexaborides}
\label{STM}
\subsection{Topographies of SmB$_6$(001) surfaces}
Many types of measurements depend on the quality of the sample surface. This
even holds for resistivity measurements, for which surfaces often need to be
prepared, e.g. by polishing or etching (see e.g. \cite{kim13,wol13,wol15,
sye15,bis17,eo18,eo19,fuh19,eo20b}). In case of surface sensitive studies like
ARPES or STM, clean sample surfaces are typically produced by {\it in situ}
cleaving. From the comparatively simple crystal structure one might expect
either B or RE terminated atomically flat surfaces. However, as a result of
the cubic structure, all bond directions are equally strong. In contrast to
materials with stronger and less-strong bond directions (an outstanding
example here are two-dimensional materials like the transition-metal
dichalcogenites \cite{hes89}), the cleaving planes in the hexaborides are much
less well defined. Additional complications stem from i) the polar nature of
B or RE surfaces and ii) the shorter distance between inter-octahedral B atoms
compared to those within the octahedra.

The latter property poses the question whether or not the B-octahedra are
broken upon cleavage. Because the intra-octahedral B distance is slightly
longer than the inter-octahedral one, cleaving may also involve
intra-octahedral bond breaking. From density functional theory (DFT)
calculations on atomically flat surfaces, the inter-octahedral cleave appears
to be slightly favorable compared to intra-octahedral bond breaking
\cite{roe16,sun18}. We note that such calculations are complicated by strong
correlations and the intermediate Sm valence in SmB$_6$. In our own
experience, cleaving at room temperature did not result in atomically flat
surfaces and frequently included intra-octahedral bond breaking. This finding
is in line with the observation of so-called doughnuts \cite{ruan14}, which
result from four square-arranged units cells all exposing B$_{(2)}$-B$_{(5)}$
atoms, see {\bf Figure} \ref{structure}. However, intra-octahedral bond
breaking has also been found after low-temperature cleaving \cite{roe16,sun18},
on which we focus in the following.

As a result of the above-mentioned properties, the vast majority of the
cleaved surfaces is rough on an atomic scale. Note that out of the more than
30 SmB$_6$ samples we cleaved so far {\it in situ} and at around 20 K, on 8
of them we were not able to find any atomically flat areas even after weeks of
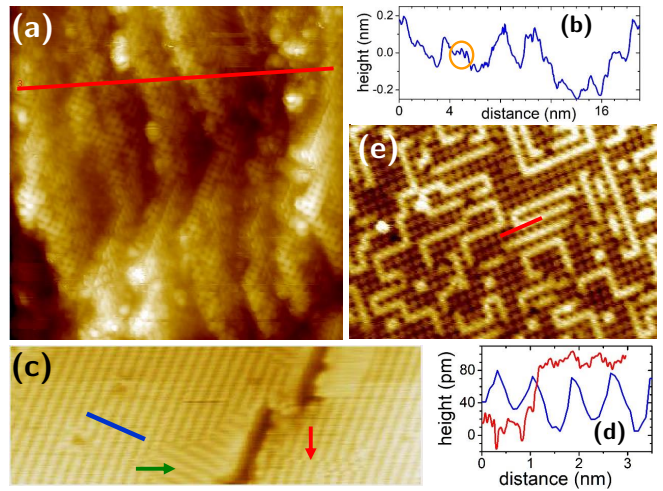
\begin{figure}[t]
\includegraphics*[width=0.48\textwidth]{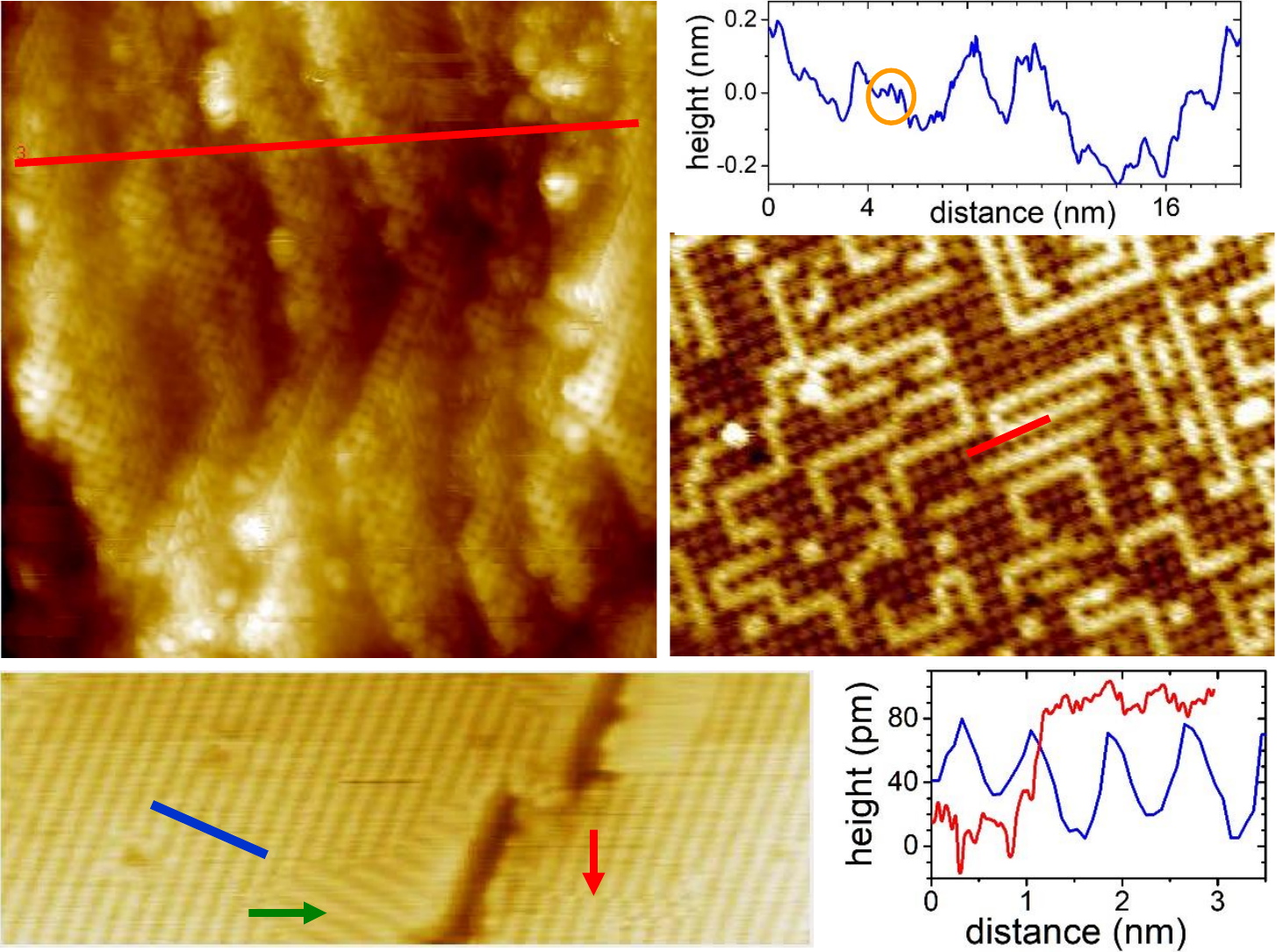}
\unitlength1cm \begin{picture}(-0.4,2)
\put(-8.7,6.0){\sffamily\bfseries\large  \textcolor{white}{(a)}}
\put(-1.4,6.07){\sffamily\bfseries (b)}
\put(-8.7,1.5){\sffamily\bfseries\large (c)}
\put(-0.96,0.65){\sffamily\bfseries (d)}
\put(-4.1,4.4){\sffamily\bfseries\large \textcolor{white}{(e)}}
\end{picture}
\caption{a) Rough topography, area 20 $\times$ 20 nm$^2$. b) Height scan along
the red line in a). Orange circle marks corrugations of distance $a$. c)
Ordered $2\times 1$ reconstructed surface, area 30 $\times$ 10 nm$^2$. Green
arrow indicates a $1 \times 2$ domain, red arrow a phase shift by $a$. d) Line
scans along blue and red lines in c) and e), respectively. e) Disordered,
reconstructed surface over area 20 $\times$ 14 nm$^2$. All topographies: $V =$
0.2 V, $I_{\rm sp} =$ 0.5 nA. c)--e) adapted under the terms of a Creative
Commons Attribution 4.0 International License \cite{wir20}. Copyright 2021
The Authors, published by Wiley.}  \label{topos}
\end{figure}
searching. A typical cleaved surface of SmB$_6$ is presented in {\bf Figure}
\ref{topos}a, obtained with bias voltage $V =$ 0.2 V and setpoint current
$I_{\rm sp} =$ 0.5 nA. The height scan shown in {\bf Figure} \ref{topos}b,
taken along the red line in (a), indicates a roughness of at least a unit cell
height $a$ and hence, both B and Sm are exposed at the surface. From the
involved height changes it appears that the crystal structure may be disturbed
within some areas. This raises the question whether such disturbed surface
areas can still be considered as weakly disordered and hence, whether the
proposed non-trivial surface states remain topologically protected
\cite{schu12}. Nonetheless, there are also some small, flat areas which exhibit
corrugations of distance $a$, see orange circle in  {\bf Figure} \ref{topos}b,
which raise hopes for atomic resolution.

Polar surfaces are prone to surface reconstructions. Indeed, upon searching,
such reconstructions can often be found \cite{yee13,roe14,pir20,her20}.
Low-energy electron diffraction (LEED) reported reconstructions \cite{aon78},
mostly of $2 \times 1$ type \cite{miy12,ram16,miy17}. The latter can be
visualized by a Sm-terminated surface with each other line of Sm atoms missing.
Conversely, a B-terminated surface with each second line of B-octahedra missing
can be imagined, see Figure 2 in \cite{sun18}. The topography presented in {\bf
Figure} \ref{topos}c is consistent with a $2 \times 1$ reconstruction, as is
obvious from the distances of $2a$ in the height scan {\bf Figure} \ref{topos}d.
The height changes seem more consistent with the former notation of Sm atomic
lines missing. In case of such a reconstruction, also $1 \times 2$ domains
should be expected and indeed, an example of such a domain is marked by a green
arrow in {\bf Figure} \ref{topos}c. As expected, we also observed an offset of
$a$ on a $2 \times 1$ reconstruction (red arrow), further supporting its
assignment as a reconstruction. Energetically, however, there is no need for
these Sm- (or B-) lines to run straight over extended distances as long as
approximately similar amounts of Sm or B-octahedra are exposed at the surface.
In result, we also observed bent lines of Sm atoms, {\bf Figure} \ref{topos}e.
Again, the height change is consistent with Sm on top of a B surface, {\bf
Figure} \ref{topos}d. The observed step heights render an assignment of the
reconstructed surfaces (be it ordered $2 \times 1$ or disordered as in
{\bf Figure} \ref{topos}e) as a pure, and hence atomically flat, Sm-terminated
surface unlikely \cite{her20} as it would require a remarkable modulation of
the local density of states (DOS). However, the assignment in \cite{her20} is
merely based on the observation of Sm-based features in tunneling spectroscopy,
and our suggestion of Sm on top of B would indeed expose a considerable amount
of Sm at the surface. Local differences in tunneling spectroscopy depending on
the exposed surface atom have been reported \cite{roe14,sun18} and will be
\begin{figure}[t]
\includegraphics*[width=0.4\textwidth]{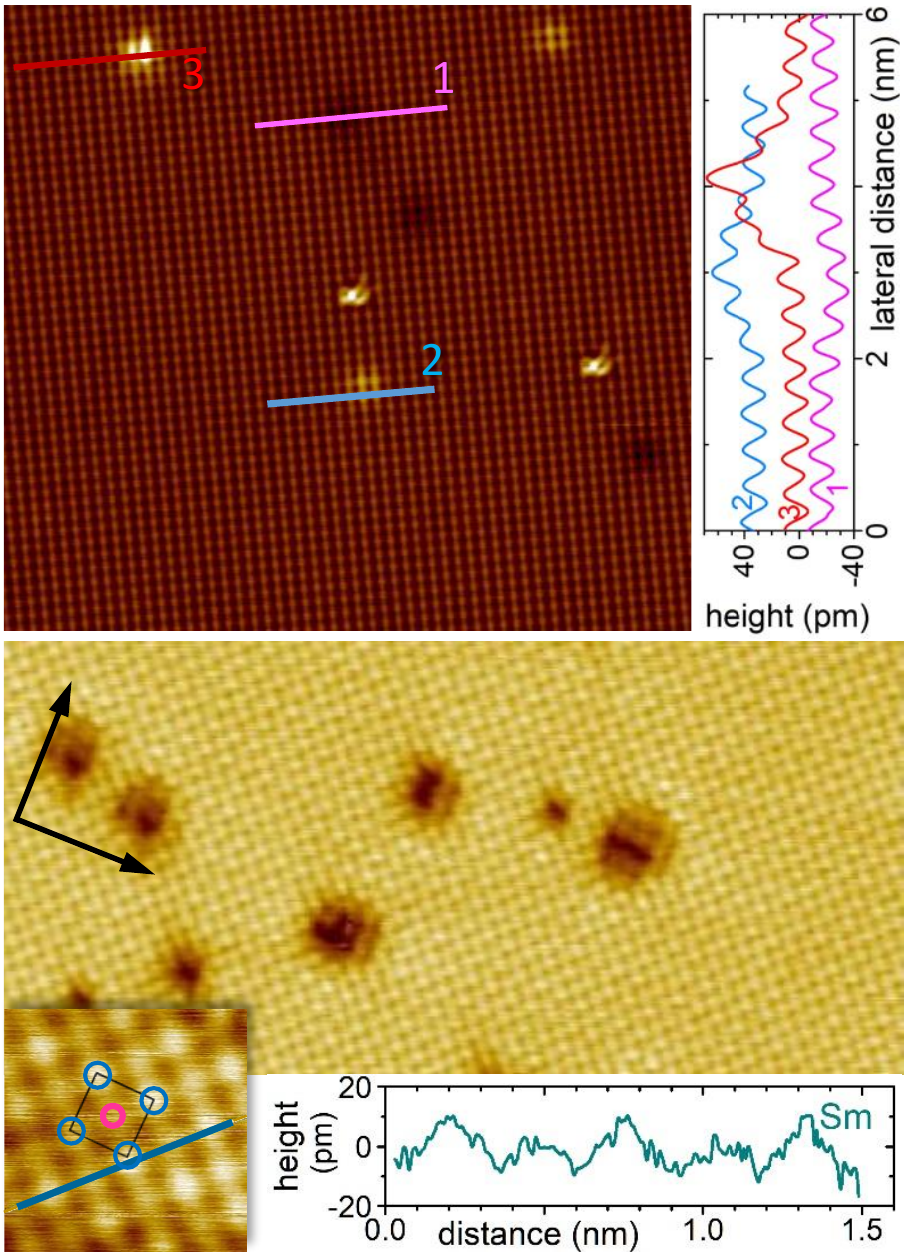}
\unitlength1cm \begin{picture}(-0.4,2)
\put(-7.2,5.2){\sffamily\bfseries\large \textcolor{white}{(a)}}
\put(-1.7,9.4){\sffamily\bfseries (b)}
\put(-7.2,2.5){\sffamily\bfseries\large (c)}
\put(-7.2,1.6){\sffamily\bfseries (d)}
\put(-3.95,0.5){\sffamily\bfseries (e)}
\end{picture}
\caption{a) SmB$_6$ topography of a B-terminated area of 22 nm $\times$ 20
nm. $V =$ 15 mV, $I_{\rm sp} =$ 0.3 nA, $T \approx$ 1.7 K. b) Height scans
along the lines shown in a). Adapted under the terms of a Creative Commons
Attribution 4.0 International License \cite{wir20}. Copyright 2021 The Authors,
published by Wiley. c) Sm-terminated area of 16 nm $\times$ 8 nm. Black arrows
indicate the $\langle$100$\rangle$ and $\langle$010$\rangle$ directions. d)
Zoomed-in area of the surface in c), 1.6 nm $\times$ 1.6 nm. A unit cell is
sketched in colors corresponding to {\bf Figure} \ref{structure}; cf.\ also
{\bf Figure} \ref{ruan14}a. e) Height scan along the line in (d). c)--e)
Adapted under the terms of a Creative Commons Attribution 4.0 International
License \cite{roe14}. Copyright 2014 The Authors, published by National
Academy of Science, USA.}
\label{Bnice}  \end{figure}
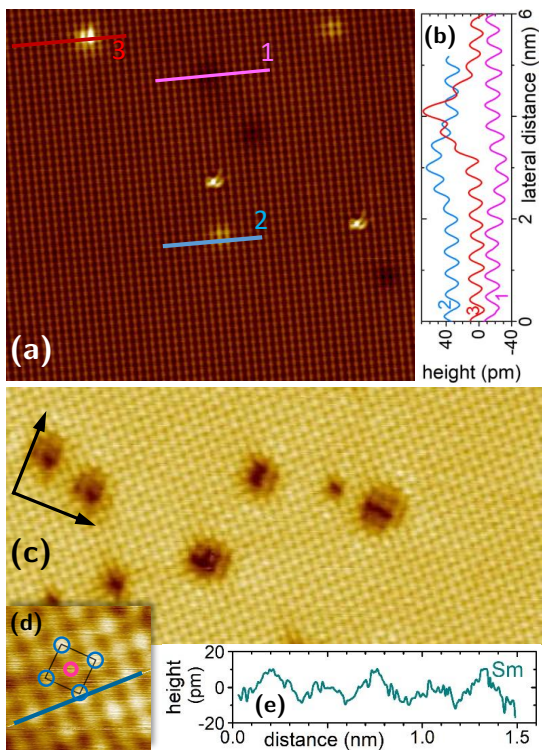
discussed below, section \ref{pureSTS}. We note that the finding of bent lines
of Sm atoms may complicate the interpretation of results which depend on
long-range order \cite{fra13}. Additional complications stem from the reported
time dependencies \cite{zhu13,ram16,zab18}. In particular, patches of $2 \times
1$ reconstructed areas observed a few days after {\it in situ} cleave could no
longer be resolved after 30 days albeit the sample was kept in UHV and below
20 K \cite{ram16}. We also wish to point out that two other reconstructions
have been reported: i) a surface with only each third line of top Sm atoms
missing was presented \cite{wir20}; ii) on a Sm-deficient sample, a new type
of reconstruction, tentatively assigned $c$($\sqrt{2} \times
3\sqrt{2}$)R45$^{\circ}$, was observed \cite{ale21}.

Surface reconstructions could pose a challenge for the study of a conducting
surface layer in SmB$_6$. As an example, the Si(111)7$\times$7 surface is
metallic, in contrast to bulk Si \cite{yoo02}. We therefore concentrate on
unreconstructed surfaces in the following.

The cubic structure of the hexaborides does not allow for a straightforward distinction of the different surface terminations, especially for atomically
flat surfaces \cite{ale21}. This becomes obvious from {\bf Figure} \ref{Bnice}a
where a nice example of a topography with corrugations spaced by $a$ is
presented, cf.\ {\bf Figure} \ref{Bnice}b. Yet, it is important to note that
these height scan lines were recorded parallel to the main
crystallographic direction $\langle$100$\rangle$ of the surface shown in
{\bf Figure} \ref{Bnice}a. In contrast, the main
crystallographic directions $\langle$100$\rangle$ and $\langle$010$\rangle$,
as indicated by black arrows, do not follow the lines of corrugations in
{\bf Figure} \ref{Bnice}c. This is better seen in {\bf Figure} \ref{Bnice}d,
which is a zoom into c), and hence, the crystallographic alignment of the
sample is identical. The height scan in d) is parallel to $\langle$110$\rangle$
and indeed, the larger protrusions are spaced by $a\sqrt{2} \approx$ 5.8 nm,
as expected. Importantly, in between smaller protrusions can be recognized. A
comparison to the crystal structure {\bf Figure} \ref{structure} suggests that
the smaller protrusions stem from the apex B$_{(1)}$-atom on a Sm-terminated
surface, cf. the unit cell indicated in {\bf Figure} \ref{Bnice}d by colors
corresponding to {\bf Figure} \ref{structure}. It should be noted that
B-terminated surfaces as in {\bf Figure} \ref{Bnice}a have been reported from
time to time \cite{yee13,sun18,her20,mat20} but Sm-terminated surfaces
analogous to {\bf Figure} \ref{Bnice}c only once \cite{ruan14}, cf.\
{\bf Figure} \ref{ruan14}a. This is consistent with our observation that the
latter can be found only extremely rarely while the former is encountered
every now and then, and conforms to findings of a x-ray absorption study
\cite{zab18}. An annealed Sm surface did not reveal the intermittent
protrusions due to remaining disorder \cite{miy17}. Compared to the B- and
Sm-terminated surfaces, the $2\times 1$ reconstruction is found somewhat more
frequently which is energetically favorable \cite{sun18,mat20}.

\begin{figure}[t]
\includegraphics*[width=0.36\textwidth]{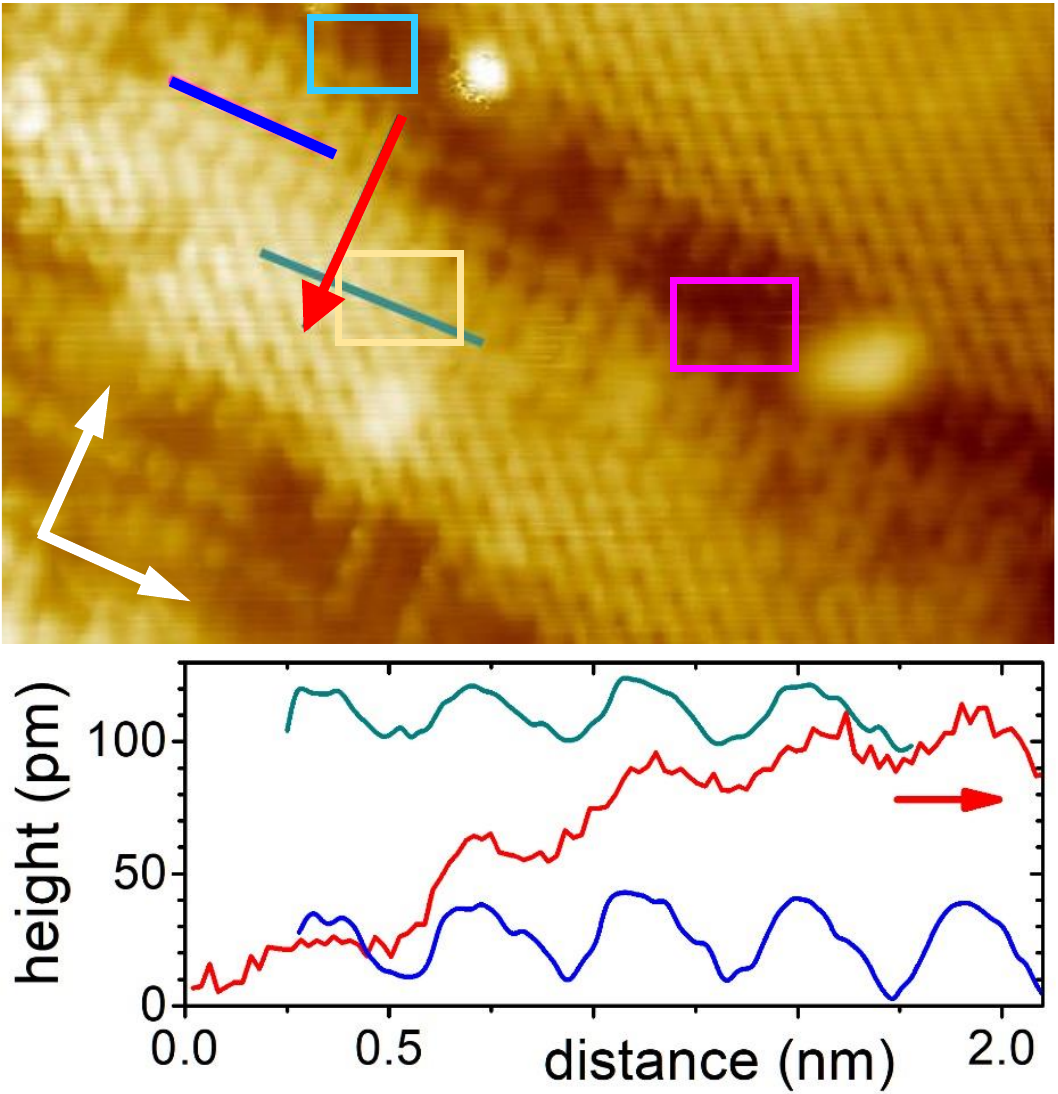}
\caption{SmB$_6$ topography of an area of 10 nm $\times$ 6 nm, which exhibits
both B- and Sm-terminated areas. $V = +0.2$ V, $I_{\rm sp} =$ 0.6 nA. Height
scans along the blue and cyan lines and red arrow are shown. White arrows
indicate the main crystallographic directions $\langle$100$\rangle$ and
$\langle$010$\rangle$. Rectangles outline areas for which spectroscopy is
reported in {\bf Figure} \ref{specSmB}a. Adapted under the terms of a Creative
Commons Attribution 4.0 International License \cite{roe14}. Copyright 2014 The
Authors, published by National Academy of Science, USA.}
\label{step}  \end{figure}

To scrutinize our assignment of surface terminations, we searched for a step
on an otherwise flat surface which exhibits both, a B- and a Sm-termination,
as presented in {\bf Figure} \ref{step}. We note that in our study of SmB$_6$, which runs for almost a decade by now, we only encountered two suitable steps
emphasizing the importance of acquiring sufficient statistics. The different
terminations are immediately apparent when keeping the crystallographic
orientation of the sample in mind, which is indicated by the white arrows.
Within the higher-up (bright) areas, e.g. below the cyan line, the
corrugations run along $\langle$110$\rangle$, while some lower (darker) areas
exhibit atomic arrangements parallel to the main axes, e.g.\ near the blue
line. The height scans along all lines reveal corrugations with periodicity
$a$. In accordance with the assignment above, the bright areas may then
correspond to Sm-terminations and the darker ones to B apex atoms. This is
nicely confirmed by the height change along the red arrow: along the arrow
(note the direction) the height increases by $\lesssim$100 pm, in good
agreement with the crystal structure {\bf Figure} \ref{structure}. The
corrugations within the B-terminated areas are even more pronounced compared
to those on Sm-terminations, cf.\ blue and cyan height scans in {\bf Figure}
\ref{step}, as one might expect.

It is important to note that our assignment of surface termination does not
directly rely on spectroscopy (of course, the density of states also enters
into height measurements). Our assignment of the B-terminated surface agrees
with other reports \cite{ruan14,sun18,her20}. Yet, the 2 $\times$ 1
reconstruction is attributed to the Sm-termination \cite{her20} primarily
based on comparison to photoemission experiments. However, in contrast to
Ref.\ \onlinecite{her20} unreconstructed areas were found as patches of some
10 nm in extent at most, 2 $\times$ 1 reconstructed areas up to some 100 nm.
Note that large-scale (several $\mu$m and beyond) STM of atomically flat areas
have not been reported so far, including \cite{her20}, in line with predictions
\cite{sch18}. It is therefore highly unlikely that identical areas can be
investigated by STM and photoemission, rather the latter most likely integrates
over differently terminated areas \cite{mat20}, possibly including rough ones.
In addition, no notion is provided \cite{her20} as to why a Sm-terminated
surface should form a superstructure. We will discuss differences in the STS
spectra on and off the Sm on 2 $\times$ 1 reconstructed surfaces below; such
a comparison is missing in Ref.\ \onlinecite{her20}. We note that the 2
$\times$ 1 reconstruction should be considered a half-Sm termination
\cite{pir20} which emphasizes the presence of Sm at the surface.

The assignment of Sm- and B-terminated, unreconstructed surfaces here differs
from those in Refs. \onlinecite{pir20,mat20}. The latter are primarily based
on band bending effects. It is argued that Sm atoms in the topmost layer
accumulate excess electrons due to the missing B layer on top \cite{mat20}.
This argument has been discussed in detail in Ref.\ \onlinecite{all16} along
with the counteracting effect of reduced hybridization at the surface (also
caused by the reduced number of neighbouring atoms). This reduced hybridization
decreases the Kondo temperature at the surface, $T_{\rm K}^{\rm S}$, relative
to the bulk and implies a shift of the Sm valence at the surface toward
Sm$^{3+}$. Such a shift has been observed in x-ray absorption spectroscopy
(XAS) \cite{lut16,uts17,zab18} and the reduced $T_{\rm K}^{\rm S}$
\cite{jiao16} conforms to the Kondo breakdown scenario \cite{ert16,pet16}. In
consequence, an assignment of the surface termination without relying on
band-bending effects appears preferable.

Further support for our assignment of terminations stems from comparison
to other hexaborides, section \ref{otherSTM}. It is also important to note
that, for a comprehensive assignment of surface terminations, \emph{all}
different terminations should be observed and compared. Differences in STM
between flux-grown and floating-zone grown samples have not been detailed.
There are several reports of surface relaxation \cite{jkim14,sch18}. The
corresponding atomic displacements in the direction of the bulk are difficult
to observe by STM.

So far, efforts on cleaving (110) surfaces did not reveal atomically flat
surface areas \cite{roe14}, again with possible consequences for the
interpretation of ARPES results \cite{den16}.

\subsection{RE substitutions and defects on SmB$_6$}
The hexaborides can not only be synthesized with a variety of elements on
the RE site, also numerous substitutions on this site have been reported
\cite{ino21}. Such an additional degree of freedom can give rise to new and
intriguing phases and phenomena as, e.g., in the case of Ce$_{1-x}$RE$_x$B$_6$
\cite{nik18}. It also allows for detailed substitution studies, including a
change of the electron count by playing with the ratio of divalent vs.
trivalent elements, and enabling element-specific measurements like
electron-spin resonance (ESR).

In case of SmB$_6$, the fate of the topological surface states upon Sm
substitution is of specific interest. In particular, the breaking of
time-reversal symmetry due to magnetic substitution can be considered a valid
route towards establishing a non-trivial topology in SmB$_6$. With the
proposal of non-trivial topology in EuB$_6$ \cite{nie20}, the series
Sm$_{1-x}$Eu$_x$B$_6$ is certainly of topical interest \cite{geb70,yeo12,
mia21}. Other magnetic substitutions include Gd \cite{geb70,kim14,fuh17,
jiao18,pir20}, Ce \cite{hat20} and Fe \cite{alt05,aki17,pir20}.

\begin{figure}[t]
\includegraphics*[width=0.44\textwidth]{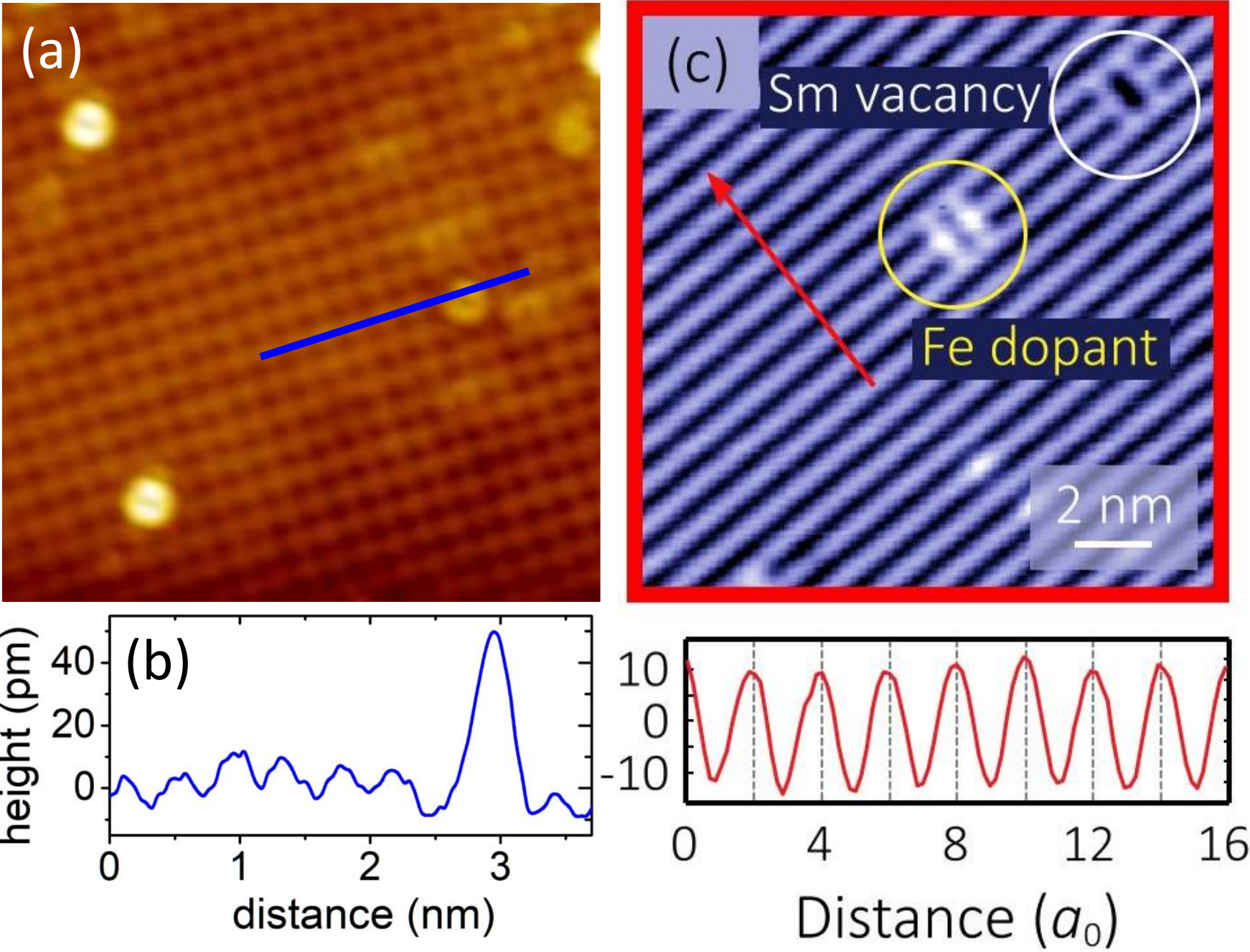}
\caption{a) Topography of Sm$_{0.995}$Gd$_{0.005}$B$_6$ of area 8 nm
$\times$ 8 nm. $V = +30$ mV, $I_{\rm sp} =$ 0.1 nA, $T =$ 0.35 K. b) Height
profile along the blue line in a). Adapted under the terms of a Creative
Commons Attribution 4.0 International License \cite{jiao18}. Copyright 2018
The Authors, published by AAAS. c) STM topography of lightly Fe-doped SmB$_6$
and corresponding height profile. c) Reproduced with permission \cite{mat20}. Copyright 2020, American Physical Society.} \label{dopants}
\end{figure}

Introducing substituents into SmB$_6$ changes the cleavage properties. For our
cases of 0.5\% and 3\% Gd as well as 3\% Y substituted samples, less force was
required for cleaving and atomically flat areas were found more easily, with
the vast majority of the investigated surface areas being reconstructed \cite{jiao18}. As for pure SmB$_6$, unreconstructed surface areas were
encountered only rarely. One such area on Sm$_{0.995}$Gd$_{0.005}$B$_6$ is
shown in {\bf Figure} \ref{dopants}a. The number of defects counted on all our
surfaces of substituted samples was typically slightly less than expected if
all defects are caused by substituents. Nonetheless, the comparison to
topographies on pure SmB$_6$ and the spectroscopic results on substituted
materials presented in section \ref{dopedSTS}, in particular the changing tip
properties on Gd-substituted SmB$_6$, suggest these defects to be caused by
substituents \cite{jiao18}. For comparison, {\bf Figure} \ref{dopants}c
\begin{figure}[b]
\includegraphics*[width=0.46\textwidth]{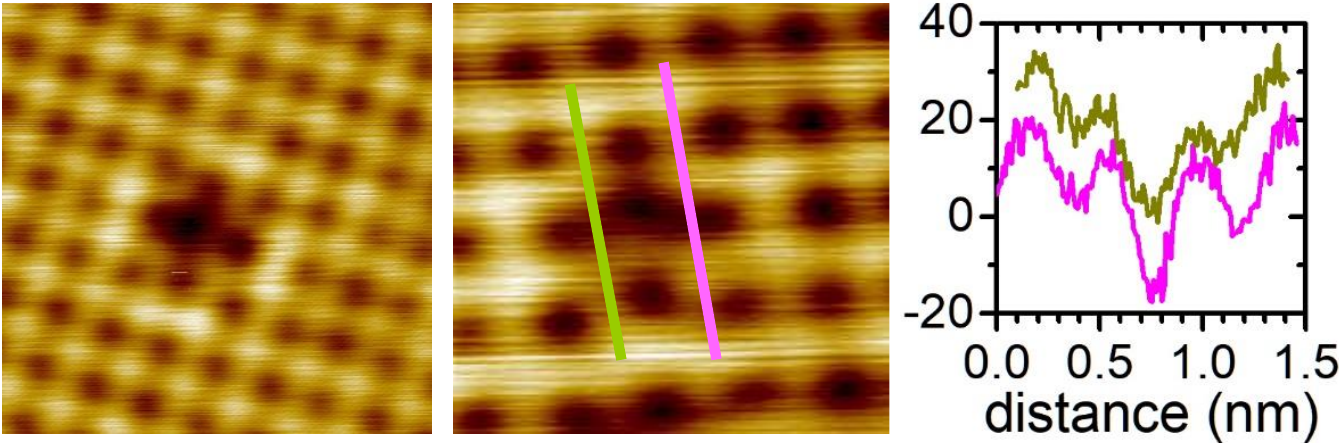}
\unitlength1cm \begin{picture}(-0.4,2)
\put(-8.3,2.3){\sffamily\bfseries\large \textcolor{white}{(a)}}
\put(-5.5,2.3){\sffamily\bfseries\large \textcolor{white}{(b)}}
\put(-2.16,1.0){\sffamily\bfseries (c)}
\end{picture}
\caption{Topography of a) pure ($3 \times 3$ nm$^2$, $V = 0.2$ V,
$I_{\rm sp} =$ 0.6 nA) and b) slightly Sm-deficient SmB$_6$ ($2 \times 2$
nm$^2$, $V = 70$ mV, $I_{\rm sp} =$ 0.3 nA) showing cross-like defects. c)
Height profile (in pm) along the lines in b). Adapted under the terms of a
Creative Commons Attribution 4.0 International License \cite{wir20}.
Copyright 2021 The Authors, published by Wiley.} \label{cross}
\end{figure}
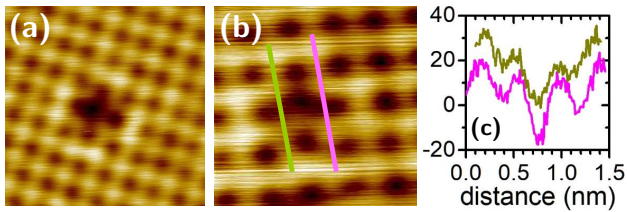
reproduces a reconstructed surface area of a Fe-substituted sample as observed
in Ref.\ \onlinecite{mat20}.

Investigation of defects can be helpful in addressing the surface termination.
Therefore, we consider Sm-deficient samples \cite{ale21}. However, the defects
can be quite different depending on the SmB$_6$ sample growth method
\cite{phe16,gab16,val16}: Compared to the floating-zone grown samples, those
grown in Al flux tend to show small crystalline Al inclusions \cite{tho19},
but show a remarkable robustness against Sm deficiency \cite{eo19,li20,ale21}.
As an example, a sample with 10\% Sm deficiency in the flux turned out with
nominal composition Sm$_{0.983}$B$_6$ \cite{ale21}. On this sample, a
characteristic type of cross-like defects was found, which was extremely rare
on surfaces of pure SmB$_6$, see {\bf Figures} \ref{cross}a and b, and hence,
was attributed to a missing Sm atom below a B-terminated surface. The four B
octahedra in the surface adjacent to the empty Sm site can then slightly move
into the bulk, which explains their slightly lesser height ($\sim$10 pm) seen
in the height profiles {\bf Figure} \ref{cross}c. This result supports our
interpretation of the involved surfaces as B-terminated.

\subsection{Comparison to other hexaborides}
\begin{figure}[t]
\includegraphics*[width=0.32\textwidth]{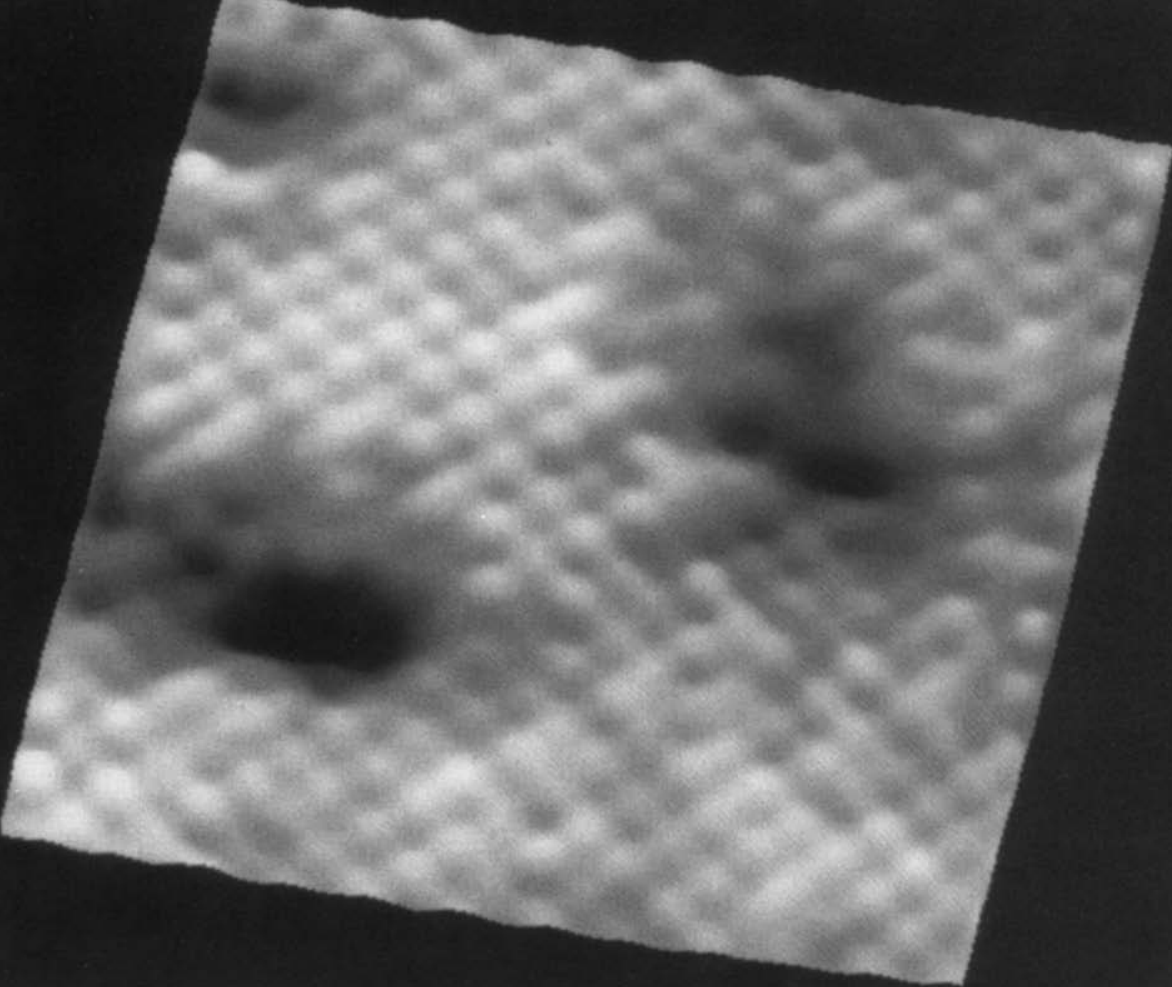}
\caption{La-terminated surface of LaB$_6$ of area 7.7 nm $\times$ 8.2 nm.
Reproduced with permission \cite{ozc92}. Copyright 1992, Elsevier.}
\label{LaB6}  \end{figure}

\labelsubsec{otherSTM}
Hexaborides have been investigated by STM very early on. In {\bf Figure}
\ref{LaB6} a La-terminated surface of LaB$_6$ is reproduced from Ref.\
\onlinecite{ozc92}. There are approximately 10\% La atoms missing. The defects
consist either of single atom sites or of clusters of more than 15 missing
atoms \cite{ozc92}. The overall appearance of this surface is remarkably
similar to the Sm-terminated surface of {\bf Figure} \ref{Bnice}c reinforcing
our claim of a Sm termination for this surface. The atomic corrugation in
{\bf Figure} \ref{LaB6} was specified to approximately 0.2 \AA\ and atomic
esolution was reported to be a rare event, both in line with a more recent
report on LaB$_6$ \cite{buc19} and our observations on SmB$_6$. In addition,
$2 \times 1$ reconstructions on LaB$_6$ were shown \cite{buc19} to closely
resemble those on SmB$_6$, {\bf Figures} \ref{topos}c and e.

\begin{figure}[t]
\includegraphics*[width=0.42\textwidth]{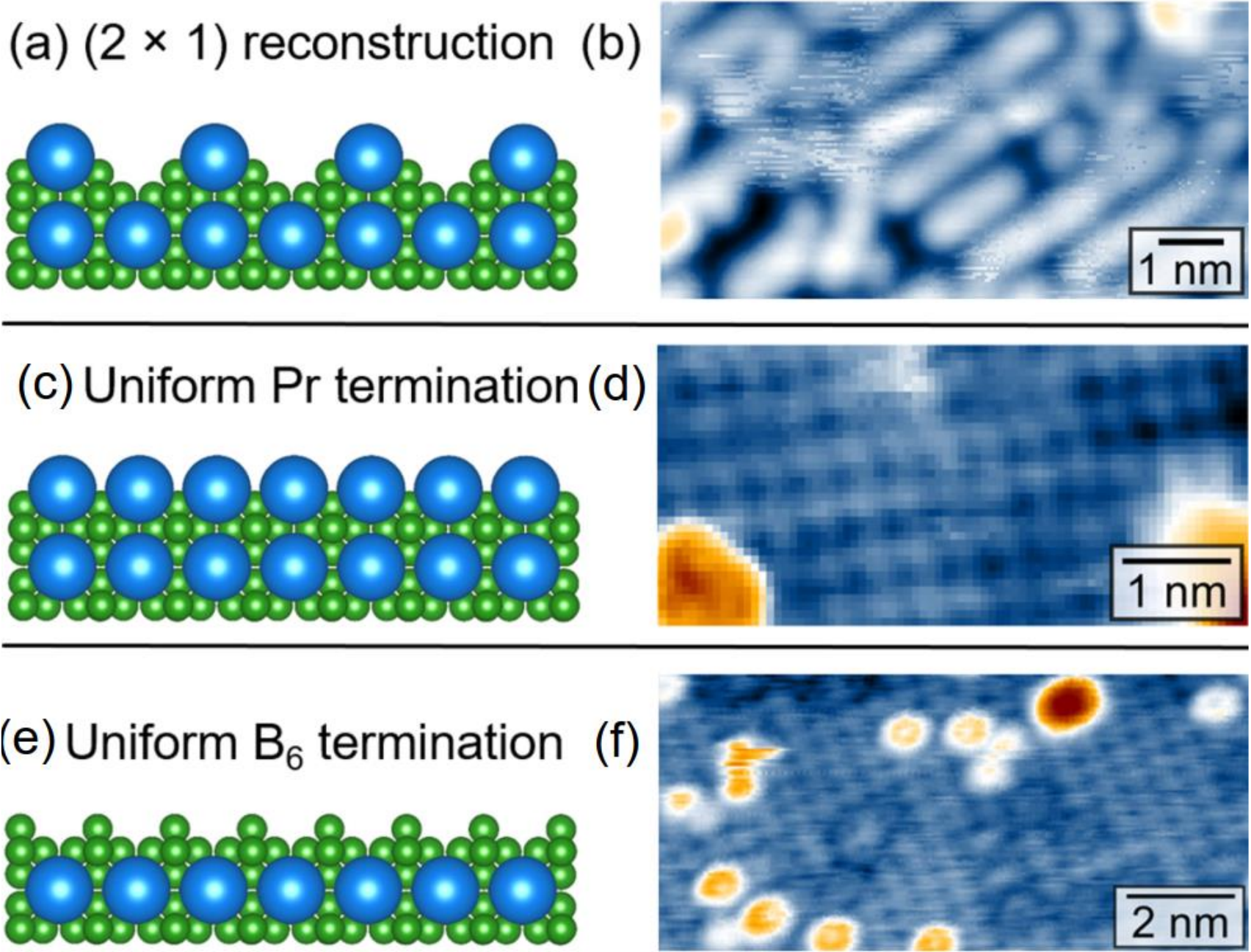}
\caption{Differently terminated surfaces of PrB$_6$ (for detailed tunneling
parameters see \cite{buc20}). Reproduced with permission \cite{buc20}.
Copyright 2020, American Physical Society.}  \label{PrB6}
\end{figure}

The afore-mentioned terminations, namely the $2 \times 1$ reconstruction, the
RE- and B-terminated surface, have also been observed on {\it in situ} cleaved
PrB$_6$ \cite{buc20}, cf. {\bf Figure} \ref{PrB6}. Again, atomically flat areas
extend only over a few (nm)$^2$. As in the case of SmB$_6$, the $2\times 1$
reconstructed surface is ascribed to parallel rows of Pr ions on top of a
B$_6$ lattice ({\bf Figures} \ref{PrB6}a and b), which avoids the buildup of an
electrical potential on otherwise polar surfaces \cite{buc20}. On some of the
$1\times 1$ surfaces, a peak is observed at around $-0.7$ V in tunneling
spectroscopy. Comparison to PES and DFT results suggests that this feature is
related to 4$f$ states and hence, the corresponding surfaces are assigned as
Pr terminated \cite{buc20}, {\bf Figures} \ref{PrB6}c and d. Unfortunately, the
intermittent B-apex atoms cannot be resolved in the topography, {\bf Figure}
\ref{PrB6}d. Surfaces, which do not feature such a peak in spectroscopy, are
considered B-terminated, {\bf Figures} \ref{PrB6}e and f.

A small patch of an atomically flat surface of CeB$_6$ (100) was reported in
Ref.\ \onlinecite{ena-the} without further assignment of its termination; yet,
a $2\times 1$ reconstruction on this particular area appears unlikely.

\begin{figure}[t]
\includegraphics*[width=0.44\textwidth]{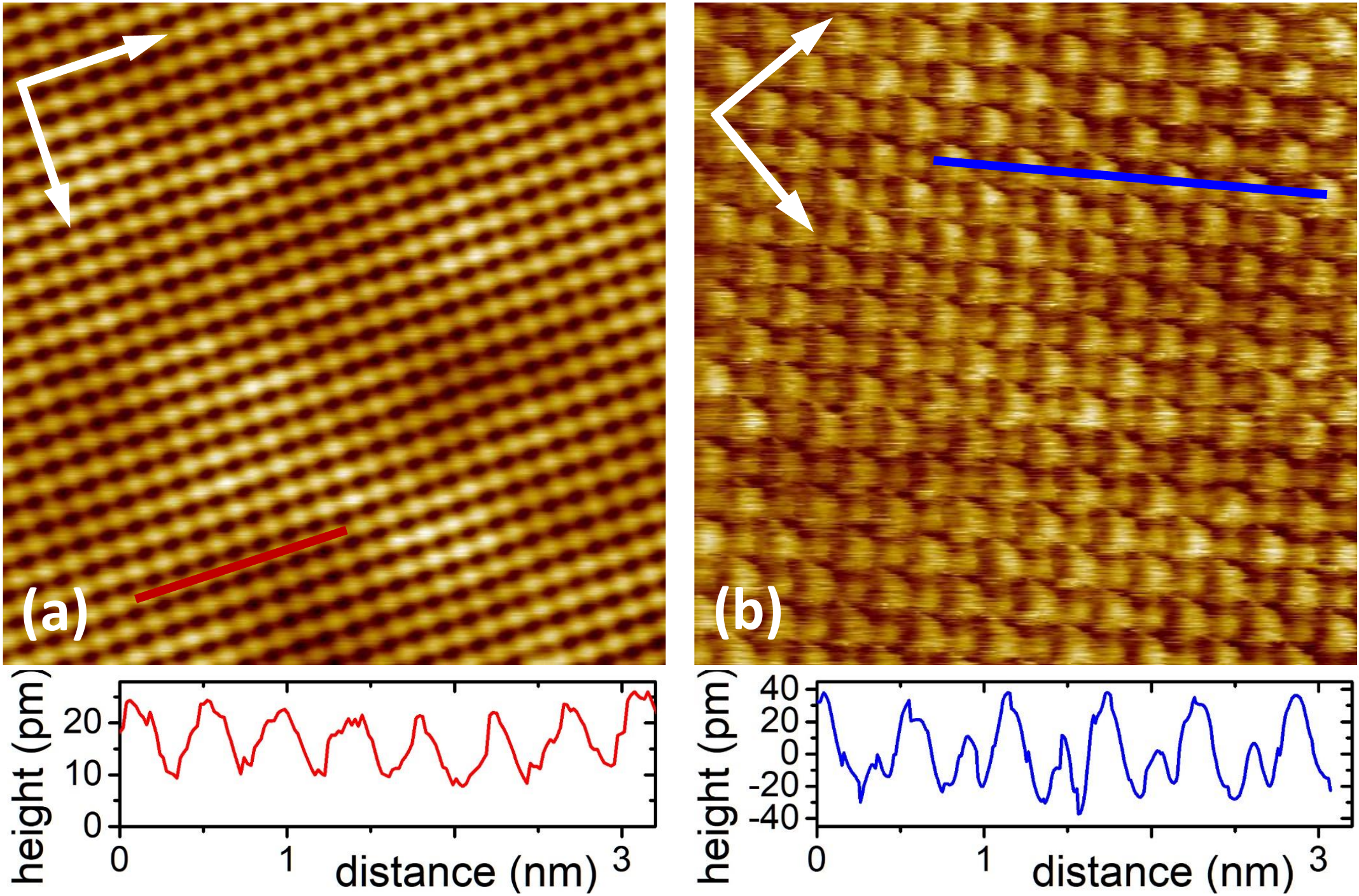}
\caption{Different surfaces of EuB$_6$. a) B-terminated area of (10 nm)$^2$
and corresponding height profile. White arrows indicate the main
crystallographic axes. b) Eu-terminated area of (5 nm)$^2$. The height
profile along the $\langle 110 \rangle$ direction clearly reveals intermittent
corrugations. $V = +0.2$ V, $I_{\rm sp} =$ 0.5 nA, $T \approx$ 6 K. Adapted
under the terms of a Creative Commons Attribution 4.0 International License
\cite{roe20}. Copyright 2020 The Authors, published by American Physical
Society.}  \label{EuB6}
\end{figure}

Similarly, for EuB$_6$ a $2\times 1$ surface reconstruction has not been
reported so far. However, both Eu- and B-terminated surfaces of up to some
ten nm in extent were observed \cite{poh18,wir20,roe20}, again after extensive
search. Note the striking similarity between EuB$_6$, {\bf Figure} \ref{EuB6},
and SmB$_6$, {\bf Figure} \ref{Bnice}, backing our findings. B-terminated areas
exhibit clear corrugations of distance $a$ along the main crystallographic
directions, {\bf Figure} \ref{EuB6}a. Some of such areas were found to be free
of any visible defect on a ten nm scale allowing for further detailed
investigation \cite{poh18,roe20}. On the surface of {\bf Figure} \ref{EuB6}b,
the bright spots of distance $a$ form a square arrangement along the main
crystallographic directions, with smaller protrusions at the square centers.
This is seen in the height profile along $\langle 110 \rangle$, which closely
resembles {\bf Figure} \ref{Bnice}e. Therefore, these smaller protrusions are
likely caused by the apex atom of the B-octahedra between Eu atoms. It should
be noted that, again, an atomically flat surface exposing the
B$_{(2)}$-B$_{(5)}$ atoms, i.e. intra-octahedral bond breaking, would result
in a regular doughnut pattern as discussed for SmB$_6$.

Impurities at the surface of EuB$_6$ can give rise to circular, large-amplitude
oscillations of the local DOS around the impurity \cite{roe20}. The
hybridization of a spin-split surface state with bulk electronic states leads
to a resonance which explains the remarkable response even in real-space
topography.

For PrB$_6$, stripe-like, or even more complex, corrugations were observed for
small bias voltages $|V| \leq$ 0.4~V \cite{buc20}. While on very rare
occasions similar complex corrugations were found on EuB$_6$, a relation to
small bias voltages (or close tip-sample distances) could not be established.
In this respect, the very close tip-sample distance used in {\bf Figure}
\ref{EuB6} ($V = +0.2$ V, $I_{\rm sp} =$ 0.5 nA) should be noted.

\section{Tunneling Spectroscopy on hexaborides}
\subsection{Pristine SmB$_6$}
\labelsubsec{pureSTS}
The interest in STS stems from the fact that the tunneling conductance
$g(V,\boldsymbol{r}) = {\rm d}I(V,\boldsymbol{r})/$d$V$ is, within simplifying
approximations, proportional to the local DOS and can, in principle, provide
insight into aspects of the electronic Green’s function \cite{kir20}. Early
reports of tunneling spectroscopy on hexaborides focused on the formation of
the zero-bias gap upon decreasing temperature \cite{gue82,bat07,ams98}. Point
contact spectroscopy in the tunneling regime of SmB$_6$ indicated a substantial
zero-bias DOS even at $T =$ 0.1 K \cite{fla01}. The concept of topologically
protected surface states, however, was applied to SmB$_6$ only about a decade
later \cite{dze10,tak11} fueling a renewed interest in spectroscopy on this
material, in particular STS and ARPES.

\begin{figure}[t]
\includegraphics*[width=0.46\textwidth]{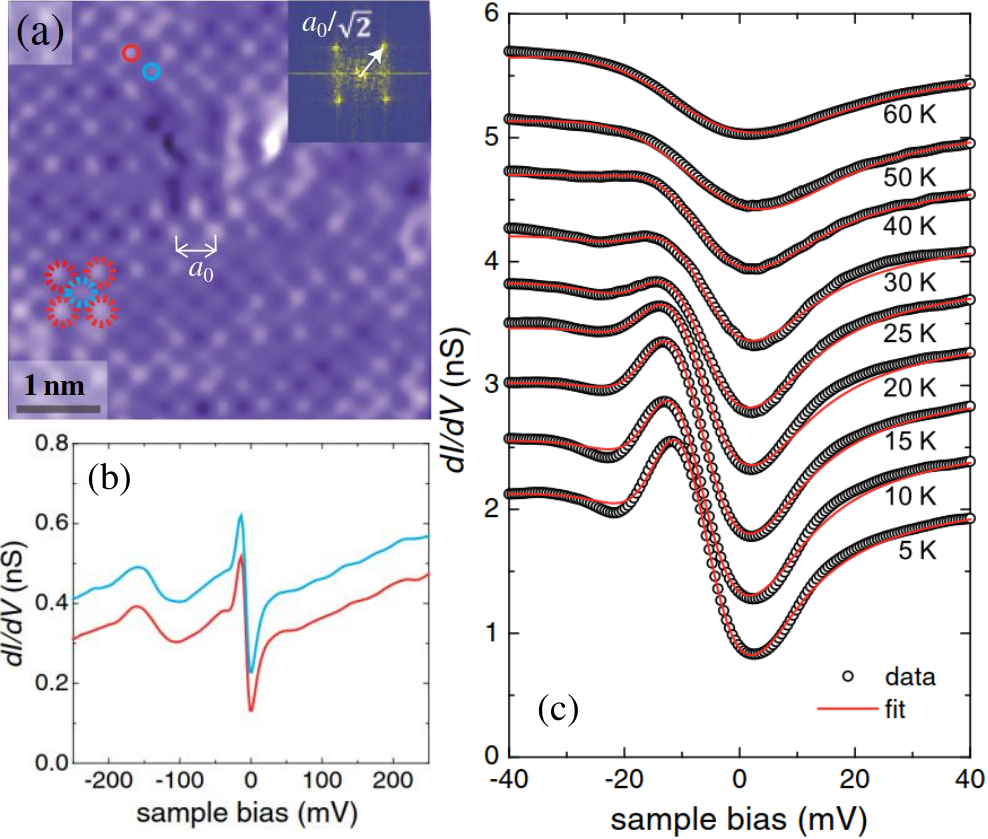}
\caption{a) Sm-terminated surface. The surface atomic structure is indicated
by dashed circles, see also {\bf Figure} \ref{Bnice}d. The upper right inset
shows the Fourier transform. b) d$I/$d$V$ spectra at $T =$ 5 K and the
positions marked by solid circles of corresponding color in a). c)
$T$-dependent d$I/$d$V$ spectra obtained on a disordered, so-called doughnut
surface. The fit is based on the co-tunneling model plus, for $T \leq$ 40 K, a
Gaussian peak. Reproduced with permission \cite{ruan14}. Copyright 2014,
American Physical Society.} \label{ruan14}
\end{figure}

For an evaluation of Kondo insulators and their suitability as strong
three-dimensional topological insulators, the bulk band structure has to be
known. In particular, the Kondo hybridization evolving at low temperature
was studied revealing a direct hybridization gap of order 20 meV \cite{gor99,
zha13,neu13,fra13,nxu13,den13,ruan14,roe14}, see {\bf Figure} \ref{ruan14}.
The gap was shown \cite{roe14} to close up with a logarithmic $T$-dependence,
which is a hallmark of Kondo physics \cite{aep92,cos00,wir16}. The Kondo
hybridization, however, complicates an interpretation of the STS spectra in
terms of a local DOS: In addition to tunneling into the conduction band there
is also the possibility to tunnel into 4$f$ quasiparticle states giving rise to
a co-tunneling phenomenon \cite{fan61,mal09,fig10,woe10}. In a simple picture,
the tunneling conductance can be described as \cite{sch00}
\begin{equation}
\frac{dI(V)}{dV} \propto \frac{(\epsilon + q)^2}{\epsilon^2 + 1} \; , \qquad
\epsilon = \frac{2(eV - E_0)}{\Gamma} \: , \label{fanf}
\end{equation}
where $\Gamma$ and $E_0$ denote the resonance width and its position in energy,
respectively. The asymmetry parameter $q$ is related to the probability of
tunneling into the 4$f$ quasiparticle states relative to the conduction band,
as well as the particle-hole asymmetry of the conduction band \cite{fig10}.
Equation \ref{fanf} could very nicely describe $dI/dV$-data obtained on
clean B-terminated surfaces \cite{roe14,jiao16}; somewhat more detailed models
\cite{mal09,fig10} were also successfully applied \cite{zha13,yee13,ruan14,
jiao16}, but require more open parameters to be adjusted. This is shown in
{\bf Figure} \ref{ruan14}c where the spectra above 40 K were fit by the mere
co-tunneling model, while at lower $T$ an additional Gaussian was introduced
\cite{ruan14}, as demonstrated for other 4$f$ Kondo systems \cite{ern11,ayn12}.
An additional contribution can also be seen, e.g., in Figure S13C of Ref.\
\onlinecite{pir20}. The hybridization gap obtained ranged between 14--20~meV
\cite{ams98,zha13,roe14,jiao16,pir20}, in line with results of other
measurements \cite{fla01,jia13,den14,neu13,fra13,nxu13,nxu14,val16,ara16,sun17,
zha18}, albeit smaller gaps were also reported \cite{ruan14,hla18,sun18}. The
observation of co-tunneling phenomena also on reconstructed \cite{yee13} or
less well-defined surfaces is consistent with the bulk nature of the
underlying Kondo effect.

\begin{figure}[t]
\includegraphics*[width=0.36\textwidth]{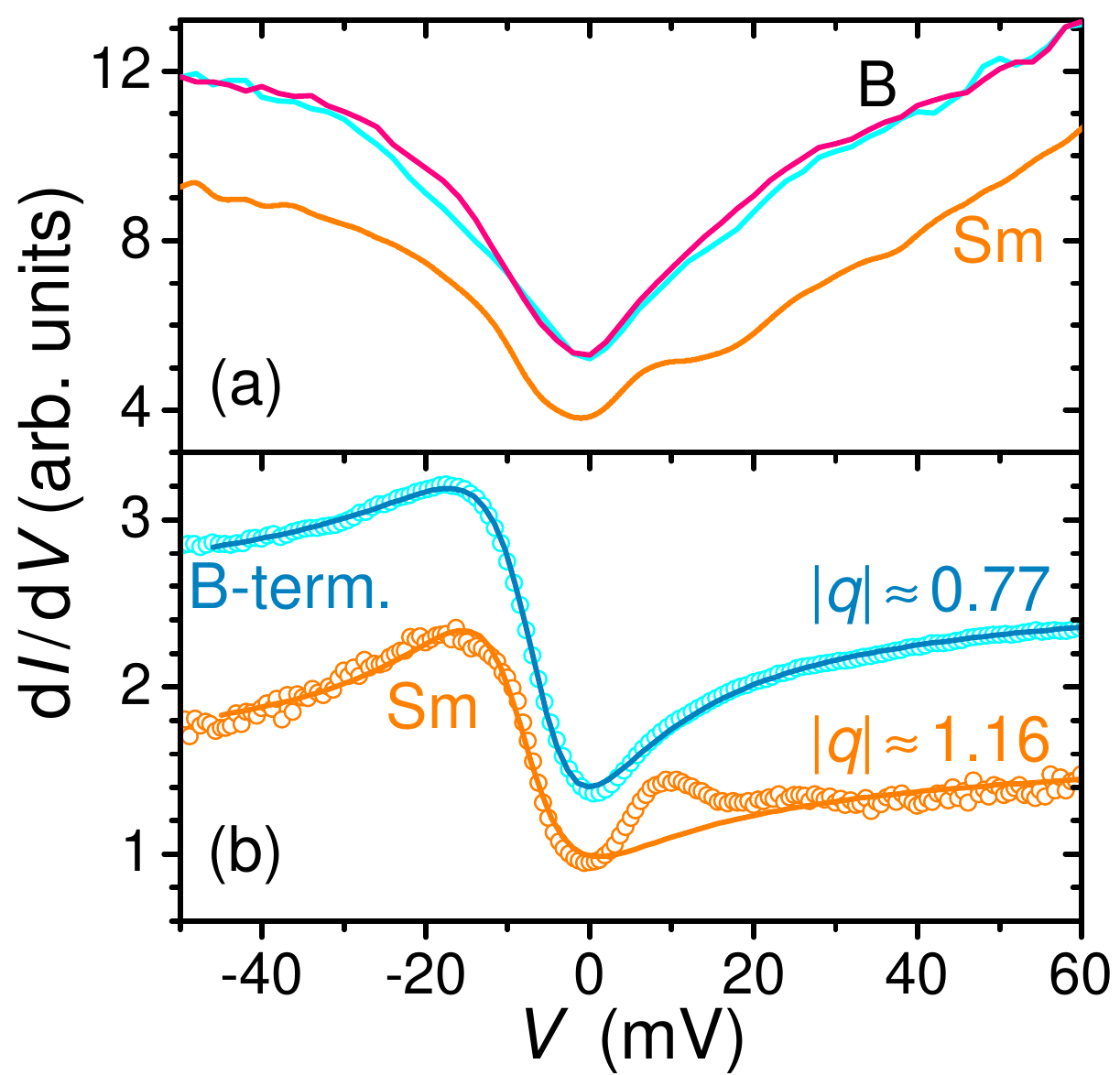}
\caption{a) Tunneling spectroscopy at $T =$ 4.6 K obtained within areas
marked by rectangles of corresponding colors in {\bf Figure} \ref{step}. $V =$
0.2 V, $I_{\rm sp} =$ 0.6 nA, $V_{\rm mod} =$ 1 mV. Adapted under the terms of
a Creative Commons Attribution 4.0 International License \cite{wir20}.
Copyright 2021 The Authors, published by Wiley. b) d$I/$d$V$ spectra (circles)
obtained on extended, clean Sm- or B-terminated areas as shown in {\bf Figures}
\ref{Bnice}a) and c). Lines are fits to eq.\ \ref{fanf}, $q$-values are
indicated. Spectra on B-terminated surfaces are offset in a) and b). Adapted
under the terms of a Creative Commons Attribution 4.0 International License
\cite{roe14}. Copyright 2014 The Authors, published by National Academy of
Science, USA.} \label{specSmB}
\end{figure}

Before discussing the d$I/$d$V$-spectra measured to lowest temperatures we
compare spectra obtained at $T =$ 4.6 K on differently terminated surfaces.
{\bf Figure} \ref{specSmB}a compares the STS data for the marked areas in
{\bf Figure} \ref{step}, which are up to 1 nm in extent. The only significant
difference is a small hump at $+10$ mV in case of the Sm-terminated surface
patch. Spectra obtained on much larger, atomically flat and clean areas (as
those shown in {\bf Figures} \ref{Bnice}) are presented in  {\bf Figure}
\ref{specSmB}b. Both curves can be fit by Equation \ref{fanf} yielding
$\Gamma \approx$ 16.5 mV, in line with the hybridization gap width. The
$q$-values 0.77/1.16 are smaller/larger than unity for B-/Sm-terminated areas
consistent with predominant tunneling into the conduction/4$f$ band. Again,
there is a peak at $+10$ mV on the Sm surface on top of the fit suggesting an
origin beyond the co-tunneling phenomenon. This peak, also adumbrated in
{\bf Figure} \ref{ruan14}b and Ref.\ \onlinecite{fla01}, may be related to an
excitation resulting from the hybridization of the 4$f$-electron wave function
\cite{ale93,ale95} as seen in inelastic neutron scattering, consistent with
an assignment to Sm-terminated surfaces.

There is a clear temperature progression of the additional peak evolving upon
lowering $T$ below 40 K, {\bf Figure} \ref{ruan14}c, see also
\cite{zha13,yee13}. Indeed, as summarized in Refs.\ \onlinecite{den13,li20},
for most samples temperatures below about 3~K suffice to access their
compelling low-temperature properties, but in some cases $T \lesssim$ 1 K is
required \cite{den13,kim19}. Apparently, there are also differences between
flux-grown and floating-zone grown samples in the very low-$T$ regime, as seen
in specific heat and quantum oscillation measurements \cite{har18,tho19}. In
consequence, STS was also conducted down to such low $T$
\cite{jiao16,sun18,pir20}.

\begin{figure}[t]
\includegraphics*[width=0.40\textwidth]{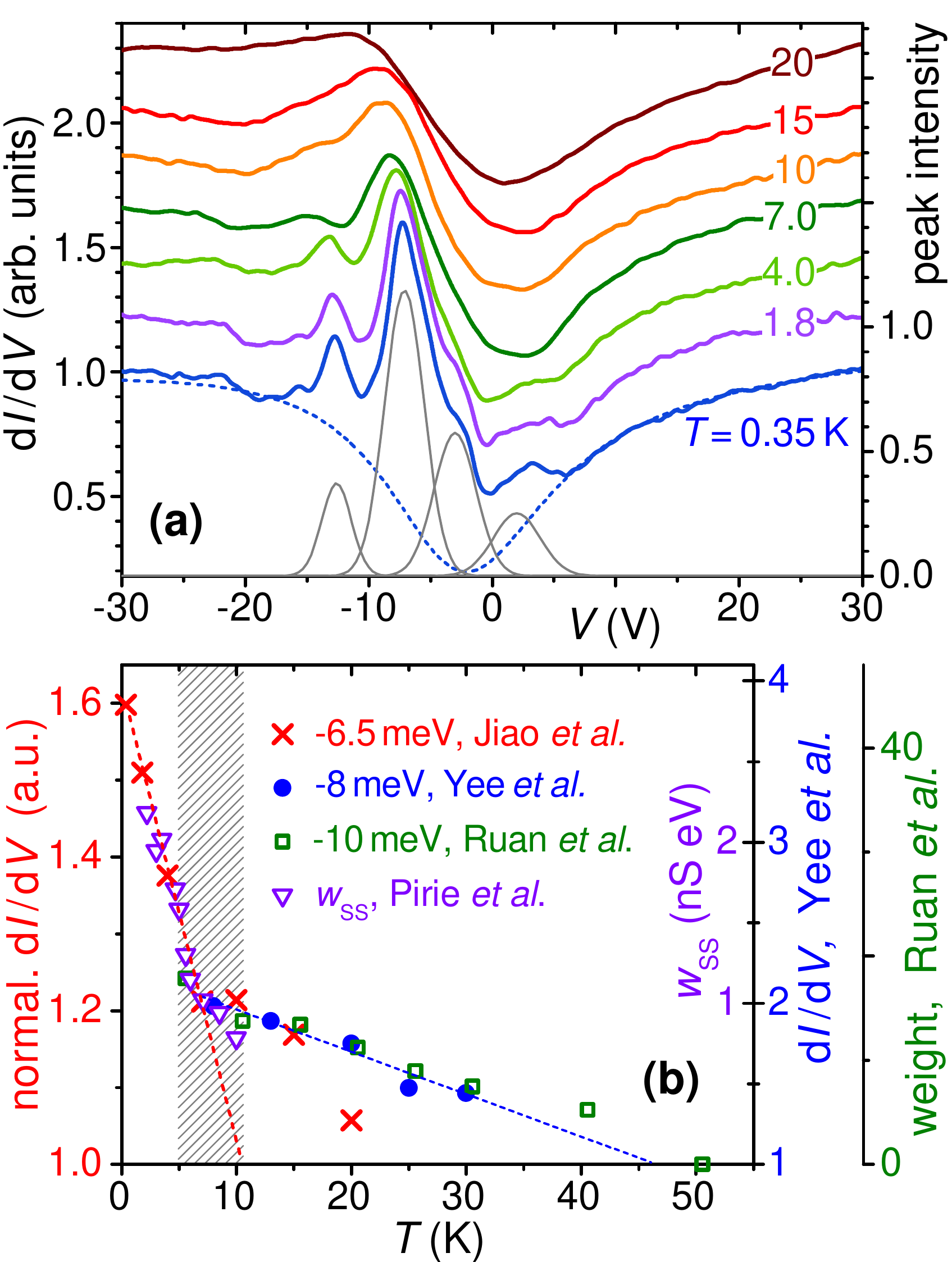}
\caption{a) Low-temperature tunneling spectroscopy on a B-terminated surface
similar to {\bf Figure} \ref{Bnice}a. The evolution from Fano-type behavior at
20 K to individually resolved peaks is evident (curves are offset for clarity).
Data at $T = 0.35$ K can well be fit by a Fano background (blue dashed line)
plus 4 peaks (gray). $V =$ 30 mV, $I_{\rm sp} =$ 100 pA, $V_{\rm mod} =$ 0.3
mV. b) $T$-evolution of the peak at $-6.5$ meV in comparison to Yee et al.\
\cite{yee13}, Ruan et al.\ \cite{ruan14} and Pirie et al.\ \cite{pir20},
$w_{\rm SS}$ - spectral weight of surface states. Adapted under the terms of a Creative Commons Attribution 4.0 International License \cite{jiao16}. Copyright
2016 The Authors, published by Springer Nature.}  \label{lowT}
\end{figure}

{\bf Figure} \ref{lowT}a reproduces $dI/dV$-data down to $T =$ 0.35 K
\cite{jiao16} taken on a B-terminated surface and far away from any defect,
cf.\ {\bf Figure} \ref{Bnice}a. The behavior is reminiscent of the one in {\bf
Figure} \ref{ruan14}c: The data can nicely be fit by Equation \ref{fanf}
at 20 K, while at lower temperature additional contributions have to be
considered. This is best seen for $dI/dV$-data at 0.35 K where at least four
Gaussian peaks (gray lines and right scale in {\bf Figure} \ref{lowT}a) can be
fitted in addition to a Fano-type background (blue dashed line) \cite{jiao16}.
Note that Gaussians were used here for simplicity as the experimental data do
not allow to distinguish between different peak shapes (Gaussian lineshapes are
usually associated with randomness, while a Lorentzian lineshape with a
relaxation phenomenon). The peak energies are approximately $-13$, $-6.5$, $-3$
and $+2$ meV. These energies do not change up to 4~K, above which temperature
specifically the latter two peaks around zero bias cannot be resolved anymore.
Along with the fact that these peaks do not change their positions in energy
even close to defects, this suggests that they contain bulk contributions. In
this respect it is reassuring that very similar energy positions of these peaks
have been observed on less clean surfaces of floating-zone grown samples at
$T =$ 0.1~K \cite{sun18}. Again, this is in line with these energy positions
being determined by the bulk properties of SmB$_6$, which apparently are very
similar for differently (flux vs.\ floating-zone) grown samples. Bulk
contributions would also explain why the main peak at $-6.5$ meV can even be
observed on reconstructed surfaces \cite{yee13,ruan14,roe14}.

Having established a relation of the energies of the main peaks with the
SmB$_6$ bulk properties, one may attempt to compare the spectra at lowest $T$
to the band structure, {\bf Figure} \ref{bandstruc}. The clear loss of spectral
weight at energies $e|V| \lesssim$ 20~meV marks the hybridization gap, as seen
by a number of spectroscopic tools and is discussed above. The most pronounced
peak at $-6.5$ meV (also observed in Ref.\ \onlinecite{yee13}) is very likely
related to the weakly dispersive $\Gamma_8^{(1)}$ band, which is strongly
renormalized. Its position in energy indicates that this peak resides inside
the (indirect) Kondo hybridization gap and therefore, it is often referred to
as in-gap state \cite{nyh97,gab99,slu00,gab01,jia13,nxu16,miy17,sun18}. As will
be argued below, its weight likely contains contributions beyond a simple
co-tunneling model. This peak certainly manifests the properties of the smaller
gap of a few meV observed in studies of thermal activation energies
\cite{men69,all80,mol82,eo19,fla01,jiao18,ale21b}, which is understood in terms
of an indirect bulk gap \cite{coo95,ris00b,chen15}, cf.\ section \ref{TKISmB6}. Such an interpretation is in line with Ref.\ \onlinecite{mal09} and supported
by the peak's temperature dependence.

The peak located at around $-13$ meV may then be related to the second
$\Gamma_8$ band \cite{lu13,jkim14,kan15,sun18}, which is brought about by the
crystal field splitting the $\Gamma_8$ quartet into two doublets away from the
$\Gamma$ point. These states hybridize only lightly. In comparison to the more
strongly hybridized states discussed above, these localized $f$-states may give
rise to only small features in the tunneling spectra \cite{ram16}. Such an
interpretation is consistent with the insensitivity of this peak to impurities
at the surface or applied magnetic fields up to 12 T \cite{jiao16} and with a
small $f$-band width of about 20 meV \cite{amo19}. Alternatively, this peak may
also be caused by flat regions in the band structure. In this case, the small
peak intensity might be related to the respective orientation in $k$ space with
respect to the surface normal. Also, the absorption of a magnetic exciton was
proposed, based on peaks observed by inelastic neutron scattering (INS)
\cite{ale95} and Raman \cite{nyh97} experiments at 14 and 16 meV, respectively.
Since the surface is metallic, the exciton at the surface has a considerable linewidth and a reduced energy \cite{kap15}. We note that other types of
measurements found features at similar energies but provided different
explanations for their origin \cite{neu13,den14,fuh15}. The peaks at $-3$ and
$+2$ meV appear only at temperatures below 7~K and are likely of different
origin, as discussed below.

As outlined above, the dominating peak at $-6.5$ meV certainly contains some
bulk contribution. It is, therefore, not surprising that its $T$-dependence
can be compared across different samples and differently terminated surfaces,
see {\bf Figure} \ref{lowT}b including data from Ref.\ \onlinecite{yee13,pir20,
ruan14}. It starts to become observable below about 50 K indicating its
relation to the evolving Kondo hybridization upon lowering $T$ \cite{zha13,
roe14}. A similar $T$-scale was found in transport \cite{all79,wol13,eo19}
and other spectroscopic measurements \cite{cal07,jia13,xu14b,par16,min17}.

At $T \lesssim$ 5~K, the peak gains in weight more rapidly compared to $T >$
5~K; a similar change of slope is observed in \cite{pir20}, yet not quite as
pronounced. At these low temperatures, the existence of metallic surface
states has been established \cite{wol13,kim13,zha13,sye15,eo19}, suggesting a
relation between the increased peak height and the surface states. As outlined
in section \ref{Kondo}, the Kondo hybridization between Sm localized 4$f$ and
itinerant 5$d$ states is a prerequisite for the existence of topologically
nontrivial surface states. However, at the surface the Kondo hybridization is
reduced due to the reduced screening of the Sm ions \cite{all16} resulting in
a suppressed Kondo temperature $T_{\rm K}^{\rm s}$ at the surface, a phenomenon
usually referred to as surface Kondo breakdown \cite{ale15,pet16}. According
to Ref.\ \onlinecite{pet16}, narrow peaks with strongly $T$-dependent STS
spectra near $E_{\rm F}$ can be regarded as smoking gun evidence for surface
Kondo breakdown scenario, cf.\ {\bf Figure} \ref{lowT}a. Below
$T_{\rm K}^{\rm s}$, the surface Kondo effect results from gradually increasing
hybridization between surface conduction electrons and one of the 4$f$-states
giving rise to a weakly dispersive band near $E_{\rm F}$ \cite{ert16}. In
consequence, we can likely associate the build-up of the Kondo effect at the
sample surface with the strong increase of the main peak height at $T < T_{\rm
K}^{\rm s} \sim$ 5--7~K in {\bf Figure} \ref{lowT}b. The development of a heavy
fermion surface state below this temperature is in line with results from
magnetothermoelectric measurements \cite{luo15}. Taken together, the main peak
at $-6.5$ mV not only contains a bulk contribution, but, at $T_{\rm K}^{\rm
s}$, there is also a surface contribution \cite{jiao16,pir20,bar14}. This will
be further elaborated on in section \ref{dopedSTS} for tunneling spectra
obtained with magnetic tips.

This interpretation naturally links the STS results with the finding of
metallic surface states below $T_{\rm K}^{\rm s}$ in numerous measurements.
In addition, it provides an explanation for the emergence of the two peaks at
$|V| =$ 2 -- 3 meV at $T \lesssim T_{\rm K}^{\rm s}$, see {\bf Figure}
\ref{lowT}a. With $T_{\rm K}^{\rm s}$ about an order of magnitude smaller than
$T_{\rm K}$ in the bulk, one may expect the Kondo hybridization gap at the
surface to be about an order of magnitude smaller than the bulk gap. With the
latter (as discussed above) found at $\sim 20$ meV, the former should be seen
at $\sim$2 meV, in good agreement with experiment. The association of these two
peaks with the surface state, along with a surface contribution to the main
peak, is supported by the sensitivity of the peak heights close to impurities,
while the peak at $-13$ mV remains largely unaffected \cite{jiao16}.

A powerful way to study the electronic structure of a given material by
tunneling spectroscopy is through quasiparticle interference (QPI)
\cite{pet98,hof11}. In case of SmB$_6$, QPI was applied to reveal the
dispersion of the expected Dirac cones. However, on non-reconstructed surfaces
no clear evolution of any QPI pattern was observed \cite{ruan14,jiao16}. A
negligible magnetic field response in Ref.\ \onlinecite{jiao16} pointed to
nonmagnetic impurities and hence, spin-dependent or spin-conserving scattering
is expected to dominate \cite{oka14}. In consequence, backscattering is
\begin{figure}[t]
\includegraphics*[width=0.32\textwidth]{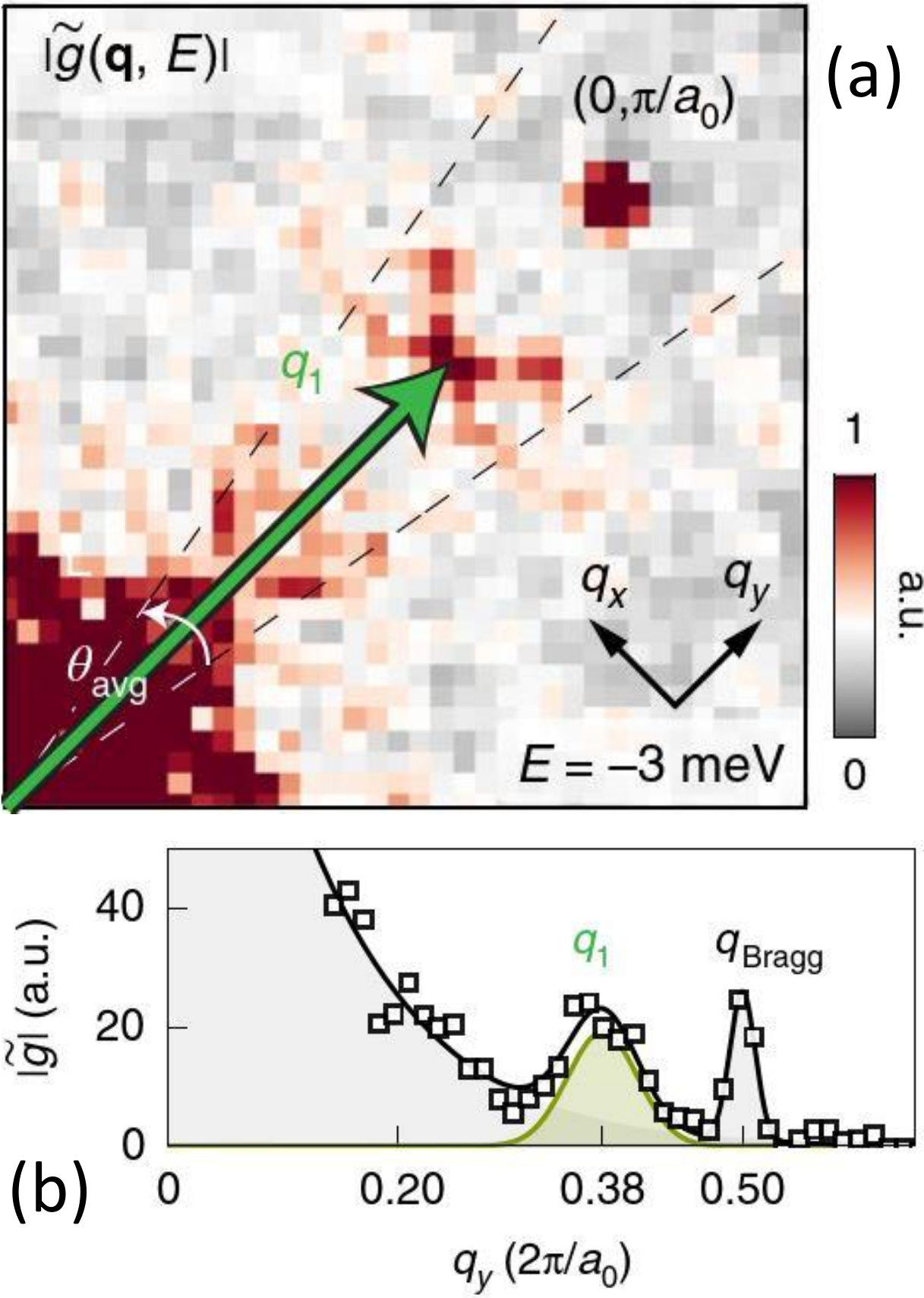}
\caption{a) Fourier transform of the local tunneling conductance within an
(30 nm)$^2$ area, which was twofold symmetrized to increase the signal-to-noise
ratio. The QPI is marked by the vector $\boldsymbol{q}_1$ corresponding to
backscattering within the Dirac cone. b) $\boldsymbol{q}_1$  is obtained from
fitting linecuts by a sum of Gaussians to the intensities averaged over a wedge
marked by dashed lines in (a). Reproduced with permission \cite{pir20}.
Copyright 2020, Springer Nature.}   \label{qpi}
\end{figure}
suppressed for topologically protected surface states with opposite, in-plane
spin directions for $\boldsymbol{k}$ and $\boldsymbol{-k}$ on the Dirac cone.
The absence of clear QPI signatures is then consistent with topologically
protected surface states \cite{zha09,guo10,bar14} and renders a scenario based
on mere Rashba splitting less likely \cite{oka14}.

This picture changes on a $2\, \times\, 1$ reconstructed surface, likely
\begin{figure*}[t]
\includegraphics*[width=0.7\textwidth]{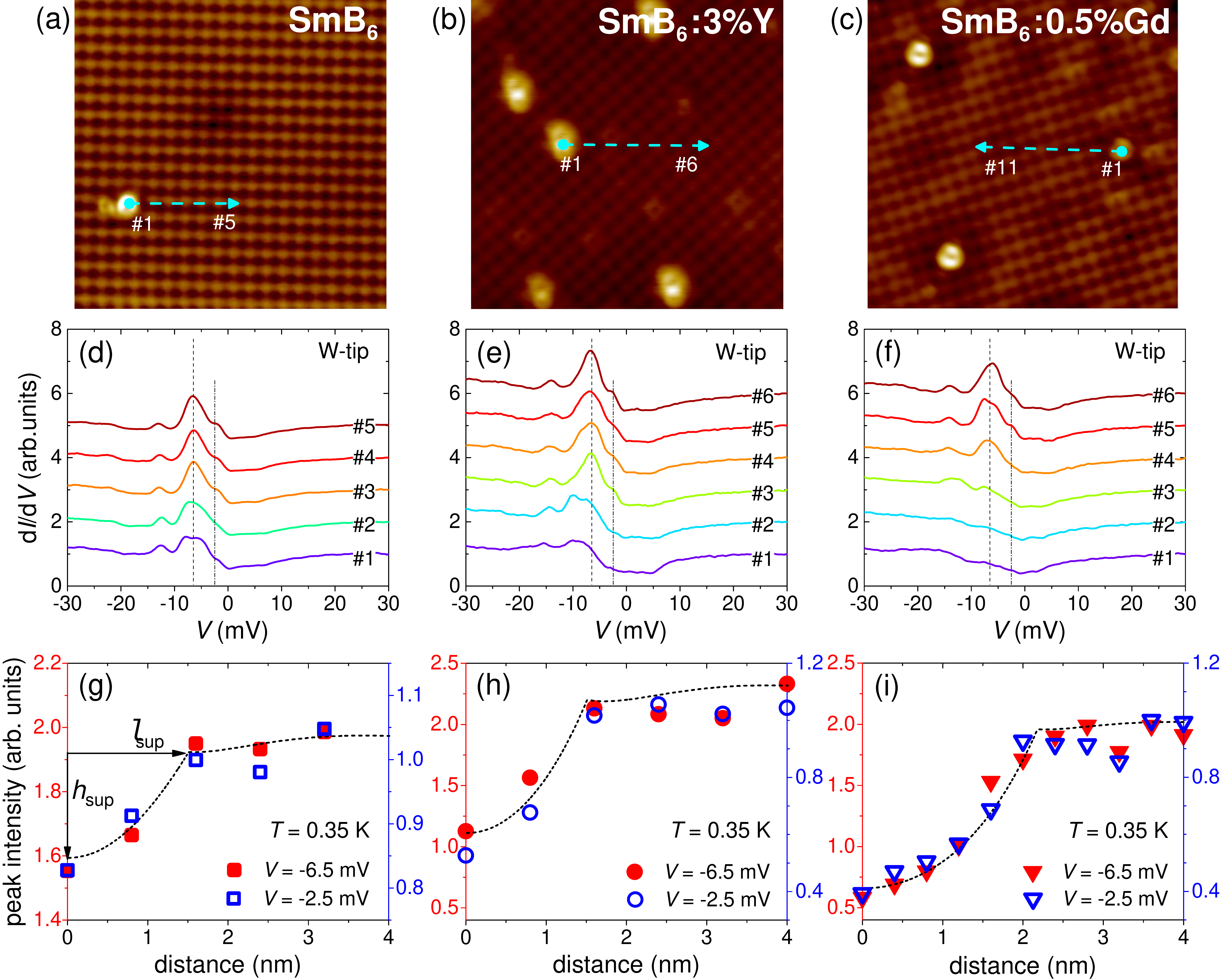}
\caption{Topographies of areas $8 \times8$ nm$^2$ on a) pure SmB$_6$ with
non-magnetic impurity, b) 3\%Y and c) 0.5\%Gd substituted SmB$_6$. d)--f) Spectroscopy along the cyan arrows in a)--c)  with \#1 recorded right on top
of the respective impurity. $V =$ 30 mV, $I_{\rm sp} =$ 100 pA. Curves offset
for clarity. g)--i) d$I/$d$V$ values at $V = -6.5$ mV (red) and $-2.5$ mV
(blue) [energies marked by dashed lines in d)--f)] in dependence on distance
from the impurity. Black dashed lines are model fits from which the suppression
of peak intensity at the impurity, $h_{\rm sup}$, and its lateral range
$\ell_{\rm sup}$ were obtained. Reproduced under the terms of a Creative
Commons Attribution 4.0 International License \cite{jiao18}. Copyright 2018
The Authors, published by AAAS.}  \label{doped}
\end{figure*}
because the defects then are mostly Sm vacancies (cf. {\bf Figure}
\ref{dopants}c) which can locally induce a magnetic moment via the Kondo-hole
effect \cite{sol91,jdt11,fig11,fuh17}. Along with a small spin polarization
out of the surface plane \cite{bau12} or a weak ferromagnetic out-of-plane
component at the surface \cite{nak15}, backscattering is then possible again
\cite{oka11}. Indeed, well-resolved QPI pattern along the Sm chains of a
$2\,\times\, 1$ reconstructed surface were reported in Ref.\
\onlinecite{pir20}, see {\bf Figure} \ref{qpi}. A scattering vector
$\boldsymbol{q}_1$ can clearly be resolved at $V = -3$ mV and is distinguished
from the Bragg peak at (0,$\pi/a_0$). It exhibits a roughly linear dispersion,
as expected for a Dirac cone located at around $-5$ meV, with an effective
Dirac mass of $(410 \pm 20)$ of the bare electron mass \cite{pir20}. A second
cone of less heavy Dirac states, also located at the same energy, is resolved
at somewhat larger bias voltages. These findings are strongly in favor of
topologically non-trivial surface states.

\subsection{Substituted SmB$_6$}
\labelsubsec{dopedSTS}
Magnetic impurities in topological insulators break time reversal symmetry
(TRS) and open a surface band gap near the Dirac point, while TRS protects the
surface states against non-magnetic ones, see reviews \cite{has10} and
\cite{ten19}. In consequence, several studies have been conducted to compare
the impact of non-magnetic and magnetic impurities on the surface states
\cite{kim14,kan16,wak16,fuh17,mia21} for insight into the surface states. It
should also be noted that SmB$_6$ doped with Gd or Eu was also used for
electron spin resonance (ESR) investigations \cite{les17,dem18}. Moreover, Gd-
and Fe-substituted samples were investigated in Ref.\ \onlinecite{pir20} to
identify the $2 \times 1$ reconstructed surface.

In {\bf Figure} \ref{doped}, topographies obtained on pristine SmB$_6$, 3\%~Y
substituted SmB$_6$ (denoted SmB$_6$:3\%Y) and 0.5\%~Gd substituted SmB$_6$
are presented \cite{jiao18}. Here, surfaces areas of (8 nm)$^2$ with only few
defects and likely of B-termination are studied. Assuming a statistical
distribution of the substituents, the defects in (b)/(c) are likely caused by
Y/Gd (for Gd, see additional argument below). Tunneling spectra taken along
the whole length of the respective blue arrows are reproduced in {\bf Figure}
\ref{doped}d--f, with spectra labelled \#1 collected right on top of the
respective defect. The most pronounced change is seen for the main peak at
$-6.5$ mV: this peak is somewhat reduced at the impurity in pure SmB$_6$, but
almost not discernable at a Gd impurity (dashed lines mark $-6.5$ and $-2.5$
mV). All spectra far away from the impurities nicely match {\bf Figure}
\ref{lowT}a. For quantification, the peak heights at $V = -6.5$ mV and
$-2.5$ mV are plotted in dependence on distance from the respective impurity
in {\bf Figures} \ref{doped}g--i. Clearly, the peaks are suppressed at the
impurity (by $h_{\rm sup}$), but completely recover in similar fashion at
distances $\ell_{\rm sup}$. Importantly, the suppression of the main peak at
$V = -6.5$ mV is significantly more pronounced in both, strength $h_{\rm sup}$
and extent $\ell_{\rm sup}$ for magnetic Gd.

The recovery of the peak height was described by a theoretical model for
surface states around magnetic impurities \cite{liu09}, which assumes Dirac electrons of the supposed topological surface state to be locally coupled to a
magnetic impurity through exchange coupling. Intuitively, it describes a local
gap opening of the topological surface states due to a magnetic impurity.
Within this model, the fit parameters are the strength of the exchange coupling
$J_z \cdot S_z$, mostly determined by $h_{\rm sup}$, its range $\ell_{\rm sup}$
and the Fermi velocity $\nu_{\rm F}$. Fits to this model, dashed lines in
{\bf Figures} \ref{doped}g--i, describe the experimentally observed behavior of
surface state suppression reasonably well. For the Gd-substituted sample, the
fit yields $J_z \cdot S_z \approx$ 1.4 meV, $\nu_{\rm F} \approx$ 3000 m/s and
$\ell_{\rm sup}$(Gd) $\approx$ 2.2 nm, while $\ell_{\rm sup}(\Box$, Y)
$\approx$ 1.5 nm on pure (with non-magnetic impurity $\Box$ \cite{jiao16}) and
Y-substituted SmB$_6$, respectively. Note that the results above for
$\nu_{\rm F}$ and $J_z \cdot S_z$ are in good agreement with \cite{pir20},
given the main peak energy of $-6.5$ mV. This, in concert with the
applicability of the exchange coupling model as such, supports an
interpretation of the surface states being of topological nature. Note
also that within the framework of this model, a peak suppression is not
expected for non-magnetic impurities; yet the latter may acquire a (small)
magnetic moment through the Kondo hole effect mentioned above.

This model also explains naturally the resistivity behavior of the substituted
samples (cf.\ {\bf Figure} \ref{resist}): SmB$_6$:3\%Y and SmB$_6$:0.5\%Gd
show a very similar resistivity plateau below about 4 K, while SmB$_6$:3\%Gd
does not exhibit any sign of a plateau down to 0.32 K \cite{kim14,jiao18}. The
resistivity plateau signals a conducting path on the surface due to the
metallic surface states. On the former two sample surfaces, the areas with
suppressed surface states are small enough in number and size to leave
sufficient surface unaffected such that conducting paths can still form and
cause the plateau. For SmB$_6$:3\%Gd with average distance between Gd
substituents of $\sim 1.3$ nm, the areas of suppressed surface states coalesce
and inhibit a conducting path. The comparison between 3\%Gd and 3\%Y clearly
shows that these areas must be larger (in size and/or number) for magnetic
impurities, as confirmed by our STS results.

There is one potential issue with the above approach: The substitution also
influences the sample's Kondo hybridization, which is a prerequisite for
\begin{figure}[t]
\includegraphics*[width=0.34\textwidth]{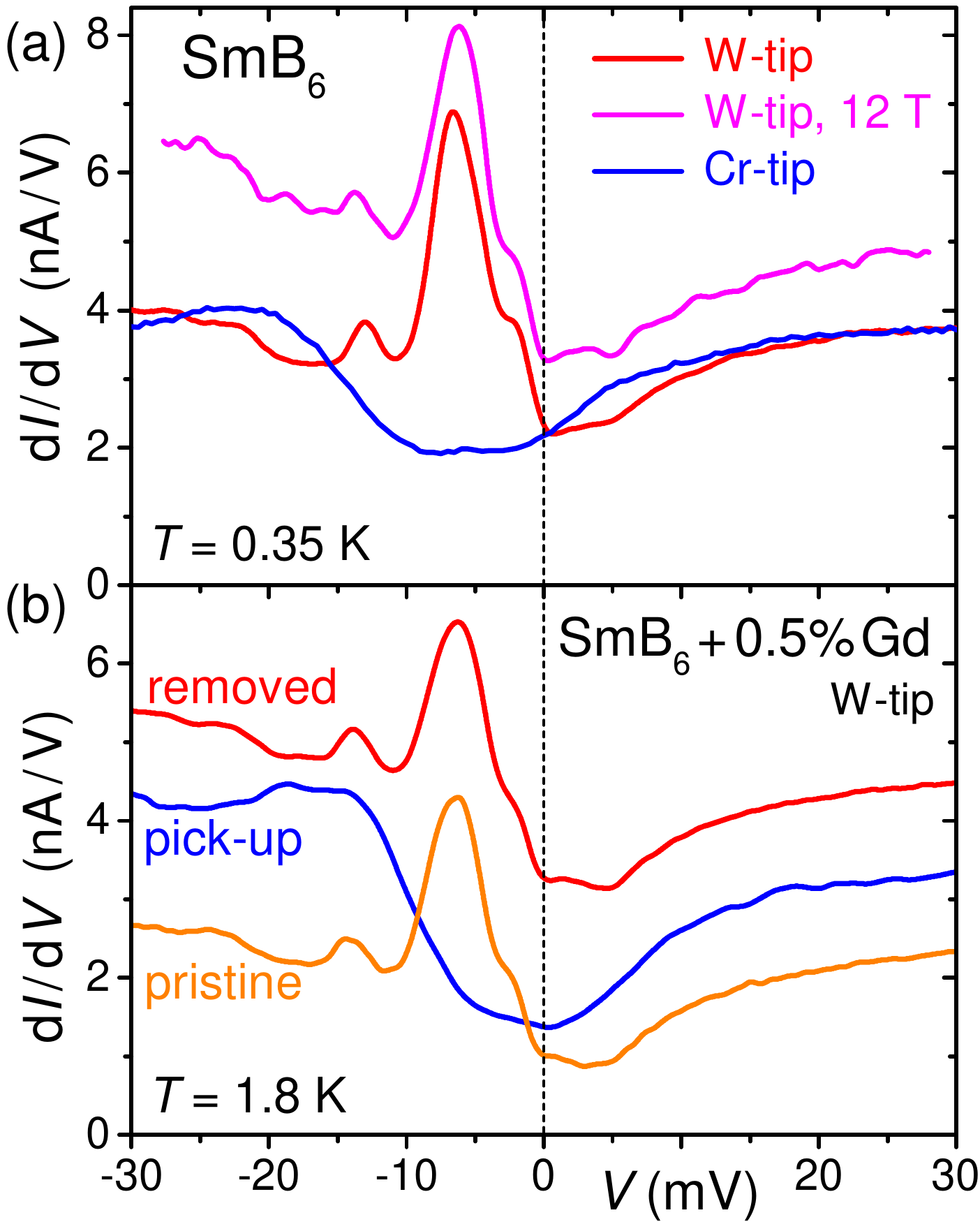}
\caption{a) Tunneling spectroscopy on pure SmB$_6$ taken with W-tip (at zero
and 12 T applied magnetic field) and with magnetic Cr-tip (blue curve). b)
Spectra obtained with W-tip before and after picking up an adatom from SmB$_6$:0.5\%Gd. After removing the picked-up atom from the tip by applying
a voltage pulse, the spectrum (red, offset for clarity) is very similar to the
original one. Reproduced under the terms of a Creative Commons Attribution 4.0
International License \cite{jiao18}. Copyright 2018 The Authors, published by
AAAS.}  \label{pick}
\end{figure}
topologically non-trivial surface states. This can, e.g., be seen in the
smaller total resistivity change upon cooling for the substituted samples
compared to pure SmB$_6$, {\bf Figure} \ref{resist} and \cite{kim14,fuh17,
jiao18,mia21}. To avoid such an influence on the samples' bulk, yet keeping
the short-range exchange interaction, the surface states should also be
suppressed by using a magnetic tip. The result is striking, {\bf Figure}
\ref{pick}a: The spectrum obtained by a Cr tip does \emph{not} exhibit any
sign of the peak at $-6.5$ mV (nor at $-2.5$ mV), which is the hallmark of
the surface states, while the hybridization still appears (albeit somewhat
reduced). This result is in line with the model above, employing an exchange
coupling between the magnetic entity (here the Cr tip) and the Dirac electrons.
Spin-polarized tunneling alone does not explain this huge reduction in peak
height \cite{bar16,kim19,oka14}. Note that a magnetic field of 12 T does not
notably influence the spectrum, ruling out any possible stray field effect of
the magnetic tip. It is also worth mentioning that the spectrum obtained with
Cr-tip is rather similar to spectra taken with W-tip at around 20 K
\cite{jiao18}, i.e. a temperature at which the surface states do not yet
manifest themselves in tunneling spectra \cite{ruan14,jiao16,pir20}. This
emphasizes that it is indeed the surface state contribution which is mainly
suppressed.

Albeit the very same surface was investigated by the two tips used in
{\bf Figure} \ref{pick}a and great care was taken, it cannot be completely
ruled out that different surface areas were investigated causing at least part
of the huge changes. In this respect, an observation came to help
\cite{jiao18}: Upon scanning a regular W tip over a surface of Gd-substituted
SmB$_6$, the tip picked up Gd on occasion, converting it into a magnetic one.
\begin{figure*}[t]
\includegraphics*[width=0.8\textwidth]{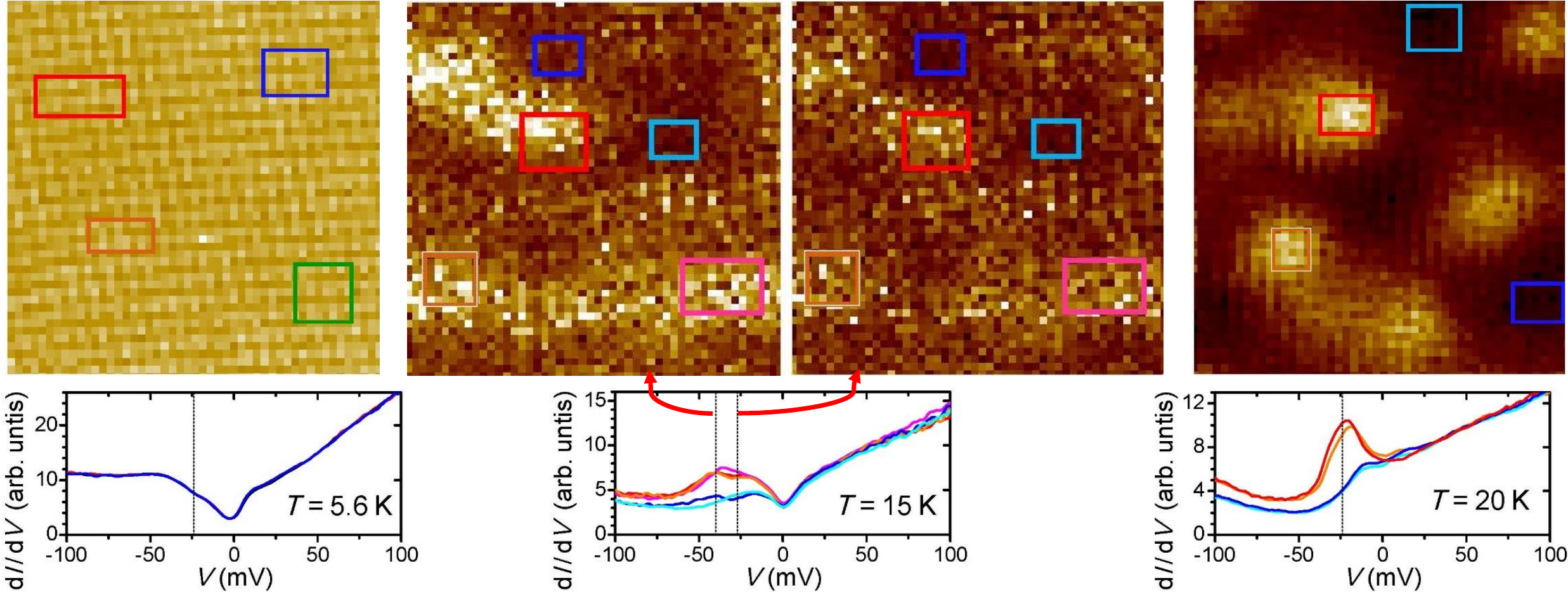}
\caption{Upper row: Spectroscopy maps of the same B-terminated surface of
EuB$_6$. Area (20 nm)$^2$, $V = -0.2$ V, $I_{\rm sp} =$ 0.5 nA. There are no
inhomogeneities at $T =$ 5.6 K, while they are clearly visible at 15 K and 20
K. Lower row: d$I/$d$V$-spectra averaged over the areas marked by rectangles
of corresponding colors in the maps above. Dashed lines mark $V$ at which the
respective maps were obtained. Adapted with permission \cite{poh18}. Copyright
2018, American Physical Society.}  \label{polaron}
\end{figure*}
A similar process was observed earlier for STM on Fe$_{1+y}$Te where excess Fe
was picked up \cite{ena14}. The process of picking up an atom is seen by a
corresponding sudden height change in topography. The fact that it was indeed
a Gd atom which was picked up is supported by the similarity between spectra
obtained with such a modified tip and with Cr-tip (cf.\ {\bf Figures}
\ref{pick}a and b) and the observation that these changes were exclusively
observed on Gd-substituted sample surfaces, but not on any of the Y-substituted
or pure SmB$_6$. What's more, the tip modification can be reversed. By applying
a voltage pulse (typically 10 V) to the tip, the picked-up atom can be removed
and the original spectrum is retained, {\bf Figure} \ref{pick}b. In this case,
the very same surface area (on atomic level) was investigated.

\subsection{STS on other hexaborides}
\labelsubsec{otherSTS}
EuB$_6$ is a ferromagnetic semimetal \cite{fis79}. Despite its simple lattice
(see {\bf Figure} \ref{structure}) and magnetic structure (the $^8S_{7/2}$
state of Eu$^{2+}$ is isotropic), EuB$_6$ shows two consecutive transitions at
about $T_{c1} =$ 15.3 K and $T_{c2} =$ 12.6 K indicating intriguingly complex
magnetic behavior \cite{sue00}. This behavior was discussed early on within
the framework of a so-called magnetic polaron \cite{urb04}, an ordered magnetic
cluster resulting from the exchange coupling between the conduction electrons
and the local magnetic moments, which gives rise to intrinsically inhomogeneous
states (see Ref.\ \onlinecite{mol01} for a review). Within a magnetic polaron,
the charge carriers are localized while spin-polarizing the local magnetic
moments over a finite distance. Once the magnetic polarons overlap, the charge
carriers can suddenly delocalize, resulting in a mobility increase and
percolation of the charges at $T_{c1}$. This intriguingly simple model can
explain the colossal magnetoresistance (CMR) effect in many materials
\cite{rao98}. In case of EuB$_6$, the magnetic polarons are believed to start
forming at around $T^* \sim$ 35--40 K, percolate at $T_{c1}$ and finally merge
at $T_{c2}$ \cite{das12,man14}. Around $T_{c1}$, EuB$_6$ exhibits a strong CMR
effect \cite{sue00} because the polarons grow in size and/or number upon
applying a magnetic field, in line with the polaronic picture.

STS can be utilized to directly visualize electronic inhomogeneity, i.e. the
polaron formation \cite{poh18}. To this end, {\bf Figure} \ref{polaron} (upper
row) exhibits STS maps on a B-terminated surface as shown in {\bf Figure}
\ref{EuB6}a at $T =$ 5.6~K, 15 K and 20 K. These temperatures were chosen to
cover the whole range, i.e., below, in between, and above $T_{c2}$ and
$T_{c1}$, respectively. As expected for the homogeneous ferromagnetic state
below $T_{c2}$, there is no sign of any electronic inhomogeneity in the
spectroscopy map at $T =$ 5.6 K. Also, the d$I/$d$V$-spectra averaged over the
different marked areas (including an average over the total area) perfectly
overlap, which evidences that homogeneous maps are observed not only at $V =
-24$~mV as shown, but rather at any bias voltage. The picture is changed
dramatically at $T =$ 20 K $> T_{c1}$ where patches of enlarged d$I/$d$V$ are
seen in the spectroscopy map at $V = -24$~mV. This voltage was chosen because
here, the largest differences in $g(V,\boldsymbol{r})$ are observed, see
d$I/$d$V$-spectra averaged over areas marked by rectangles of corresponding
color in the map. The enlarged local DOS is attributed to the magnetic
polarons. At 20 K, these polarons are well separated and are about 3–-4 nm in
extent. Note that polarons of very similar size have been visualized by STM
in manganites \cite{ron06}.

At intermediate $T =$ 15~K, the picture is not as clear-cut, mostly because
the local differences in $g(V,\boldsymbol{r})$ are not as pronounced (see
d$I/$d$V$-spectra). This is likely related to the system being close to the
magnetic transition $T_{c2}$ at which any inhomogeneity is expected to vanish.
Nonetheless, there appear to be extended regions of enlarged d$I/$d$V$
(specifically at $V = -40$ mV) which may indicate the expected percolating
conducting paths between $T_{c2}$ and $T_{c1}$. Note that the STS measurements
shown here are corroborated by local Hall effect measurements \cite{poh18}.

While in the above investigations it was important to look at clean surfaces,
surfaces with impurities showed d$I/$d$V$-oscillations of large amplitude due
to elastic scattering of electrons \cite{roe20}. Fourier transform STS and
DFT slab calculations point to a band with hole-like dispersion lying
approximately 68 meV below $E_{\rm F}$ and a spin-split surface state band,
which stems from the dangling $p_z$ orbitals of the apical boron atom and
hybridizes with bulk states near the $\bar{\Gamma}$ point.

Similarly, in LaB$_6$ a surface resonance at $-0.2$ eV is expected from DFT
slab calculations which is composed of B dangling bonds. However, these states
turn out difficult to image. At positive bias, the La $d$-orbitals contribute
to d$I/$d$V$ \cite{buc19}. The B dangling bonds are speculated to also lead to
a feature at $-0.2$ eV on PrB$_6$ surfaces \cite{buc20}. In addition, a peak
at $-0.7$ eV is observed in d$I/$d$V$ for the Pr-terminated surface of
{\bf Figure} \ref{PrB6}d, which is related to the 4$f$ states.

Although there is a number of ARPES studies on YbB$_6$ and CeB$_6$ (e.g.\
Ref.\ \onlinecite{ram16}), STM investigations on these materials are scarce
\cite{ena-the,buc-the}. Planar tunnel junctions on SmB$_6$, EuB$_6$, CeB$_6$,
CaB$_6$ and SrB$_6$ evidenced the formation of partial gaps in the DOS at low
$T$ for all investigated materials \cite{ams98}.

\subsection{$I(z)$-spectroscopy}
\labelsubsec{Iz}
LaB$_6$ is well-known for its very low work function of about 2.7 eV
\cite{eto85} and therefore, is an established cathode material. LaB$_6$ was
even suggested as a substrate for STM on weakly conducting thin films
\cite{mar98}. The work function $\Phi_s$ of a clean sample surface can be
estimated from the tunneling barrier height $\Phi$ involved in STM (note that
also the tip work function $\Phi_t$ enters into $\Phi$). To this end, the
dependence of the tunneling current $I$ on the tip displacement with respect
to the sample $\Delta z$ has to be measured, which are related by
\begin{equation}\label{eqIz}
I(\Delta  z) \propto \exp (-2\kappa \Delta z) \quad {\rm where} \quad
\kappa^2 = \textstyle \frac{2m_e}{\hbar^2} \displaystyle \Phi \; ,
\end{equation}
where $m_e$ is the bare electron mass and $V \ll \Phi_{s,t}$. Indeed, a value
of $\Phi \approx$ 3~eV was obtained on a $2 \times 1$ reconstructed LaB$_6$
surface for a bias voltage set point of 0.2 V \cite{buc19}; a smaller set
point resulted in a smaller $\Phi$ \cite{nag20}. Unfortunately, no data were
provided on the studied $\Delta z$-range.

On B-terminated surfaces of SmB$_6$, $\Phi_s$-values between 4.5 eV and 6.7
eV were found \cite{ale21}. A somewhat larger range of $\Phi_s$ from order of
4 eV up to 7 eV was reported for small (order of 1 nm) patches,  with the
larger values likely obtained on defects \cite{sun18}. Based on DFT
calculations, the work function of Sm-terminated surfaces should be of order
2~eV, while being at least twice as high on B-terminated surfaces \cite{sun18},
supporting the findings and assignments. These data are confirmed by an early
report on SmB$_6$ with very little Sm in the surface layer (4.2~eV,
\cite{aon79}) and more recent ARPES measurements (4.5~eV, \cite{neu13}). For
comparison, the work function of pure Sm is close to 2.7 eV and for pure B
about 4.45 eV \cite{mic77}.

For B-terminated EuB$_6$, values of $\Phi_s =$ 4.7~eV and 5.6~eV were
determined \cite{wir20}, in good agreement with B-terminated SmB$_6$.
Unfortunately, there are no values for RE-terminated surfaces given.

\section{Modification of the S\MakeLowercase{m}B$_6$ surface}
For many investigations and applications it is useful to prepare materials
in thin film form. Attempts to grow films of hexaborides were made early on,
see e.g.\ Ref.\ \onlinecite{mit97,shi14,ino21}. Special efforts have been made
for LaB$_6$ because of its low work function, and for SmB$_6$ subsequent to
the proposal of non-trivial surface states \cite{yon14,li18,bat18,liu18}. One
particularly interesting application of a thin film of SmB$_6$ makes use of
the proximity effect \cite{lee16}. Here, superconductivity in SmB$_6$ is
\begin{figure}[t]
\includegraphics*[width=0.48\textwidth]{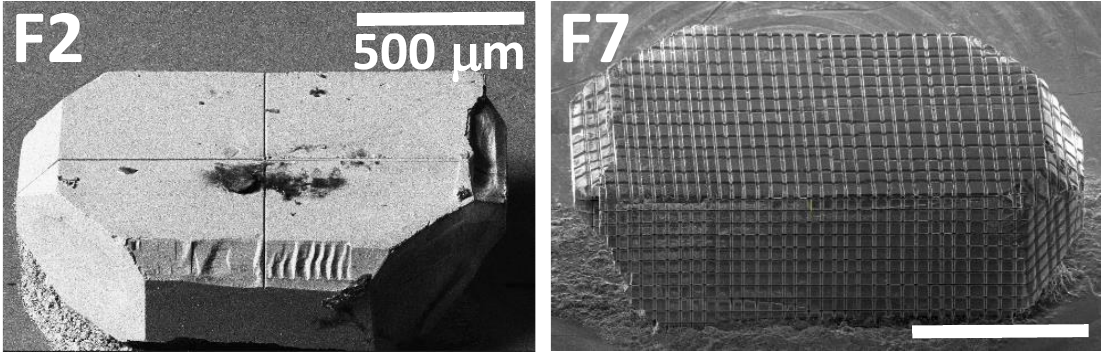}
\caption{SEM images of an SmB$_6$ sample after cutting increasing numbers of
lines by FIB. Left (F2): only one cross of lines was cut; right (F7): average
line distance is about 40 $\mu$m. The depth of FIB-cut lines is 7--10 $\mu$m,
their widths are less than 10 $\mu$m. Scale bars: 500 $\mu$m. Adapted under
the terms of a Creative Commons Attribution 4.0 International License
\cite{ale21b}. Copyright 2021 The Authors, published by American Physical
Society.}  \label{fibbed}
\end{figure}
induced by fabricating bilayers of the latter and superconducting Nb thin
films. The thickness of the surface state was estimated to be $\approx 6$~nm.

The surface of SmB$_6$ can be damaged by ion irradiation to different depths
\cite{wak15,wak16}. Subsequent resistivity measurements indicated that the
surface states form anew in a layer below the damaged material, in line with
its non-trivial topology. Using a focused ion beam (FIB), the ion damage can
be inflicted in a locally more controlled way \cite{ale21b}, see examples in
{\bf Figure} \ref{fibbed}. With increasing number of FIB-cut lines, the low-$T$
resistivity plateau kicks in at lower temperature and shifts to higher
resistivity values likely indicating a confinement of the persisting surface
states \cite{miy17} despite possible disorder effects \cite{sen20,abe20} due
to the FIB treatment.

The surface states can also be influenced in a controlled way by applying
strain to the SmB$_6$ sample \cite{ste17}. This can, in principle, be utilized
{\it in situ} in STM \cite{nah13}.

\section{Conclusions and Outlook}
The exciting variety of physical phenomena observed in hexaborides effectively
calls for investigations by such dedicated tools like STM, which provides
insight into the density of states, or more generally, into the Green's
function. In particular, requirements like low temperatures, energy resolution
on a meV-scale or the possibility of applying magnetic fields can be met by
STM. Unfortunately, the cubic crystal structure does not provide a preferred
cleaving plane and hence, atomically flat surface areas have to be searched
for extensively, and are usually only up a few 100 nm in extent. Both
restrictions are even more true if unreconstructed surfaces are strived for.
While this behavior is a severe obstacle for STM, it should also be kept in
mind in any other surface sensitive technique, e.g.\ ARPES or point contact
spectroscopy.

Nonetheless, atomically flat surface areas can be found, and we here provided
an overview of the different surface terminations encountered. Given the common
crystal structure, the topographies of the different hexaborides can be nicely
compared and, with a sufficiently broad database, the surface terminations can
reasonably be assigned. Knowing the surface termination of atomically flat
surfaces, i.e.\ B- vs.\ RE-terminated or $2 \times 1$ reconstructed, is
certainly helpful for any interpretation of STS data.

Tunneling spectroscopy confirmed the Kondo effect in SmB$_6$. At temperatures
below a few K, individual peaks beyond a co-tunneling effect can be resolved
and a Kondo breakdown at the surface, in line with predictions, is hinted at.
Magnetic substituents locally suppress the formation of the surface state,
again in line with theory, which is a strong indication for a topologically
non-trivial nature of these surface states. Investigating the quasiparticle
interference pattern on reconstructed surfaces revealed a linear dispersion
as expected for a Dirac cone \cite{pir20}. On clean EuB$_6$ surfaces, the
formation of polarons near the metal-semiconductor transition could directly
be visualized.

Finding atomically flat surface areas may fuel renewed interest in STS on other
hexaborides. Specifically, CeB$_6$ is certainly worthwhile studying provided
sufficiently low temperatures (below about 2 K) are achieved and, preferably,
magnetic fields can be applied.

With the availability of hexaboride thin films not only the issues related to
the cleavage of hexaborides could be circumvented, but also new phenomena and
applications become available for study. An example here is the SmB$_6$/YB$_6$
heterostructure for which Klein tunneling was observed \cite{lee19}. STM is
certainly also helpful if hexaboride nanowires are to be investigated.
One may even imagine using sharply pointed SmB$_6$ single crystals as tunneling
tips, with the unique spin-momentum locking in the TKI SmB$_6$ likely giving
rise to interesting tunneling phenomena.

In addition to materials development, new functionality can also be added to
STM. For example, {\it in situ} strain devices have been implemented into STM
\cite{yim18}. Such an additional experimental degree of freedom may also
provide new insight into hexaborides.

\section{Acknowledgements}
The authors gratefully acknowledge collaborations and fruitful discussions 
with Victoria Ale Crivilero, Z. Fisk, Lin Jiao, S. Kirchner, J. M\"{u}ller, 
Sahana R\"{o}{\ss}ler, Priscila Rosa, F. Steglich and L. Hao Tjeng. SW 
acknowledges support by the Deutsche Forschungsgemeinschaft through WI 
1324/5-1.


\begin{thebibliography}{240}
\expandafter\ifx\csname natexlab\endcsname\relax\def\natexlab#1{#1}\fi
\expandafter\ifx\csname bibnamefont\endcsname\relax
  \def\bibnamefont#1{#1}\fi
\expandafter\ifx\csname bibfnamefont\endcsname\relax
  \def\bibfnamefont#1{#1}\fi
\expandafter\ifx\csname citenamefont\endcsname\relax
  \def\citenamefont#1{#1}\fi
\expandafter\ifx\csname url\endcsname\relax
  \def\url#1{\texttt{#1}}\fi
\expandafter\ifx\csname urlprefix\endcsname\relax\def\urlprefix{URL }\fi
\providecommand{\bibinfo}[2]{#2}
\providecommand{\eprint}[2][]{\url{#2}}

\bibitem[{\citenamefont{Moissan and Williams}(1897)}]{moi1897}
\bibinfo{author}{\bibfnamefont{H.}~\bibnamefont{Moissan}} \bibnamefont{and}
  \bibinfo{author}{\bibfnamefont{P.}~\bibnamefont{Williams}},
  \bibinfo{journal}{C.\ R.\ Acad.\ Sci.\ (Paris)}
  \textbf{\bibinfo{volume}{125}}, \bibinfo{pages}{629} (\bibinfo{year}{1897}).

\bibitem[{\citenamefont{Pauling and Weinbaum}(1934)}]{pau34}
\bibinfo{author}{\bibfnamefont{L.}~\bibnamefont{Pauling}} \bibnamefont{and}
  \bibinfo{author}{\bibfnamefont{S.}~\bibnamefont{Weinbaum}},
  \bibinfo{journal}{Z.\ Kristallogr.} \textbf{\bibinfo{volume}{87}},
  \bibinfo{pages}{181} (\bibinfo{year}{1934}).

\bibitem[{\citenamefont{Geballe et~al.}(1968)\citenamefont{Geballe, Matthias,
  Andres, Maita, Cooper, and Corenzwit}}]{geb68}
\bibinfo{author}{\bibfnamefont{T.~H.} \bibnamefont{Geballe}},
  \bibinfo{author}{\bibfnamefont{B.~T.} \bibnamefont{Matthias}},
  \bibinfo{author}{\bibfnamefont{K.}~\bibnamefont{Andres}},
  \bibinfo{author}{\bibfnamefont{J.~P.} \bibnamefont{Maita}},
  \bibinfo{author}{\bibfnamefont{A.~S.} \bibnamefont{Cooper}},
  \bibnamefont{and}
  \bibinfo{author}{\bibfnamefont{E.}~\bibnamefont{Corenzwit}},
  \bibinfo{journal}{Science} \textbf{\bibinfo{volume}{160}},
  \bibinfo{pages}{1443} (\bibinfo{year}{1968}).

\bibitem[{\citenamefont{Nyhus et~al.}(1997{\natexlab{a}})\citenamefont{Nyhus,
  Yoon, Kauffman, Cooper, Fisk, and Sarrao}}]{nyh97b}
\bibinfo{author}{\bibfnamefont{P.}~\bibnamefont{Nyhus}},
  \bibinfo{author}{\bibfnamefont{S.}~\bibnamefont{Yoon}},
  \bibinfo{author}{\bibfnamefont{M.}~\bibnamefont{Kauffman}},
  \bibinfo{author}{\bibfnamefont{S.~L.} \bibnamefont{Cooper}},
  \bibinfo{author}{\bibfnamefont{Z.}~\bibnamefont{Fisk}}, \bibnamefont{and}
  \bibinfo{author}{\bibfnamefont{J.}~\bibnamefont{Sarrao}},
  \bibinfo{journal}{Phys.\ Rev.\ B} \textbf{\bibinfo{volume}{56}},
  \bibinfo{pages}{2717} (\bibinfo{year}{1997}{\natexlab{a}}).

\bibitem[{\citenamefont{Urbano et~al.}(2004)\citenamefont{Urbano, Pagliuso,
  Rettori, Oseroff, Sarrao, Schlottmann, and Fisk}}]{urb04}
\bibinfo{author}{\bibfnamefont{R.~R.} \bibnamefont{Urbano}},
  \bibinfo{author}{\bibfnamefont{P.~G.} \bibnamefont{Pagliuso}},
  \bibinfo{author}{\bibfnamefont{C.}~\bibnamefont{Rettori}},
  \bibinfo{author}{\bibfnamefont{S.~B.} \bibnamefont{Oseroff}},
  \bibinfo{author}{\bibfnamefont{J.~L.} \bibnamefont{Sarrao}},
  \bibinfo{author}{\bibfnamefont{P.}~\bibnamefont{Schlottmann}},
  \bibnamefont{and} \bibinfo{author}{\bibfnamefont{Z.}~\bibnamefont{Fisk}},
  \bibinfo{journal}{Phys.\ Rev.\ B} \textbf{\bibinfo{volume}{70}},
  \bibinfo{pages}{140401(R)} (\bibinfo{year}{2004}).

\bibitem[{\citenamefont{Schell et~al.}(1982)\citenamefont{Schell, Winter,
  Rietschel, and Gompf}}]{sch82}
\bibinfo{author}{\bibfnamefont{G.}~\bibnamefont{Schell}},
  \bibinfo{author}{\bibfnamefont{H.}~\bibnamefont{Winter}},
  \bibinfo{author}{\bibfnamefont{H.}~\bibnamefont{Rietschel}},
  \bibnamefont{and} \bibinfo{author}{\bibfnamefont{F.}~\bibnamefont{Gompf}},
  \bibinfo{journal}{Phys.\ Rev.\ B} \textbf{\bibinfo{volume}{25}},
  \bibinfo{pages}{1589} (\bibinfo{year}{1982}).

\bibitem[{\citenamefont{Effantin et~al.}(1985)\citenamefont{Effantin,
  {Rossat-Mignod}, Burlet, Bartholin, Kunii, and Kasuya}}]{eff85}
\bibinfo{author}{\bibfnamefont{J.~M.} \bibnamefont{Effantin}},
  \bibinfo{author}{\bibfnamefont{J.}~\bibnamefont{{Rossat-Mignod}}},
  \bibinfo{author}{\bibfnamefont{P.}~\bibnamefont{Burlet}},
  \bibinfo{author}{\bibfnamefont{H.}~\bibnamefont{Bartholin}},
  \bibinfo{author}{\bibfnamefont{S.}~\bibnamefont{Kunii}}, \bibnamefont{and}
  \bibinfo{author}{\bibfnamefont{T.}~\bibnamefont{Kasuya}},
  \bibinfo{journal}{J. Magn. Magn. Mater.} \textbf{\bibinfo{volume}{47-48}},
  \bibinfo{pages}{145} (\bibinfo{year}{1985}).

\bibitem[{\citenamefont{Zhou et~al.}(2015)\citenamefont{Zhou, Kim, Rosa, Wu,
  Guo, Zhang, Wang, Kang, Zhang, Yi et~al.}}]{zho15}
\bibinfo{author}{\bibfnamefont{Y.}~\bibnamefont{Zhou}},
  \bibinfo{author}{\bibfnamefont{D.-J.} \bibnamefont{Kim}},
  \bibinfo{author}{\bibfnamefont{P.~F.~S.} \bibnamefont{Rosa}},
  \bibinfo{author}{\bibfnamefont{Q.}~\bibnamefont{Wu}},
  \bibinfo{author}{\bibfnamefont{J.}~\bibnamefont{Guo}},
  \bibinfo{author}{\bibfnamefont{S.}~\bibnamefont{Zhang}},
  \bibinfo{author}{\bibfnamefont{Z.}~\bibnamefont{Wang}},
  \bibinfo{author}{\bibfnamefont{D.}~\bibnamefont{Kang}},
  \bibinfo{author}{\bibfnamefont{C.}~\bibnamefont{Zhang}},
  \bibinfo{author}{\bibfnamefont{W.}~\bibnamefont{Yi}}, \bibnamefont{et~al.},
  \bibinfo{journal}{Phys.\ Rev.\ B} \textbf{\bibinfo{volume}{92}},
  \bibinfo{pages}{241118(R)} (\bibinfo{year}{2015}).

\bibitem[{\citenamefont{Deng et~al.}(2013)\citenamefont{Deng, Haule, and
  Kotliar}}]{deng13}
\bibinfo{author}{\bibfnamefont{X.}~\bibnamefont{Deng}},
  \bibinfo{author}{\bibfnamefont{K.}~\bibnamefont{Haule}}, \bibnamefont{and}
  \bibinfo{author}{\bibfnamefont{G.}~\bibnamefont{Kotliar}},
  \bibinfo{journal}{Phys.\ Rev.\ Lett.} \textbf{\bibinfo{volume}{111}},
  \bibinfo{pages}{176404} (\bibinfo{year}{2013}).

\bibitem[{\citenamefont{Dzero et~al.}(2010)\citenamefont{Dzero, Sun, Galitski,
  and Coleman}}]{dze10}
\bibinfo{author}{\bibfnamefont{M.}~\bibnamefont{Dzero}},
  \bibinfo{author}{\bibfnamefont{K.}~\bibnamefont{Sun}},
  \bibinfo{author}{\bibfnamefont{V.}~\bibnamefont{Galitski}}, \bibnamefont{and}
  \bibinfo{author}{\bibfnamefont{P.}~\bibnamefont{Coleman}},
  \bibinfo{journal}{Phys.\ Rev.\ Lett.} \textbf{\bibinfo{volume}{104}},
  \bibinfo{pages}{106408} (\bibinfo{year}{2010}).

\bibitem[{\citenamefont{Hasan and Kane}(2010)}]{has10}
\bibinfo{author}{\bibfnamefont{M.~Z.} \bibnamefont{Hasan}} \bibnamefont{and}
  \bibinfo{author}{\bibfnamefont{C.~L.} \bibnamefont{Kane}},
  \bibinfo{journal}{Rev.\ Mod.\ Phys.} \textbf{\bibinfo{volume}{82}},
  \bibinfo{pages}{3045} (\bibinfo{year}{2010}).

\bibitem[{\citenamefont{Aeppli and Fisk}(1992)}]{aep92}
\bibinfo{author}{\bibfnamefont{G.}~\bibnamefont{Aeppli}} \bibnamefont{and}
  \bibinfo{author}{\bibfnamefont{Z.}~\bibnamefont{Fisk}},
  \bibinfo{journal}{Comments Cond.\ Mat.\ Phys.} \textbf{\bibinfo{volume}{16}},
  \bibinfo{pages}{155} (\bibinfo{year}{1992}).

\bibitem[{\citenamefont{Riseborough}(2000{\natexlab{a}})}]{ris00}
\bibinfo{author}{\bibfnamefont{P.~S.} \bibnamefont{Riseborough}},
  \bibinfo{journal}{Adv.\ Phys.} \textbf{\bibinfo{volume}{49}},
  \bibinfo{pages}{257} (\bibinfo{year}{2000}{\natexlab{a}}).

\bibitem[{\citenamefont{Allen et~al.}(1979)\citenamefont{Allen, Batlogg, and
  Wachter}}]{all79}
\bibinfo{author}{\bibfnamefont{J.~W.} \bibnamefont{Allen}},
  \bibinfo{author}{\bibfnamefont{B.}~\bibnamefont{Batlogg}}, \bibnamefont{and}
  \bibinfo{author}{\bibfnamefont{P.}~\bibnamefont{Wachter}},
  \bibinfo{journal}{Phys.\ Rev.\ B} \textbf{\bibinfo{volume}{20}},
  \bibinfo{pages}{4807} (\bibinfo{year}{1979}).

\bibitem[{\citenamefont{Kirchner et~al.}(2020)\citenamefont{Kirchner, Paschen,
  Chen, Wirth, Feng, Thompson, and Si}}]{kir20}
\bibinfo{author}{\bibfnamefont{S.}~\bibnamefont{Kirchner}},
  \bibinfo{author}{\bibfnamefont{S.}~\bibnamefont{Paschen}},
  \bibinfo{author}{\bibfnamefont{Q.}~\bibnamefont{Chen}},
  \bibinfo{author}{\bibfnamefont{S.}~\bibnamefont{Wirth}},
  \bibinfo{author}{\bibfnamefont{D.}~\bibnamefont{Feng}},
  \bibinfo{author}{\bibfnamefont{J.~D.} \bibnamefont{Thompson}},
  \bibnamefont{and} \bibinfo{author}{\bibfnamefont{Q.}~\bibnamefont{Si}},
  \bibinfo{journal}{Rev.\ Mod.\ Phys.} \textbf{\bibinfo{volume}{92}},
  \bibinfo{pages}{011002} (\bibinfo{year}{2020}).

\bibitem[{\citenamefont{Neupane et~al.}(2013)\citenamefont{Neupane, Alidoust,
  Xu, Kondo, Ishida, Kim, Liu, Belopolski, Jo, Chang et~al.}}]{neu13}
\bibinfo{author}{\bibfnamefont{M.}~\bibnamefont{Neupane}},
  \bibinfo{author}{\bibfnamefont{N.}~\bibnamefont{Alidoust}},
  \bibinfo{author}{\bibfnamefont{S.-Y.} \bibnamefont{Xu}},
  \bibinfo{author}{\bibfnamefont{T.}~\bibnamefont{Kondo}},
  \bibinfo{author}{\bibfnamefont{Y.}~\bibnamefont{Ishida}},
  \bibinfo{author}{\bibfnamefont{D.~J.} \bibnamefont{Kim}},
  \bibinfo{author}{\bibfnamefont{C.}~\bibnamefont{Liu}},
  \bibinfo{author}{\bibfnamefont{I.}~\bibnamefont{Belopolski}},
  \bibinfo{author}{\bibfnamefont{Y.~J.} \bibnamefont{Jo}},
  \bibinfo{author}{\bibfnamefont{T.-R.} \bibnamefont{Chang}},
  \bibnamefont{et~al.}, \bibinfo{journal}{Nat.\ Commun.}
  \textbf{\bibinfo{volume}{4}}, \bibinfo{pages}{2991} (\bibinfo{year}{2013}).

\bibitem[{\citenamefont{Jiang et~al.}(2013)\citenamefont{Jiang, Li, Zhang, Sun,
  Chen, Ye, Xu, Ge, Tan, Niu et~al.}}]{jia13}
\bibinfo{author}{\bibfnamefont{J.}~\bibnamefont{Jiang}},
  \bibinfo{author}{\bibfnamefont{S.}~\bibnamefont{Li}},
  \bibinfo{author}{\bibfnamefont{T.}~\bibnamefont{Zhang}},
  \bibinfo{author}{\bibfnamefont{Z.}~\bibnamefont{Sun}},
  \bibinfo{author}{\bibfnamefont{F.}~\bibnamefont{Chen}},
  \bibinfo{author}{\bibfnamefont{Z.~R.} \bibnamefont{Ye}},
  \bibinfo{author}{\bibfnamefont{M.}~\bibnamefont{Xu}},
  \bibinfo{author}{\bibfnamefont{Q.~Q.} \bibnamefont{Ge}},
  \bibinfo{author}{\bibfnamefont{S.~Y.} \bibnamefont{Tan}},
  \bibinfo{author}{\bibfnamefont{X.~H.} \bibnamefont{Niu}},
  \bibnamefont{et~al.}, \bibinfo{journal}{Nat.\ Commun.}
  \textbf{\bibinfo{volume}{4}}, \bibinfo{pages}{3010} (\bibinfo{year}{2013}).

\bibitem[{\citenamefont{Xu et~al.}(2013)\citenamefont{Xu, Shi, Biswas, Matt,
  Dhaka, Huang, Plumb, Radovi{\'c}, Dil, Pomjakushina et~al.}}]{nxu13}
\bibinfo{author}{\bibfnamefont{N.}~\bibnamefont{Xu}},
  \bibinfo{author}{\bibfnamefont{X.}~\bibnamefont{Shi}},
  \bibinfo{author}{\bibfnamefont{P.~K.} \bibnamefont{Biswas}},
  \bibinfo{author}{\bibfnamefont{C.~E.} \bibnamefont{Matt}},
  \bibinfo{author}{\bibfnamefont{R.~S.} \bibnamefont{Dhaka}},
  \bibinfo{author}{\bibfnamefont{Y.}~\bibnamefont{Huang}},
  \bibinfo{author}{\bibfnamefont{N.~C.} \bibnamefont{Plumb}},
  \bibinfo{author}{\bibfnamefont{M.}~\bibnamefont{Radovi{\'c}}},
  \bibinfo{author}{\bibfnamefont{J.~H.} \bibnamefont{Dil}},
  \bibinfo{author}{\bibfnamefont{E.}~\bibnamefont{Pomjakushina}},
  \bibnamefont{et~al.}, \bibinfo{journal}{Phys.\ Rev.\ B}
  \textbf{\bibinfo{volume}{88}}, \bibinfo{pages}{121102}
  (\bibinfo{year}{2013}).

\bibitem[{\citenamefont{Xu et~al.}(2014{\natexlab{a}})\citenamefont{Xu, Matt,
  Pomjakushina, Shi, Dhaka, Plumb, Radovi{\'c}, Biswas, Evtushinsky, Zabolotnyy
  et~al.}}]{xu14b}
\bibinfo{author}{\bibfnamefont{N.}~\bibnamefont{Xu}},
  \bibinfo{author}{\bibfnamefont{C.~E.} \bibnamefont{Matt}},
  \bibinfo{author}{\bibfnamefont{E.}~\bibnamefont{Pomjakushina}},
  \bibinfo{author}{\bibfnamefont{X.}~\bibnamefont{Shi}},
  \bibinfo{author}{\bibfnamefont{R.~S.} \bibnamefont{Dhaka}},
  \bibinfo{author}{\bibfnamefont{N.~C.} \bibnamefont{Plumb}},
  \bibinfo{author}{\bibfnamefont{M.}~\bibnamefont{Radovi{\'c}}},
  \bibinfo{author}{\bibfnamefont{P.~K.} \bibnamefont{Biswas}},
  \bibinfo{author}{\bibfnamefont{D.}~\bibnamefont{Evtushinsky}},
  \bibinfo{author}{\bibfnamefont{V.}~\bibnamefont{Zabolotnyy}},
  \bibnamefont{et~al.}, \bibinfo{journal}{Phys.\ Rev.\ B}
  \textbf{\bibinfo{volume}{90}}, \bibinfo{pages}{085148}
  (\bibinfo{year}{2014}{\natexlab{a}}).

\bibitem[{\citenamefont{Denlinger et~al.}(2014)\citenamefont{Denlinger, Allen,
  Kang, Sun, Min, Kim, and Fisk}}]{den14}
\bibinfo{author}{\bibfnamefont{J.~D.} \bibnamefont{Denlinger}},
  \bibinfo{author}{\bibfnamefont{J.~W.} \bibnamefont{Allen}},
  \bibinfo{author}{\bibfnamefont{J.-S.} \bibnamefont{Kang}},
  \bibinfo{author}{\bibfnamefont{K.}~\bibnamefont{Sun}},
  \bibinfo{author}{\bibfnamefont{B.-I.} \bibnamefont{Min}},
  \bibinfo{author}{\bibfnamefont{D.-J.} \bibnamefont{Kim}}, \bibnamefont{and}
  \bibinfo{author}{\bibfnamefont{Z.}~\bibnamefont{Fisk}}, \bibinfo{journal}{JPS
  Conf.\ Proc.} \textbf{\bibinfo{volume}{3}}, \bibinfo{pages}{017038}
  (\bibinfo{year}{2014}).

\bibitem[{\citenamefont{Xu et~al.}(2014{\natexlab{b}})\citenamefont{Xu, Biswas,
  Dil, Dhaka, Landolt, Muff, Matt, Shi, Plumb, Radovi{\'c} et~al.}}]{nxu14}
\bibinfo{author}{\bibfnamefont{N.}~\bibnamefont{Xu}},
  \bibinfo{author}{\bibfnamefont{P.~K.} \bibnamefont{Biswas}},
  \bibinfo{author}{\bibfnamefont{J.~H.} \bibnamefont{Dil}},
  \bibinfo{author}{\bibfnamefont{R.~S.} \bibnamefont{Dhaka}},
  \bibinfo{author}{\bibfnamefont{G.}~\bibnamefont{Landolt}},
  \bibinfo{author}{\bibfnamefont{S.}~\bibnamefont{Muff}},
  \bibinfo{author}{\bibfnamefont{C.~E.} \bibnamefont{Matt}},
  \bibinfo{author}{\bibfnamefont{X.}~\bibnamefont{Shi}},
  \bibinfo{author}{\bibfnamefont{N.~C.} \bibnamefont{Plumb}},
  \bibinfo{author}{\bibfnamefont{M.}~\bibnamefont{Radovi{\'c}}},
  \bibnamefont{et~al.}, \bibinfo{journal}{Nat.\ Commun.}
  \textbf{\bibinfo{volume}{5}}, \bibinfo{pages}{4566}
  (\bibinfo{year}{2014}{\natexlab{b}}).

\bibitem[{\citenamefont{Suga et~al.}(2014)\citenamefont{Suga, Sakamoto, Okuda,
  Miyamoto, Kuroda, Sekiyama, Yamaguchi, Fujiwara, Irizawa, Ito
  et~al.}}]{sug14}
\bibinfo{author}{\bibfnamefont{S.}~\bibnamefont{Suga}},
  \bibinfo{author}{\bibfnamefont{K.}~\bibnamefont{Sakamoto}},
  \bibinfo{author}{\bibfnamefont{T.}~\bibnamefont{Okuda}},
  \bibinfo{author}{\bibfnamefont{K.}~\bibnamefont{Miyamoto}},
  \bibinfo{author}{\bibfnamefont{K.}~\bibnamefont{Kuroda}},
  \bibinfo{author}{\bibfnamefont{A.}~\bibnamefont{Sekiyama}},
  \bibinfo{author}{\bibfnamefont{J.}~\bibnamefont{Yamaguchi}},
  \bibinfo{author}{\bibfnamefont{H.}~\bibnamefont{Fujiwara}},
  \bibinfo{author}{\bibfnamefont{A.}~\bibnamefont{Irizawa}},
  \bibinfo{author}{\bibfnamefont{T.}~\bibnamefont{Ito}}, \bibnamefont{et~al.},
  \bibinfo{journal}{J.\ Phys.\ Soc.\ Jpn.} \textbf{\bibinfo{volume}{83}},
  \bibinfo{pages}{014705} (\bibinfo{year}{2014}).

\bibitem[{\citenamefont{Hlawenka et~al.}(2018)\citenamefont{Hlawenka,
  Siemensmeyer, Weschke, Varykhalov, S\'{a}nchez-Barriga, Shitsevalova,
  Dukhnenko, Filipov, Gab\'{a}ni, Flachbart et~al.}}]{hla18}
\bibinfo{author}{\bibfnamefont{P.}~\bibnamefont{Hlawenka}},
  \bibinfo{author}{\bibfnamefont{K.}~\bibnamefont{Siemensmeyer}},
  \bibinfo{author}{\bibfnamefont{E.}~\bibnamefont{Weschke}},
  \bibinfo{author}{\bibfnamefont{A.}~\bibnamefont{Varykhalov}},
  \bibinfo{author}{\bibfnamefont{J.}~\bibnamefont{S\'{a}nchez-Barriga}},
  \bibinfo{author}{\bibfnamefont{N.~Y.} \bibnamefont{Shitsevalova}},
  \bibinfo{author}{\bibfnamefont{A.~V.} \bibnamefont{Dukhnenko}},
  \bibinfo{author}{\bibfnamefont{V.~B.} \bibnamefont{Filipov}},
  \bibinfo{author}{\bibfnamefont{S.}~\bibnamefont{Gab\'{a}ni}},
  \bibinfo{author}{\bibfnamefont{K.}~\bibnamefont{Flachbart}},
  \bibnamefont{et~al.}, \bibinfo{journal}{Nat.\ Commun.}
  \textbf{\bibinfo{volume}{9}}, \bibinfo{pages}{517} (\bibinfo{year}{2018}).

\bibitem[{\citenamefont{Takimoto}(2011)}]{tak11}
\bibinfo{author}{\bibfnamefont{T.}~\bibnamefont{Takimoto}},
  \bibinfo{journal}{J.\ Phys.\ Soc.\ Jpn.} \textbf{\bibinfo{volume}{80}},
  \bibinfo{pages}{123710} (\bibinfo{year}{2011}).

\bibitem[{\citenamefont{Lu et~al.}(2013)\citenamefont{Lu, Zhao, Weng, Fang, and
  Dai}}]{lu13}
\bibinfo{author}{\bibfnamefont{F.}~\bibnamefont{Lu}},
  \bibinfo{author}{\bibfnamefont{J.}~\bibnamefont{Zhao}},
  \bibinfo{author}{\bibfnamefont{H.}~\bibnamefont{Weng}},
  \bibinfo{author}{\bibfnamefont{Z.}~\bibnamefont{Fang}}, \bibnamefont{and}
  \bibinfo{author}{\bibfnamefont{X.}~\bibnamefont{Dai}},
  \bibinfo{journal}{Phys.\ Rev.\ Lett.} \textbf{\bibinfo{volume}{110}},
  \bibinfo{pages}{096401} (\bibinfo{year}{2013}).

\bibitem[{\citenamefont{Alexandrov et~al.}(2013)\citenamefont{Alexandrov,
  Dzero, and Coleman}}]{ale13}
\bibinfo{author}{\bibfnamefont{V.}~\bibnamefont{Alexandrov}},
  \bibinfo{author}{\bibfnamefont{M.}~\bibnamefont{Dzero}}, \bibnamefont{and}
  \bibinfo{author}{\bibfnamefont{P.}~\bibnamefont{Coleman}},
  \bibinfo{journal}{Phys.\ Rev.\ Lett.} \textbf{\bibinfo{volume}{111}},
  \bibinfo{pages}{226403} (\bibinfo{year}{2013}).

\bibitem[{\citenamefont{Baruselli and Vojta}(2014)}]{bar14}
\bibinfo{author}{\bibfnamefont{P.~P.} \bibnamefont{Baruselli}}
  \bibnamefont{and} \bibinfo{author}{\bibfnamefont{M.}~\bibnamefont{Vojta}},
  \bibinfo{journal}{Phys.\ Rev.\ B} \textbf{\bibinfo{volume}{90}},
  \bibinfo{pages}{201106} (\bibinfo{year}{2014}).

\bibitem[{\citenamefont{Baruselli and Vojta}(2015)}]{bar15}
\bibinfo{author}{\bibfnamefont{P.~P.} \bibnamefont{Baruselli}}
  \bibnamefont{and} \bibinfo{author}{\bibfnamefont{M.}~\bibnamefont{Vojta}},
  \bibinfo{journal}{Phys.\ Rev.\ Lett.} \textbf{\bibinfo{volume}{115}},
  \bibinfo{pages}{156404} (\bibinfo{year}{2015}).

\bibitem[{\citenamefont{Yu et~al.}(2015)\citenamefont{Yu, Weng, Hu, Fang, and
  Dai}}]{yu15}
\bibinfo{author}{\bibfnamefont{R.}~\bibnamefont{Yu}},
  \bibinfo{author}{\bibfnamefont{H.~M.} \bibnamefont{Weng}},
  \bibinfo{author}{\bibfnamefont{X.}~\bibnamefont{Hu}},
  \bibinfo{author}{\bibfnamefont{Z.}~\bibnamefont{Fang}}, \bibnamefont{and}
  \bibinfo{author}{\bibfnamefont{X.}~\bibnamefont{Dai}}, \bibinfo{journal}{New
  J.\ Phys.} \textbf{\bibinfo{volume}{17}}, \bibinfo{pages}{023012}
  (\bibinfo{year}{2015}).

\bibitem[{\citenamefont{Kim et~al.}(2014{\natexlab{a}})\citenamefont{Kim, Kim,
  Kang, Kim, Choi, Kang, Denlinger, and Min}}]{jkim14}
\bibinfo{author}{\bibfnamefont{J.}~\bibnamefont{Kim}},
  \bibinfo{author}{\bibfnamefont{K.}~\bibnamefont{Kim}},
  \bibinfo{author}{\bibfnamefont{C.-J.} \bibnamefont{Kang}},
  \bibinfo{author}{\bibfnamefont{S.}~\bibnamefont{Kim}},
  \bibinfo{author}{\bibfnamefont{H.~C.} \bibnamefont{Choi}},
  \bibinfo{author}{\bibfnamefont{J.-S.} \bibnamefont{Kang}},
  \bibinfo{author}{\bibfnamefont{J.~D.} \bibnamefont{Denlinger}},
  \bibnamefont{and} \bibinfo{author}{\bibfnamefont{B.~I.} \bibnamefont{Min}},
  \bibinfo{journal}{Phys. Rev. B} \textbf{\bibinfo{volume}{90}},
  \bibinfo{pages}{075131} (\bibinfo{year}{2014}{\natexlab{a}}).

\bibitem[{\citenamefont{Kang et~al.}(2015)\citenamefont{Kang, Kim, Kim, Kang,
  Denlinger, and Min}}]{kan15}
\bibinfo{author}{\bibfnamefont{C.-J.} \bibnamefont{Kang}},
  \bibinfo{author}{\bibfnamefont{J.}~\bibnamefont{Kim}},
  \bibinfo{author}{\bibfnamefont{K.}~\bibnamefont{Kim}},
  \bibinfo{author}{\bibfnamefont{J.}~\bibnamefont{Kang}},
  \bibinfo{author}{\bibfnamefont{J.~D.} \bibnamefont{Denlinger}},
  \bibnamefont{and} \bibinfo{author}{\bibfnamefont{B.~I.} \bibnamefont{Min}},
  \bibinfo{journal}{J.\ Phys.\ Soc.\ Jpn.} \textbf{\bibinfo{volume}{84}},
  \bibinfo{pages}{024722} (\bibinfo{year}{2015}).

\bibitem[{\citenamefont{Hundley et~al.}(1990)\citenamefont{Hundley, Canfield,
  Thompson, Fisk, and Lawrence}}]{hun90}
\bibinfo{author}{\bibfnamefont{M.~F.} \bibnamefont{Hundley}},
  \bibinfo{author}{\bibfnamefont{P.~C.} \bibnamefont{Canfield}},
  \bibinfo{author}{\bibfnamefont{J.~D.} \bibnamefont{Thompson}},
  \bibinfo{author}{\bibfnamefont{Z.}~\bibnamefont{Fisk}}, \bibnamefont{and}
  \bibinfo{author}{\bibfnamefont{J.~M.} \bibnamefont{Lawrence}},
  \bibinfo{journal}{Phys.\ Rev.\ B} \textbf{\bibinfo{volume}{42}},
  \bibinfo{pages}{6842} (\bibinfo{year}{1990}).

\bibitem[{\citenamefont{Meisner et~al.}(1985)\citenamefont{Meisner,
  Torikachvili, Yang, Maple, and Guertin}}]{mei85}
\bibinfo{author}{\bibfnamefont{G.~P.} \bibnamefont{Meisner}},
  \bibinfo{author}{\bibfnamefont{M.~S.} \bibnamefont{Torikachvili}},
  \bibinfo{author}{\bibfnamefont{K.~N.} \bibnamefont{Yang}},
  \bibinfo{author}{\bibfnamefont{M.~B.} \bibnamefont{Maple}}, \bibnamefont{and}
  \bibinfo{author}{\bibfnamefont{R.}~\bibnamefont{Guertin}},
  \bibinfo{journal}{J.\ Appl.\ Phys.} \textbf{\bibinfo{volume}{57}},
  \bibinfo{pages}{3073} (\bibinfo{year}{1985}).

\bibitem[{\citenamefont{Paschen et~al.}(2010)\citenamefont{Paschen, Winkler,
  Nezu, Kriegisch, Hilschner, Custers, Prokofiev, and Strydom}}]{pas10}
\bibinfo{author}{\bibfnamefont{S.}~\bibnamefont{Paschen}},
  \bibinfo{author}{\bibfnamefont{H.}~\bibnamefont{Winkler}},
  \bibinfo{author}{\bibfnamefont{T.}~\bibnamefont{Nezu}},
  \bibinfo{author}{\bibfnamefont{M.}~\bibnamefont{Kriegisch}},
  \bibinfo{author}{\bibfnamefont{G.}~\bibnamefont{Hilschner}},
  \bibinfo{author}{\bibfnamefont{J.}~\bibnamefont{Custers}},
  \bibinfo{author}{\bibfnamefont{A.}~\bibnamefont{Prokofiev}},
  \bibnamefont{and} \bibinfo{author}{\bibfnamefont{A.}~\bibnamefont{Strydom}},
  \bibinfo{journal}{J.\ Phys.: Conf.\ Series} \textbf{\bibinfo{volume}{200}},
  \bibinfo{pages}{012156} (\bibinfo{year}{2010}).

\bibitem[{\citenamefont{Iga et~al.}(1988)\citenamefont{Iga, Kasaya, and
  Kasuya}}]{iga88}
\bibinfo{author}{\bibfnamefont{F.}~\bibnamefont{Iga}},
  \bibinfo{author}{\bibfnamefont{M.}~\bibnamefont{Kasaya}}, \bibnamefont{and}
  \bibinfo{author}{\bibfnamefont{T.}~\bibnamefont{Kasuya}},
  \bibinfo{journal}{J.\ Magn.\ Magn.\ Mater.} \textbf{\bibinfo{volume}{76-77}},
  \bibinfo{pages}{156} (\bibinfo{year}{1988}).

\bibitem[{\citenamefont{DiTusa et~al.}(1997)\citenamefont{DiTusa, Friemelt,
  Bucher, Aeppli, and Ramirez}}]{dit97}
\bibinfo{author}{\bibfnamefont{J.~F.} \bibnamefont{DiTusa}},
  \bibinfo{author}{\bibfnamefont{K.}~\bibnamefont{Friemelt}},
  \bibinfo{author}{\bibfnamefont{E.}~\bibnamefont{Bucher}},
  \bibinfo{author}{\bibfnamefont{G.}~\bibnamefont{Aeppli}}, \bibnamefont{and}
  \bibinfo{author}{\bibfnamefont{A.~P.} \bibnamefont{Ramirez}},
  \bibinfo{journal}{Phys.\ Rev.\ Lett.} \textbf{\bibinfo{volume}{78}},
  \bibinfo{pages}{2831} (\bibinfo{year}{1997}).

\bibitem[{\citenamefont{Menth et~al.}(1969)\citenamefont{Menth, Buehler, and
  Geballe}}]{men69}
\bibinfo{author}{\bibfnamefont{A.}~\bibnamefont{Menth}},
  \bibinfo{author}{\bibfnamefont{E.}~\bibnamefont{Buehler}}, \bibnamefont{and}
  \bibinfo{author}{\bibfnamefont{T.~H.} \bibnamefont{Geballe}},
  \bibinfo{journal}{Phys.\ Rev.\ Lett.} \textbf{\bibinfo{volume}{22}},
  \bibinfo{pages}{295} (\bibinfo{year}{1969}).

\bibitem[{\citenamefont{Maple and Wohlleben}(1971)}]{map71}
\bibinfo{author}{\bibfnamefont{M.~B.} \bibnamefont{Maple}} \bibnamefont{and}
  \bibinfo{author}{\bibfnamefont{D.}~\bibnamefont{Wohlleben}},
  \bibinfo{journal}{Phys.\ Rev.\ Lett.} \textbf{\bibinfo{volume}{27}},
  \bibinfo{pages}{511} (\bibinfo{year}{1971}).

\bibitem[{\citenamefont{Takabatake et~al.}(1992)\citenamefont{Takabatake,
  Nagasawa, Fujii, Hido, Nohara, Nishigori, Suzuki, Fujita, Helfrich, Ahlheim
  et~al.}}]{tak92}
\bibinfo{author}{\bibfnamefont{T.}~\bibnamefont{Takabatake}},
  \bibinfo{author}{\bibfnamefont{M.}~\bibnamefont{Nagasawa}},
  \bibinfo{author}{\bibfnamefont{H.}~\bibnamefont{Fujii}},
  \bibinfo{author}{\bibfnamefont{G.}~\bibnamefont{Hido}},
  \bibinfo{author}{\bibfnamefont{M.}~\bibnamefont{Nohara}},
  \bibinfo{author}{\bibfnamefont{S.}~\bibnamefont{Nishigori}},
  \bibinfo{author}{\bibfnamefont{T.}~\bibnamefont{Suzuki}},
  \bibinfo{author}{\bibfnamefont{T.}~\bibnamefont{Fujita}},
  \bibinfo{author}{\bibfnamefont{R.}~\bibnamefont{Helfrich}},
  \bibinfo{author}{\bibfnamefont{U.}~\bibnamefont{Ahlheim}},
  \bibnamefont{et~al.}, \bibinfo{journal}{Phys.\ Rev.\ B}
  \textbf{\bibinfo{volume}{45}}, \bibinfo{pages}{5740(R)}
  (\bibinfo{year}{1992}).

\bibitem[{\citenamefont{Izawa et~al.}(1999)\citenamefont{Izawa, Suzuki, Fujita,
  Takabatake, Nakamoto, Fujii, and Maezawa}}]{iza99}
\bibinfo{author}{\bibfnamefont{K.}~\bibnamefont{Izawa}},
  \bibinfo{author}{\bibfnamefont{T.}~\bibnamefont{Suzuki}},
  \bibinfo{author}{\bibfnamefont{T.}~\bibnamefont{Fujita}},
  \bibinfo{author}{\bibfnamefont{T.}~\bibnamefont{Takabatake}},
  \bibinfo{author}{\bibfnamefont{G.}~\bibnamefont{Nakamoto}},
  \bibinfo{author}{\bibfnamefont{H.}~\bibnamefont{Fujii}}, \bibnamefont{and}
  \bibinfo{author}{\bibfnamefont{K.}~\bibnamefont{Maezawa}},
  \bibinfo{journal}{Phys.\ Rev.\ B} \textbf{\bibinfo{volume}{59}},
  \bibinfo{pages}{2599} (\bibinfo{year}{1999}).

\bibitem[{\citenamefont{Schlesinger et~al.}(1993)\citenamefont{Schlesinger,
  Fisk, Zhang, Maple, Di{T}usa, and Aeppli}}]{sch93}
\bibinfo{author}{\bibfnamefont{Z.}~\bibnamefont{Schlesinger}},
  \bibinfo{author}{\bibfnamefont{Z.}~\bibnamefont{Fisk}},
  \bibinfo{author}{\bibfnamefont{H.~T.} \bibnamefont{Zhang}},
  \bibinfo{author}{\bibfnamefont{M.~B.} \bibnamefont{Maple}},
  \bibinfo{author}{\bibfnamefont{J.~F.} \bibnamefont{Di{T}usa}},
  \bibnamefont{and} \bibinfo{author}{\bibfnamefont{G.}~\bibnamefont{Aeppli}},
  \bibinfo{journal}{Phys.\ Rev.\ Lett.} \textbf{\bibinfo{volume}{71}},
  \bibinfo{pages}{1748} (\bibinfo{year}{1993}).

\bibitem[{\citenamefont{Cooley et~al.}(1999)\citenamefont{Cooley, Mielke,
  Hults, Goettee, Honold, Modler, Lacerda, Rickel, and Smith}}]{coo99}
\bibinfo{author}{\bibfnamefont{J.~C.} \bibnamefont{Cooley}},
  \bibinfo{author}{\bibfnamefont{C.~H.} \bibnamefont{Mielke}},
  \bibinfo{author}{\bibfnamefont{W.~L.} \bibnamefont{Hults}},
  \bibinfo{author}{\bibfnamefont{J.~D.} \bibnamefont{Goettee}},
  \bibinfo{author}{\bibfnamefont{M.~M.} \bibnamefont{Honold}},
  \bibinfo{author}{\bibfnamefont{R.~M.} \bibnamefont{Modler}},
  \bibinfo{author}{\bibfnamefont{A.}~\bibnamefont{Lacerda}},
  \bibinfo{author}{\bibfnamefont{D.~G.} \bibnamefont{Rickel}},
  \bibnamefont{and} \bibinfo{author}{\bibfnamefont{J.~L.} \bibnamefont{Smith}},
  \bibinfo{journal}{J.\ Supercond.} \textbf{\bibinfo{volume}{12}},
  \bibinfo{pages}{171} (\bibinfo{year}{1999}).

\bibitem[{\citenamefont{Jaime et~al.}(2000)\citenamefont{Jaime, Movshovich,
  Stewart, Beyermann, {Gomez Berisso}, Hundley, Canfield, and Sarrao}}]{jai00}
\bibinfo{author}{\bibfnamefont{M.}~\bibnamefont{Jaime}},
  \bibinfo{author}{\bibfnamefont{R.}~\bibnamefont{Movshovich}},
  \bibinfo{author}{\bibfnamefont{G.~R.} \bibnamefont{Stewart}},
  \bibinfo{author}{\bibfnamefont{W.~P.} \bibnamefont{Beyermann}},
  \bibinfo{author}{\bibfnamefont{M.}~\bibnamefont{{Gomez Berisso}}},
  \bibinfo{author}{\bibfnamefont{M.~F.} \bibnamefont{Hundley}},
  \bibinfo{author}{\bibfnamefont{P.~C.} \bibnamefont{Canfield}},
  \bibnamefont{and} \bibinfo{author}{\bibfnamefont{J.~L.}
  \bibnamefont{Sarrao}}, \bibinfo{journal}{Nature}
  \textbf{\bibinfo{volume}{405}}, \bibinfo{pages}{160} (\bibinfo{year}{2000}).

\bibitem[{\citenamefont{Sugiyama et~al.}(1988)\citenamefont{Sugiyama, Iga,
  Kasaya, Kasuya, and Date}}]{sug88}
\bibinfo{author}{\bibfnamefont{K.}~\bibnamefont{Sugiyama}},
  \bibinfo{author}{\bibfnamefont{F.}~\bibnamefont{Iga}},
  \bibinfo{author}{\bibfnamefont{M.}~\bibnamefont{Kasaya}},
  \bibinfo{author}{\bibfnamefont{T.}~\bibnamefont{Kasuya}}, \bibnamefont{and}
  \bibinfo{author}{\bibfnamefont{M.}~\bibnamefont{Date}}, \bibinfo{journal}{J.\
  Phys.\ Soc.\ Jpn.} \textbf{\bibinfo{volume}{57}}, \bibinfo{pages}{3946}
  (\bibinfo{year}{1988}).

\bibitem[{\citenamefont{Cooley et~al.}(1995{\natexlab{a}})\citenamefont{Cooley,
  Aronson, Fisk, and Canfield}}]{coo95}
\bibinfo{author}{\bibfnamefont{J.~C.} \bibnamefont{Cooley}},
  \bibinfo{author}{\bibfnamefont{M.~C.} \bibnamefont{Aronson}},
  \bibinfo{author}{\bibfnamefont{Z.}~\bibnamefont{Fisk}}, \bibnamefont{and}
  \bibinfo{author}{\bibfnamefont{P.~C.} \bibnamefont{Canfield}},
  \bibinfo{journal}{Phys.\ Rev.\ Lett.} \textbf{\bibinfo{volume}{74}},
  \bibinfo{pages}{1629} (\bibinfo{year}{1995}{\natexlab{a}}).

\bibitem[{\citenamefont{Cooley et~al.}(1995{\natexlab{b}})\citenamefont{Cooley,
  Aronson, Lacerda, Fisk, Canfield, and Guertin}}]{coo95b}
\bibinfo{author}{\bibfnamefont{J.~C.} \bibnamefont{Cooley}},
  \bibinfo{author}{\bibfnamefont{M.~C.} \bibnamefont{Aronson}},
  \bibinfo{author}{\bibfnamefont{A.}~\bibnamefont{Lacerda}},
  \bibinfo{author}{\bibfnamefont{Z.}~\bibnamefont{Fisk}},
  \bibinfo{author}{\bibfnamefont{P.~C.} \bibnamefont{Canfield}},
  \bibnamefont{and} \bibinfo{author}{\bibfnamefont{R.~P.}
  \bibnamefont{Guertin}}, \bibinfo{journal}{Phys.\ Rev.\ B}
  \textbf{\bibinfo{volume}{52}}, \bibinfo{pages}{7322}
  (\bibinfo{year}{1995}{\natexlab{b}}).

\bibitem[{\citenamefont{Barla et~al.}(2005)\citenamefont{Barla, Derr, Sanchez,
  Salce, Lapertot, Doyle, R{\"u}ffer, Lengsdorf, Abd-Elmeguid, and
  Flouquet}}]{bar05}
\bibinfo{author}{\bibfnamefont{A.}~\bibnamefont{Barla}},
  \bibinfo{author}{\bibfnamefont{J.}~\bibnamefont{Derr}},
  \bibinfo{author}{\bibfnamefont{J.~P.} \bibnamefont{Sanchez}},
  \bibinfo{author}{\bibfnamefont{B.}~\bibnamefont{Salce}},
  \bibinfo{author}{\bibfnamefont{G.}~\bibnamefont{Lapertot}},
  \bibinfo{author}{\bibfnamefont{B.~P.} \bibnamefont{Doyle}},
  \bibinfo{author}{\bibfnamefont{R.}~\bibnamefont{R{\"u}ffer}},
  \bibinfo{author}{\bibfnamefont{R.}~\bibnamefont{Lengsdorf}},
  \bibinfo{author}{\bibfnamefont{M.~M.} \bibnamefont{Abd-Elmeguid}},
  \bibnamefont{and} \bibinfo{author}{\bibfnamefont{J.}~\bibnamefont{Flouquet}},
  \bibinfo{journal}{Phys.\ Rev.\ Lett.} \textbf{\bibinfo{volume}{94}},
  \bibinfo{pages}{166401} (\bibinfo{year}{2005}).

\bibitem[{\citenamefont{Barla et~al.}(2004)\citenamefont{Barla, Sanchez, Haga,
  Lapertot, Doyle, Leupold, R{\"u}ffer, Abd-Elmeguid, Lengsdorf, and
  Flouquet}}]{bar04}
\bibinfo{author}{\bibfnamefont{A.}~\bibnamefont{Barla}},
  \bibinfo{author}{\bibfnamefont{J.~P.} \bibnamefont{Sanchez}},
  \bibinfo{author}{\bibfnamefont{Y.}~\bibnamefont{Haga}},
  \bibinfo{author}{\bibfnamefont{G.}~\bibnamefont{Lapertot}},
  \bibinfo{author}{\bibfnamefont{B.~P.} \bibnamefont{Doyle}},
  \bibinfo{author}{\bibfnamefont{O.}~\bibnamefont{Leupold}},
  \bibinfo{author}{\bibfnamefont{R.}~\bibnamefont{R{\"u}ffer}},
  \bibinfo{author}{\bibfnamefont{M.~M.} \bibnamefont{Abd-Elmeguid}},
  \bibinfo{author}{\bibfnamefont{R.}~\bibnamefont{Lengsdorf}},
  \bibnamefont{and} \bibinfo{author}{\bibfnamefont{J.}~\bibnamefont{Flouquet}},
  \bibinfo{journal}{Phys.\ Rev.\ Lett.} \textbf{\bibinfo{volume}{92}},
  \bibinfo{pages}{066401} (\bibinfo{year}{2004}).

\bibitem[{\citenamefont{{von Moln{\'a}r} et~al.}(1982)\citenamefont{{von
  Moln{\'a}r}, Theis, Benoit, Briggs, Flouquet, Ravex, and Fisk}}]{mol82}
\bibinfo{author}{\bibfnamefont{S.}~\bibnamefont{{von Moln{\'a}r}}},
  \bibinfo{author}{\bibfnamefont{T.}~\bibnamefont{Theis}},
  \bibinfo{author}{\bibfnamefont{A.}~\bibnamefont{Benoit}},
  \bibinfo{author}{\bibfnamefont{A.}~\bibnamefont{Briggs}},
  \bibinfo{author}{\bibfnamefont{J.}~\bibnamefont{Flouquet}},
  \bibinfo{author}{\bibfnamefont{J.}~\bibnamefont{Ravex}}, \bibnamefont{and}
  \bibinfo{author}{\bibfnamefont{Z.}~\bibnamefont{Fisk}}, in
  \emph{\bibinfo{booktitle}{Valence Instabilities}}, edited by
  \bibinfo{editor}{\bibfnamefont{P.}~\bibnamefont{Wachter}} \bibnamefont{and}
  \bibinfo{editor}{\bibfnamefont{H.}~\bibnamefont{Boppart}}
  (\bibinfo{publisher}{North-Holland, Amsterdam}, \bibinfo{year}{1982}), pp.
  \bibinfo{pages}{389--395}.

\bibitem[{\citenamefont{Nanba et~al.}(1993)\citenamefont{Nanba, Ohta, Motokawa,
  Kimura, Kunii, and Kasuya}}]{nan93}
\bibinfo{author}{\bibfnamefont{T.}~\bibnamefont{Nanba}},
  \bibinfo{author}{\bibfnamefont{H.}~\bibnamefont{Ohta}},
  \bibinfo{author}{\bibfnamefont{M.}~\bibnamefont{Motokawa}},
  \bibinfo{author}{\bibfnamefont{S.}~\bibnamefont{Kimura}},
  \bibinfo{author}{\bibfnamefont{S.}~\bibnamefont{Kunii}}, \bibnamefont{and}
  \bibinfo{author}{\bibfnamefont{T.}~\bibnamefont{Kasuya}},
  \bibinfo{journal}{Physica B} \textbf{\bibinfo{volume}{186-8}},
  \bibinfo{pages}{440} (\bibinfo{year}{1993}).

\bibitem[{\citenamefont{Gorshunov et~al.}(1999)\citenamefont{Gorshunov,
  Sluchanko, Volkov, Dressel, Knebel, Loidl, and Kunii}}]{gor99}
\bibinfo{author}{\bibfnamefont{B.}~\bibnamefont{Gorshunov}},
  \bibinfo{author}{\bibfnamefont{N.}~\bibnamefont{Sluchanko}},
  \bibinfo{author}{\bibfnamefont{A.}~\bibnamefont{Volkov}},
  \bibinfo{author}{\bibfnamefont{M.}~\bibnamefont{Dressel}},
  \bibinfo{author}{\bibfnamefont{G.}~\bibnamefont{Knebel}},
  \bibinfo{author}{\bibfnamefont{A.}~\bibnamefont{Loidl}}, \bibnamefont{and}
  \bibinfo{author}{\bibfnamefont{S.}~\bibnamefont{Kunii}},
  \bibinfo{journal}{Phys.\ Rev.\ B} \textbf{\bibinfo{volume}{59}},
  \bibinfo{pages}{1808} (\bibinfo{year}{1999}).

\bibitem[{\citenamefont{Nyhus et~al.}(1997{\natexlab{b}})\citenamefont{Nyhus,
  Cooper, Fisk, and Sarrao}}]{nyh97}
\bibinfo{author}{\bibfnamefont{P.}~\bibnamefont{Nyhus}},
  \bibinfo{author}{\bibfnamefont{S.~L.} \bibnamefont{Cooper}},
  \bibinfo{author}{\bibfnamefont{Z.}~\bibnamefont{Fisk}}, \bibnamefont{and}
  \bibinfo{author}{\bibfnamefont{J.}~\bibnamefont{Sarrao}},
  \bibinfo{journal}{Phys.\ Rev.\ B} \textbf{\bibinfo{volume}{55}},
  \bibinfo{pages}{12488} (\bibinfo{year}{1997}{\natexlab{b}}).

\bibitem[{\citenamefont{Alekseev
  et~al.}(1993{\natexlab{a}})\citenamefont{Alekseev, Mignot, Rossat-Mignod,
  Lazukov, and Sadikov}}]{ale93b}
\bibinfo{author}{\bibfnamefont{P.~A.} \bibnamefont{Alekseev}},
  \bibinfo{author}{\bibfnamefont{J.~M.} \bibnamefont{Mignot}},
  \bibinfo{author}{\bibfnamefont{J.}~\bibnamefont{Rossat-Mignod}},
  \bibinfo{author}{\bibfnamefont{V.~N.} \bibnamefont{Lazukov}},
  \bibnamefont{and} \bibinfo{author}{\bibfnamefont{I.~P.}
  \bibnamefont{Sadikov}}, \bibinfo{journal}{Physica B}
  \textbf{\bibinfo{volume}{186-8}}, \bibinfo{pages}{384}
  (\bibinfo{year}{1993}{\natexlab{a}}).

\bibitem[{\citenamefont{Alekseev et~al.}(1995)\citenamefont{Alekseev, Mignot,
  Rossat-Mignod, Lazukov, Sadikov, Konovalova, and Paderno}}]{ale95}
\bibinfo{author}{\bibfnamefont{P.~A.} \bibnamefont{Alekseev}},
  \bibinfo{author}{\bibfnamefont{J.~M.} \bibnamefont{Mignot}},
  \bibinfo{author}{\bibfnamefont{J.}~\bibnamefont{Rossat-Mignod}},
  \bibinfo{author}{\bibfnamefont{V.~N.} \bibnamefont{Lazukov}},
  \bibinfo{author}{\bibfnamefont{I.~P.} \bibnamefont{Sadikov}},
  \bibinfo{author}{\bibfnamefont{E.~S.} \bibnamefont{Konovalova}},
  \bibnamefont{and} \bibinfo{author}{\bibfnamefont{Y.~B.}
  \bibnamefont{Paderno}}, \bibinfo{journal}{J.\ Phys.\ Condens.\ Matter}
  \textbf{\bibinfo{volume}{7}}, \bibinfo{pages}{289} (\bibinfo{year}{1995}).

\bibitem[{\citenamefont{Yeo et~al.}(2003)\citenamefont{Yeo, Nakatsuji, Bianchi,
  Schlottmann, Fisk, Balicas, Stampe, and Kennedy}}]{yeo03}
\bibinfo{author}{\bibfnamefont{S.}~\bibnamefont{Yeo}},
  \bibinfo{author}{\bibfnamefont{S.}~\bibnamefont{Nakatsuji}},
  \bibinfo{author}{\bibfnamefont{A.~D.} \bibnamefont{Bianchi}},
  \bibinfo{author}{\bibfnamefont{P.}~\bibnamefont{Schlottmann}},
  \bibinfo{author}{\bibfnamefont{Z.}~\bibnamefont{Fisk}},
  \bibinfo{author}{\bibfnamefont{L.}~\bibnamefont{Balicas}},
  \bibinfo{author}{\bibfnamefont{P.~A.} \bibnamefont{Stampe}},
  \bibnamefont{and} \bibinfo{author}{\bibfnamefont{R.~J.}
  \bibnamefont{Kennedy}}, \bibinfo{journal}{Phys.\ Rev.\ Lett.}
  \textbf{\bibinfo{volume}{91}}, \bibinfo{pages}{046401}
  (\bibinfo{year}{2003}).

\bibitem[{\citenamefont{Sluchanko et~al.}(2002)\citenamefont{Sluchanko,
  Glushkov, Demishev, Menovsky, Weckhuysen, and Moshchalkov}}]{slu02}
\bibinfo{author}{\bibfnamefont{N.~E.} \bibnamefont{Sluchanko}},
  \bibinfo{author}{\bibfnamefont{V.~V.} \bibnamefont{Glushkov}},
  \bibinfo{author}{\bibfnamefont{S.~V.} \bibnamefont{Demishev}},
  \bibinfo{author}{\bibfnamefont{A.~A.} \bibnamefont{Menovsky}},
  \bibinfo{author}{\bibfnamefont{L.}~\bibnamefont{Weckhuysen}},
  \bibnamefont{and} \bibinfo{author}{\bibfnamefont{V.~V.}
  \bibnamefont{Moshchalkov}}, \bibinfo{journal}{Phys.\ Rev.\ B}
  \textbf{\bibinfo{volume}{65}}, \bibinfo{pages}{064404}
  (\bibinfo{year}{2002}).

\bibitem[{\citenamefont{Bouvet et~al.}(1998)\citenamefont{Bouvet, Kasuya,
  Bonnet, Regnault, Rossat-Mignod, Iga, F{\aa}k, and Severing}}]{bou98}
\bibinfo{author}{\bibfnamefont{A.}~\bibnamefont{Bouvet}},
  \bibinfo{author}{\bibfnamefont{T.}~\bibnamefont{Kasuya}},
  \bibinfo{author}{\bibfnamefont{M.}~\bibnamefont{Bonnet}},
  \bibinfo{author}{\bibfnamefont{L.~P.} \bibnamefont{Regnault}},
  \bibinfo{author}{\bibfnamefont{J.}~\bibnamefont{Rossat-Mignod}},
  \bibinfo{author}{\bibfnamefont{F.}~\bibnamefont{Iga}},
  \bibinfo{author}{\bibfnamefont{B.}~\bibnamefont{F{\aa}k}}, \bibnamefont{and}
  \bibinfo{author}{\bibfnamefont{A.}~\bibnamefont{Severing}},
  \bibinfo{journal}{J.\ Phys.: Condens.\ Matter.}
  \textbf{\bibinfo{volume}{10}}, \bibinfo{pages}{5667} (\bibinfo{year}{1998}).

\bibitem[{\citenamefont{Nefeodova et~al.}(1999)\citenamefont{Nefeodova,
  Alekseev, Mignot, Lazukov, Sadikov, Paderno, Shitsevalova, and
  Eccleston}}]{nef99}
\bibinfo{author}{\bibfnamefont{E.~V.} \bibnamefont{Nefeodova}},
  \bibinfo{author}{\bibfnamefont{P.~A.} \bibnamefont{Alekseev}},
  \bibinfo{author}{\bibfnamefont{J.-M.} \bibnamefont{Mignot}},
  \bibinfo{author}{\bibfnamefont{V.~N.} \bibnamefont{Lazukov}},
  \bibinfo{author}{\bibfnamefont{I.~P.} \bibnamefont{Sadikov}},
  \bibinfo{author}{\bibfnamefont{Y.~B.} \bibnamefont{Paderno}},
  \bibinfo{author}{\bibfnamefont{N.~Y.} \bibnamefont{Shitsevalova}},
  \bibnamefont{and} \bibinfo{author}{\bibfnamefont{R.~S.}
  \bibnamefont{Eccleston}}, \bibinfo{journal}{Phys.\ Rev.\ B}
  \textbf{\bibinfo{volume}{60}}, \bibinfo{pages}{13507} (\bibinfo{year}{1999}).

\bibitem[{\citenamefont{Riseborough}(2001)}]{ale01}
\bibinfo{author}{\bibfnamefont{P.~S.} \bibnamefont{Riseborough}},
  \bibinfo{journal}{Phys.\ Rev.\ B} \textbf{\bibinfo{volume}{63}},
  \bibinfo{pages}{064411} (\bibinfo{year}{2001}).

\bibitem[{\citenamefont{Riseborough}(2003)}]{ris03}
\bibinfo{author}{\bibfnamefont{P.~S.} \bibnamefont{Riseborough}},
  \bibinfo{journal}{Phys.\ Rev.\ B} \textbf{\bibinfo{volume}{69}},
  \bibinfo{pages}{235213} (\bibinfo{year}{2003}).

\bibitem[{\citenamefont{Sollie and Schlottmann}(1991)}]{sol91}
\bibinfo{author}{\bibfnamefont{R.}~\bibnamefont{Sollie}} \bibnamefont{and}
  \bibinfo{author}{\bibfnamefont{P.}~\bibnamefont{Schlottmann}},
  \bibinfo{journal}{J.\ Appl.\ Phys.} \textbf{\bibinfo{volume}{69}},
  \bibinfo{pages}{5478} (\bibinfo{year}{1991}).

\bibitem[{\citenamefont{Schlottmann}(1992)}]{sch92}
\bibinfo{author}{\bibfnamefont{P.}~\bibnamefont{Schlottmann}},
  \bibinfo{journal}{Phys.\ Rev.\ B} \textbf{\bibinfo{volume}{46}},
  \bibinfo{pages}{998} (\bibinfo{year}{1992}).

\bibitem[{\citenamefont{Kasuya}(1996)}]{kas96}
\bibinfo{author}{\bibfnamefont{T.}~\bibnamefont{Kasuya}}, \bibinfo{journal}{J.\
  Phys.\ Soc.\ Jpn.} \textbf{\bibinfo{volume}{65}}, \bibinfo{pages}{2548}
  (\bibinfo{year}{1996}).

\bibitem[{\citenamefont{Riseborough}(2000{\natexlab{b}})}]{ris00b}
\bibinfo{author}{\bibfnamefont{P.~S.} \bibnamefont{Riseborough}},
  \bibinfo{journal}{Ann.\ Phys.} \textbf{\bibinfo{volume}{9}},
  \bibinfo{pages}{813} (\bibinfo{year}{2000}{\natexlab{b}}).

\bibitem[{\citenamefont{Moore and Balents}(2007)}]{moo07}
\bibinfo{author}{\bibfnamefont{J.~E.} \bibnamefont{Moore}} \bibnamefont{and}
  \bibinfo{author}{\bibfnamefont{L.}~\bibnamefont{Balents}},
  \bibinfo{journal}{Phys.\ Rev.\ B} \textbf{\bibinfo{volume}{75}},
  \bibinfo{pages}{121306(R)} (\bibinfo{year}{2007}).

\bibitem[{\citenamefont{Fu et~al.}(2007)\citenamefont{Fu, Kane, and
  Mele}}]{fuk07}
\bibinfo{author}{\bibfnamefont{L.}~\bibnamefont{Fu}},
  \bibinfo{author}{\bibfnamefont{C.~L.} \bibnamefont{Kane}}, \bibnamefont{and}
  \bibinfo{author}{\bibfnamefont{E.~J.} \bibnamefont{Mele}},
  \bibinfo{journal}{Phys.\ Rev.\ Lett.} \textbf{\bibinfo{volume}{98}},
  \bibinfo{pages}{106803} (\bibinfo{year}{2007}).

\bibitem[{\citenamefont{Hsieh et~al.}(2008)\citenamefont{Hsieh, Qian, Wray,
  Xia, Hor, Cava, and Hasan}}]{hsi08}
\bibinfo{author}{\bibfnamefont{D.}~\bibnamefont{Hsieh}},
  \bibinfo{author}{\bibfnamefont{D.}~\bibnamefont{Qian}},
  \bibinfo{author}{\bibfnamefont{L.}~\bibnamefont{Wray}},
  \bibinfo{author}{\bibfnamefont{Y.}~\bibnamefont{Xia}},
  \bibinfo{author}{\bibfnamefont{Y.~S.} \bibnamefont{Hor}},
  \bibinfo{author}{\bibfnamefont{R.~J.} \bibnamefont{Cava}}, \bibnamefont{and}
  \bibinfo{author}{\bibfnamefont{M.~Z.} \bibnamefont{Hasan}},
  \bibinfo{journal}{Nature} \textbf{\bibinfo{volume}{452}},
  \bibinfo{pages}{970} (\bibinfo{year}{2008}).

\bibitem[{\citenamefont{Roy}(2009)}]{roy09}
\bibinfo{author}{\bibfnamefont{R.}~\bibnamefont{Roy}}, \bibinfo{journal}{Phys.\
  Rev.\ B} \textbf{\bibinfo{volume}{79}}, \bibinfo{pages}{195321}
  (\bibinfo{year}{2009}).

\bibitem[{\citenamefont{Dzero et~al.}(2016)\citenamefont{Dzero, Xia, Galitski,
  and Coleman}}]{dze16}
\bibinfo{author}{\bibfnamefont{M.}~\bibnamefont{Dzero}},
  \bibinfo{author}{\bibfnamefont{J.}~\bibnamefont{Xia}},
  \bibinfo{author}{\bibfnamefont{V.}~\bibnamefont{Galitski}}, \bibnamefont{and}
  \bibinfo{author}{\bibfnamefont{P.}~\bibnamefont{Coleman}},
  \bibinfo{journal}{Annu.\ Rev.\ Condens.\ Matter Phys.}
  \textbf{\bibinfo{volume}{7}}, \bibinfo{pages}{249} (\bibinfo{year}{2016}).

\bibitem[{\citenamefont{Wakeham
  et~al.}(2016{\natexlab{a}})\citenamefont{Wakeham, Rosa, Wang, Kang, Fisk,
  Ronning, and Thompson}}]{wak16b}
\bibinfo{author}{\bibfnamefont{N.}~\bibnamefont{Wakeham}},
  \bibinfo{author}{\bibfnamefont{P.~F.~S.} \bibnamefont{Rosa}},
  \bibinfo{author}{\bibfnamefont{Y.~Q.} \bibnamefont{Wang}},
  \bibinfo{author}{\bibfnamefont{M.}~\bibnamefont{Kang}},
  \bibinfo{author}{\bibfnamefont{Z.}~\bibnamefont{Fisk}},
  \bibinfo{author}{\bibfnamefont{F.}~\bibnamefont{Ronning}}, \bibnamefont{and}
  \bibinfo{author}{\bibfnamefont{J.~D.} \bibnamefont{Thompson}},
  \bibinfo{journal}{Phys.\ Rev.\ B} \textbf{\bibinfo{volume}{94}},
  \bibinfo{pages}{035127} (\bibinfo{year}{2016}{\natexlab{a}}).

\bibitem[{\citenamefont{Hagiwara et~al.}(2016)\citenamefont{Hagiwara, Ohtsubo,
  Matsunami, Ideta, Tanaka, Miyazaki, Rault, {Le F{\`e}vre}, Bertran,
  {Taleb-Ibrahimi} et~al.}}]{hag16}
\bibinfo{author}{\bibfnamefont{K.}~\bibnamefont{Hagiwara}},
  \bibinfo{author}{\bibfnamefont{Y.}~\bibnamefont{Ohtsubo}},
  \bibinfo{author}{\bibfnamefont{M.}~\bibnamefont{Matsunami}},
  \bibinfo{author}{\bibfnamefont{S.-i.} \bibnamefont{Ideta}},
  \bibinfo{author}{\bibfnamefont{K.}~\bibnamefont{Tanaka}},
  \bibinfo{author}{\bibfnamefont{H.}~\bibnamefont{Miyazaki}},
  \bibinfo{author}{\bibfnamefont{J.~E.} \bibnamefont{Rault}},
  \bibinfo{author}{\bibfnamefont{P.}~\bibnamefont{{Le F{\`e}vre}}},
  \bibinfo{author}{\bibfnamefont{F.}~\bibnamefont{Bertran}},
  \bibinfo{author}{\bibfnamefont{A.}~\bibnamefont{{Taleb-Ibrahimi}}},
  \bibnamefont{et~al.}, \bibinfo{journal}{Nat.\ Commun.}
  \textbf{\bibinfo{volume}{7}}, \bibinfo{pages}{12690} (\bibinfo{year}{2016}).

\bibitem[{\citenamefont{Jiao et~al.}(2018)\citenamefont{Jiao, R{\"o}{\ss}ler,
  Kasinathan, Rosa, Guo, Yuan, Liu, Fisk, Steglich, and Wirth}}]{jiao18}
\bibinfo{author}{\bibfnamefont{L.}~\bibnamefont{Jiao}},
  \bibinfo{author}{\bibfnamefont{S.}~\bibnamefont{R{\"o}{\ss}ler}},
  \bibinfo{author}{\bibfnamefont{D.}~\bibnamefont{Kasinathan}},
  \bibinfo{author}{\bibfnamefont{P.~F.~S.} \bibnamefont{Rosa}},
  \bibinfo{author}{\bibfnamefont{C.}~\bibnamefont{Guo}},
  \bibinfo{author}{\bibfnamefont{H.}~\bibnamefont{Yuan}},
  \bibinfo{author}{\bibfnamefont{C.-X.} \bibnamefont{Liu}},
  \bibinfo{author}{\bibfnamefont{Z.}~\bibnamefont{Fisk}},
  \bibinfo{author}{\bibfnamefont{F.}~\bibnamefont{Steglich}}, \bibnamefont{and}
  \bibinfo{author}{\bibfnamefont{S.}~\bibnamefont{Wirth}},
  \bibinfo{journal}{Sci. Adv.} \textbf{\bibinfo{volume}{4}},
  \bibinfo{pages}{eaau4886} (\bibinfo{year}{2018}).

\bibitem[{\citenamefont{Vainshtein et~al.}(1965)\citenamefont{Vainshtein,
  Blokhin, and Paderno}}]{vai64}
\bibinfo{author}{\bibfnamefont{E.~E.} \bibnamefont{Vainshtein}},
  \bibinfo{author}{\bibfnamefont{S.~M.} \bibnamefont{Blokhin}},
  \bibnamefont{and} \bibinfo{author}{\bibfnamefont{Y.~B.}
  \bibnamefont{Paderno}}, \bibinfo{journal}{Sov.\ Phys.-Solid State}
  \textbf{\bibinfo{volume}{6}}, \bibinfo{pages}{2318} (\bibinfo{year}{1965}).

\bibitem[{\citenamefont{Sluchanko et~al.}(1999)\citenamefont{Sluchanko, Volkov,
  Glushkov, Gorshunov, Demishev, Kondrin, Pronin, and Samarin}}]{slu99}
\bibinfo{author}{\bibfnamefont{N.~E.} \bibnamefont{Sluchanko}},
  \bibinfo{author}{\bibfnamefont{A.~A.} \bibnamefont{Volkov}},
  \bibinfo{author}{\bibfnamefont{V.~V.} \bibnamefont{Glushkov}},
  \bibinfo{author}{\bibfnamefont{B.~P.} \bibnamefont{Gorshunov}},
  \bibinfo{author}{\bibfnamefont{S.~V.} \bibnamefont{Demishev}},
  \bibinfo{author}{\bibfnamefont{M.~V.} \bibnamefont{Kondrin}},
  \bibinfo{author}{\bibfnamefont{A.~A.} \bibnamefont{Pronin}},
  \bibnamefont{and} \bibinfo{author}{\bibfnamefont{N.~A.}
  \bibnamefont{Samarin}}, \bibinfo{journal}{J.\ Exp.\ Theor.\ Phys.}
  \textbf{\bibinfo{volume}{88}}, \bibinfo{pages}{533} (\bibinfo{year}{1999}).

\bibitem[{\citenamefont{Cohen et~al.}(1970)\citenamefont{Cohen, Eibsch{\"u}tz,
  and West}}]{coh70}
\bibinfo{author}{\bibfnamefont{R.~L.} \bibnamefont{Cohen}},
  \bibinfo{author}{\bibfnamefont{M.}~\bibnamefont{Eibsch{\"u}tz}},
  \bibnamefont{and} \bibinfo{author}{\bibfnamefont{K.~W.} \bibnamefont{West}},
  \bibinfo{journal}{Phys.\ Rev.\ Lett.} \textbf{\bibinfo{volume}{24}},
  \bibinfo{pages}{383} (\bibinfo{year}{1970}).

\bibitem[{\citenamefont{Takigawa et~al.}(1981)\citenamefont{Takigawa, Yasuoka,
  Kitaoka, Tanaka, Nozaki, and Ishizawa}}]{tak81}
\bibinfo{author}{\bibfnamefont{M.}~\bibnamefont{Takigawa}},
  \bibinfo{author}{\bibfnamefont{H.}~\bibnamefont{Yasuoka}},
  \bibinfo{author}{\bibfnamefont{Y.}~\bibnamefont{Kitaoka}},
  \bibinfo{author}{\bibfnamefont{T.}~\bibnamefont{Tanaka}},
  \bibinfo{author}{\bibfnamefont{H.}~\bibnamefont{Nozaki}}, \bibnamefont{and}
  \bibinfo{author}{\bibfnamefont{Y.}~\bibnamefont{Ishizawa}},
  \bibinfo{journal}{J.\ Phys.\ Soc.\ Jpn.} \textbf{\bibinfo{volume}{50}},
  \bibinfo{pages}{2525} (\bibinfo{year}{1981}).

\bibitem[{\citenamefont{Schlottmann}(2014)}]{sch14}
\bibinfo{author}{\bibfnamefont{P.}~\bibnamefont{Schlottmann}},
  \bibinfo{journal}{Phys.\ Rev.\ B} \textbf{\bibinfo{volume}{90}},
  \bibinfo{pages}{165127} (\bibinfo{year}{2014}).

\bibitem[{\citenamefont{Glushkov et~al.}(2006)\citenamefont{Glushkov,
  Kuznetsov, Churkin, Demishev, Paderno, Shitsevalova, and Sluchanko}}]{glu06}
\bibinfo{author}{\bibfnamefont{V.~V.} \bibnamefont{Glushkov}},
  \bibinfo{author}{\bibfnamefont{A.~V.} \bibnamefont{Kuznetsov}},
  \bibinfo{author}{\bibfnamefont{O.~A.} \bibnamefont{Churkin}},
  \bibinfo{author}{\bibfnamefont{S.~V.} \bibnamefont{Demishev}},
  \bibinfo{author}{\bibfnamefont{Y.~B.} \bibnamefont{Paderno}},
  \bibinfo{author}{\bibfnamefont{N.~Y.} \bibnamefont{Shitsevalova}},
  \bibnamefont{and} \bibinfo{author}{\bibfnamefont{N.~E.}
  \bibnamefont{Sluchanko}}, \bibinfo{journal}{Physica B}
  \textbf{\bibinfo{volume}{378-380}}, \bibinfo{pages}{614}
  (\bibinfo{year}{2006}).

\bibitem[{\citenamefont{Thomas et~al.}(2016)\citenamefont{Thomas, Kim, Chung,
  Grant, Fisk, and Xia}}]{tho16}
\bibinfo{author}{\bibfnamefont{S.}~\bibnamefont{Thomas}},
  \bibinfo{author}{\bibfnamefont{D.~J.} \bibnamefont{Kim}},
  \bibinfo{author}{\bibfnamefont{S.~B.} \bibnamefont{Chung}},
  \bibinfo{author}{\bibfnamefont{T.}~\bibnamefont{Grant}},
  \bibinfo{author}{\bibfnamefont{Z.}~\bibnamefont{Fisk}}, \bibnamefont{and}
  \bibinfo{author}{\bibfnamefont{J.}~\bibnamefont{Xia}},
  \bibinfo{journal}{Phys.\ Rev.\ B} \textbf{\bibinfo{volume}{94}},
  \bibinfo{pages}{205114} (\bibinfo{year}{2016}).

\bibitem[{\citenamefont{Zhu et~al.}(2013)\citenamefont{Zhu, Nicolaou, Levy,
  Butch, Syers, Wang, Paglione, Sawatzky, Elfimov, and Damascelli}}]{zhu13}
\bibinfo{author}{\bibfnamefont{Z.-H.} \bibnamefont{Zhu}},
  \bibinfo{author}{\bibfnamefont{A.}~\bibnamefont{Nicolaou}},
  \bibinfo{author}{\bibfnamefont{G.}~\bibnamefont{Levy}},
  \bibinfo{author}{\bibfnamefont{N.~P.} \bibnamefont{Butch}},
  \bibinfo{author}{\bibfnamefont{P.}~\bibnamefont{Syers}},
  \bibinfo{author}{\bibfnamefont{X.~F.} \bibnamefont{Wang}},
  \bibinfo{author}{\bibfnamefont{J.}~\bibnamefont{Paglione}},
  \bibinfo{author}{\bibfnamefont{G.~A.} \bibnamefont{Sawatzky}},
  \bibinfo{author}{\bibfnamefont{I.~S.} \bibnamefont{Elfimov}},
  \bibnamefont{and}
  \bibinfo{author}{\bibfnamefont{A.}~\bibnamefont{Damascelli}},
  \bibinfo{journal}{Phys.\ Rev.\ Lett.} \textbf{\bibinfo{volume}{111}},
  \bibinfo{pages}{216402} (\bibinfo{year}{2013}).

\bibitem[{\citenamefont{Denlinger et~al.}(2013)\citenamefont{Denlinger, Allen,
  Kang, Sun, Kim, Shim, Min, Kim, and Fisk}}]{den13}
\bibinfo{author}{\bibfnamefont{J.~D.} \bibnamefont{Denlinger}},
  \bibinfo{author}{\bibfnamefont{J.~W.} \bibnamefont{Allen}},
  \bibinfo{author}{\bibfnamefont{J.-S.} \bibnamefont{Kang}},
  \bibinfo{author}{\bibfnamefont{K.}~\bibnamefont{Sun}},
  \bibinfo{author}{\bibfnamefont{J.-W.} \bibnamefont{Kim}},
  \bibinfo{author}{\bibfnamefont{J.~H.} \bibnamefont{Shim}},
  \bibinfo{author}{\bibfnamefont{B.~I.} \bibnamefont{Min}},
  \bibinfo{author}{\bibfnamefont{D.-J.} \bibnamefont{Kim}}, \bibnamefont{and}
  \bibinfo{author}{\bibfnamefont{Z.}~\bibnamefont{Fisk}},
  \emph{\bibinfo{title}{Temperature dependence of linked gap and surface state
  evolution in the mixed valent topological insulator {S}m{B}$_6$}},
  \bibinfo{howpublished}{({P}reprint) {a}rXiv:1312.6637}
  (\bibinfo{year}{2013}).

\bibitem[{\citenamefont{Hsieh et~al.}(2009)\citenamefont{Hsieh, Xia, Qian,
  Wray, Dil, Meier, Osterwalder, Patthey, Checkelsky, Ong et~al.}}]{hsi09}
\bibinfo{author}{\bibfnamefont{D.}~\bibnamefont{Hsieh}},
  \bibinfo{author}{\bibfnamefont{Y.}~\bibnamefont{Xia}},
  \bibinfo{author}{\bibfnamefont{D.}~\bibnamefont{Qian}},
  \bibinfo{author}{\bibfnamefont{L.}~\bibnamefont{Wray}},
  \bibinfo{author}{\bibfnamefont{J.~H.} \bibnamefont{Dil}},
  \bibinfo{author}{\bibfnamefont{F.}~\bibnamefont{Meier}},
  \bibinfo{author}{\bibfnamefont{J.}~\bibnamefont{Osterwalder}},
  \bibinfo{author}{\bibfnamefont{L.}~\bibnamefont{Patthey}},
  \bibinfo{author}{\bibfnamefont{J.~G.} \bibnamefont{Checkelsky}},
  \bibinfo{author}{\bibfnamefont{N.~P.} \bibnamefont{Ong}},
  \bibnamefont{et~al.}, \bibinfo{journal}{Nature}
  \textbf{\bibinfo{volume}{460}}, \bibinfo{pages}{1101} (\bibinfo{year}{2009}).

\bibitem[{\citenamefont{Wolgast et~al.}(2013)\citenamefont{Wolgast, Kurdak,
  Sun, Allen, Kim, and Fisk}}]{wol13}
\bibinfo{author}{\bibfnamefont{S.}~\bibnamefont{Wolgast}},
  \bibinfo{author}{\bibfnamefont{C.}~\bibnamefont{Kurdak}},
  \bibinfo{author}{\bibfnamefont{K.}~\bibnamefont{Sun}},
  \bibinfo{author}{\bibfnamefont{J.~W.} \bibnamefont{Allen}},
  \bibinfo{author}{\bibfnamefont{D.-J.} \bibnamefont{Kim}}, \bibnamefont{and}
  \bibinfo{author}{\bibfnamefont{Z.}~\bibnamefont{Fisk}},
  \bibinfo{journal}{Phys.\ Rev.\ B} \textbf{\bibinfo{volume}{88}},
  \bibinfo{pages}{180405(R)} (\bibinfo{year}{2013}).

\bibitem[{\citenamefont{Kim et~al.}(2013)\citenamefont{Kim, Thomas, Grant,
  Botimer, Fisk, and Xia}}]{kim13}
\bibinfo{author}{\bibfnamefont{D.~J.} \bibnamefont{Kim}},
  \bibinfo{author}{\bibfnamefont{S.}~\bibnamefont{Thomas}},
  \bibinfo{author}{\bibfnamefont{T.}~\bibnamefont{Grant}},
  \bibinfo{author}{\bibfnamefont{J.}~\bibnamefont{Botimer}},
  \bibinfo{author}{\bibfnamefont{Z.}~\bibnamefont{Fisk}}, \bibnamefont{and}
  \bibinfo{author}{\bibfnamefont{J.}~\bibnamefont{Xia}},
  \bibinfo{journal}{Sci.\ Rep.} \textbf{\bibinfo{volume}{3}},
  \bibinfo{pages}{3150} (\bibinfo{year}{2013}).

\bibitem[{\citenamefont{Eo et~al.}(2018)\citenamefont{Eo, Sun, Kurdak, Kim, and
  Fisk}}]{eo18}
\bibinfo{author}{\bibfnamefont{Y.~S.} \bibnamefont{Eo}},
  \bibinfo{author}{\bibfnamefont{K.}~\bibnamefont{Sun}},
  \bibinfo{author}{\bibfnamefont{C.}~\bibnamefont{Kurdak}},
  \bibinfo{author}{\bibfnamefont{D.-J.} \bibnamefont{Kim}}, \bibnamefont{and}
  \bibinfo{author}{\bibfnamefont{Z.}~\bibnamefont{Fisk}},
  \bibinfo{journal}{Phys.\ Rev.\ Appl.} \textbf{\bibinfo{volume}{9}},
  \bibinfo{pages}{044006} (\bibinfo{year}{2018}).

\bibitem[{\citenamefont{Eo et~al.}(2019)\citenamefont{Eo, Rakoski, Lucien,
  Mihaliov, Kurdak, Rosa, and Fisk}}]{eo19}
\bibinfo{author}{\bibfnamefont{Y.~S.} \bibnamefont{Eo}},
  \bibinfo{author}{\bibfnamefont{A.}~\bibnamefont{Rakoski}},
  \bibinfo{author}{\bibfnamefont{J.}~\bibnamefont{Lucien}},
  \bibinfo{author}{\bibfnamefont{D.}~\bibnamefont{Mihaliov}},
  \bibinfo{author}{\bibfnamefont{C.}~\bibnamefont{Kurdak}},
  \bibinfo{author}{\bibfnamefont{P.~F.~S.} \bibnamefont{Rosa}},
  \bibnamefont{and} \bibinfo{author}{\bibfnamefont{Z.}~\bibnamefont{Fisk}},
  \bibinfo{journal}{Proc.\ Natl.\ Acad.\ Sci.\ USA}
  \textbf{\bibinfo{volume}{116}}, \bibinfo{pages}{12638}
  (\bibinfo{year}{2019}).

\bibitem[{\citenamefont{Kim et~al.}(2012)\citenamefont{Kim, Grant, and
  Fisk}}]{kim12}
\bibinfo{author}{\bibfnamefont{D.~J.} \bibnamefont{Kim}},
  \bibinfo{author}{\bibfnamefont{T.}~\bibnamefont{Grant}}, \bibnamefont{and}
  \bibinfo{author}{\bibfnamefont{Z.}~\bibnamefont{Fisk}},
  \bibinfo{journal}{Phys.\ Rev.\ Lett.} \textbf{\bibinfo{volume}{109}},
  \bibinfo{pages}{096601} (\bibinfo{year}{2012}).

\bibitem[{\citenamefont{Stern et~al.}(2016)\citenamefont{Stern, Efimkin,
  Galitski, Fisk, and Xia}}]{ste16}
\bibinfo{author}{\bibfnamefont{A.}~\bibnamefont{Stern}},
  \bibinfo{author}{\bibfnamefont{D.~K.} \bibnamefont{Efimkin}},
  \bibinfo{author}{\bibfnamefont{V.}~\bibnamefont{Galitski}},
  \bibinfo{author}{\bibfnamefont{Z.}~\bibnamefont{Fisk}}, \bibnamefont{and}
  \bibinfo{author}{\bibfnamefont{J.}~\bibnamefont{Xia}},
  \bibinfo{journal}{Phys.\ Rev.\ Lett.} \textbf{\bibinfo{volume}{116}},
  \bibinfo{pages}{166603} (\bibinfo{year}{2016}).

\bibitem[{\citenamefont{Casas et~al.}(2018)\citenamefont{Casas, Stern, Efimkin,
  Fisk, and Xia}}]{cas18}
\bibinfo{author}{\bibfnamefont{B.}~\bibnamefont{Casas}},
  \bibinfo{author}{\bibfnamefont{A.}~\bibnamefont{Stern}},
  \bibinfo{author}{\bibfnamefont{D.~K.} \bibnamefont{Efimkin}},
  \bibinfo{author}{\bibfnamefont{Z.}~\bibnamefont{Fisk}}, \bibnamefont{and}
  \bibinfo{author}{\bibfnamefont{J.}~\bibnamefont{Xia}},
  \bibinfo{journal}{Phys.\ Rev.\ B} \textbf{\bibinfo{volume}{97}},
  \bibinfo{pages}{035121} (\bibinfo{year}{2018}).

\bibitem[{\citenamefont{Stankiewicz et~al.}(2021)\citenamefont{Stankiewicz,
  Blasco, Schlottmann, {Ciomaga Hatnean}, and Balakrishnan}}]{sta21}
\bibinfo{author}{\bibfnamefont{J.}~\bibnamefont{Stankiewicz}},
  \bibinfo{author}{\bibfnamefont{J.}~\bibnamefont{Blasco}},
  \bibinfo{author}{\bibfnamefont{P.}~\bibnamefont{Schlottmann}},
  \bibinfo{author}{\bibfnamefont{M.}~\bibnamefont{{Ciomaga Hatnean}}},
  \bibnamefont{and}
  \bibinfo{author}{\bibfnamefont{G.}~\bibnamefont{Balakrishnan}},
  \emph{\bibinfo{title}{Impedance spectroscopy of {S}m{B}$_6$ single
  crystals}}, \bibinfo{howpublished}{({P}reprint) {a}rXiv:2109.05313}
  (\bibinfo{year}{2021}).

\bibitem[{\citenamefont{Cahill and Graeve}(2019)}]{cah19}
\bibinfo{author}{\bibfnamefont{J.~T.} \bibnamefont{Cahill}} \bibnamefont{and}
  \bibinfo{author}{\bibfnamefont{O.~A.} \bibnamefont{Graeve}},
  \bibinfo{journal}{J.\ Mater.\ Res.\ Technol.} \textbf{\bibinfo{volume}{8}},
  \bibinfo{pages}{6321} (\bibinfo{year}{2019}).

\bibitem[{\citenamefont{Berrada et~al.}(1978)\citenamefont{Berrada, Mercurio,
  Etourneau, Hagenmuller, and Shroff}}]{ber78}
\bibinfo{author}{\bibfnamefont{A.}~\bibnamefont{Berrada}},
  \bibinfo{author}{\bibfnamefont{J.~P.} \bibnamefont{Mercurio}},
  \bibinfo{author}{\bibfnamefont{J.}~\bibnamefont{Etourneau}},
  \bibinfo{author}{\bibfnamefont{P.}~\bibnamefont{Hagenmuller}},
  \bibnamefont{and} \bibinfo{author}{\bibfnamefont{A.~M.}
  \bibnamefont{Shroff}}, \bibinfo{journal}{J. Less-Common Met.}
  \textbf{\bibinfo{volume}{59}}, \bibinfo{pages}{7} (\bibinfo{year}{1978}).

\bibitem[{\citenamefont{Mercurio et~al.}(1976)\citenamefont{Mercurio,
  Etourneau, Naslain, and Hagenmuller}}]{mer76}
\bibinfo{author}{\bibfnamefont{J.~P.} \bibnamefont{Mercurio}},
  \bibinfo{author}{\bibfnamefont{J.}~\bibnamefont{Etourneau}},
  \bibinfo{author}{\bibfnamefont{R.}~\bibnamefont{Naslain}}, \bibnamefont{and}
  \bibinfo{author}{\bibfnamefont{P.}~\bibnamefont{Hagenmuller}},
  \bibinfo{journal}{J. Less-Common Met.} \textbf{\bibinfo{volume}{47}},
  \bibinfo{pages}{175} (\bibinfo{year}{1976}).

\bibitem[{\citenamefont{Massidda et~al.}(1996)\citenamefont{Massidda,
  Continenza, {de Pascale}, and Monnier}}]{mas96}
\bibinfo{author}{\bibfnamefont{S.}~\bibnamefont{Massidda}},
  \bibinfo{author}{\bibfnamefont{A.}~\bibnamefont{Continenza}},
  \bibinfo{author}{\bibfnamefont{T.~M.} \bibnamefont{{de Pascale}}},
  \bibnamefont{and} \bibinfo{author}{\bibfnamefont{R.}~\bibnamefont{Monnier}},
  \bibinfo{journal}{Z.\ Phys.\ B} \textbf{\bibinfo{volume}{102}},
  \bibinfo{pages}{83} (\bibinfo{year}{1996}).

\bibitem[{\citenamefont{Ahmed and Broers}(1972)}]{ahm72}
\bibinfo{author}{\bibfnamefont{H.}~\bibnamefont{Ahmed}} \bibnamefont{and}
  \bibinfo{author}{\bibfnamefont{A.~N.} \bibnamefont{Broers}},
  \bibinfo{journal}{J.\ Appl.\ Phys.} \textbf{\bibinfo{volume}{43}},
  \bibinfo{pages}{2185} (\bibinfo{year}{1972}).

\bibitem[{\citenamefont{Canfield and Fisk}(1992)}]{can92}
\bibinfo{author}{\bibfnamefont{P.~C.} \bibnamefont{Canfield}} \bibnamefont{and}
  \bibinfo{author}{\bibfnamefont{Z.}~\bibnamefont{Fisk}},
  \bibinfo{journal}{Philos.\ Mag.\ B} \textbf{\bibinfo{volume}{65}},
  \bibinfo{pages}{1117} (\bibinfo{year}{1992}).

\bibitem[{\citenamefont{Rosa and Fisk}(2018)}]{rosa18}
\bibinfo{author}{\bibfnamefont{P.~F.~S.} \bibnamefont{Rosa}} \bibnamefont{and}
  \bibinfo{author}{\bibfnamefont{Z.}~\bibnamefont{Fisk}}, in
  \emph{\bibinfo{booktitle}{Crystal Growth of Intermetallics}}, edited by
  \bibinfo{editor}{\bibfnamefont{P.}~\bibnamefont{Gille}} \bibnamefont{and}
  \bibinfo{editor}{\bibfnamefont{Y.}~\bibnamefont{Grin}}
  (\bibinfo{publisher}{Berlin, Boston: De Gruyter}, \bibinfo{year}{2018}), pp.
  \bibinfo{pages}{49--60}.

\bibitem[{\citenamefont{Tanaka et~al.}(1980)\citenamefont{Tanaka, Nishitani,
  Oshima, Bannai, and Kawai}}]{tan80}
\bibinfo{author}{\bibfnamefont{T.}~\bibnamefont{Tanaka}},
  \bibinfo{author}{\bibfnamefont{R.}~\bibnamefont{Nishitani}},
  \bibinfo{author}{\bibfnamefont{C.}~\bibnamefont{Oshima}},
  \bibinfo{author}{\bibfnamefont{E.}~\bibnamefont{Bannai}}, \bibnamefont{and}
  \bibinfo{author}{\bibfnamefont{S.}~\bibnamefont{Kawai}},
  \bibinfo{journal}{J.\ Appl.\ Phys.} \textbf{\bibinfo{volume}{51}},
  \bibinfo{pages}{3877} (\bibinfo{year}{1980}).

\bibitem[{\citenamefont{Paderno et~al.}(1981)\citenamefont{Paderno, Lazorenko,
  and Kovalev}}]{pad81}
\bibinfo{author}{\bibfnamefont{Y.~B.} \bibnamefont{Paderno}},
  \bibinfo{author}{\bibfnamefont{V.~I.} \bibnamefont{Lazorenko}},
  \bibnamefont{and} \bibinfo{author}{\bibfnamefont{A.~V.}
  \bibnamefont{Kovalev}}, \bibinfo{journal}{Powder Metall.\ Met.\ Ceram.}
  \textbf{\bibinfo{volume}{20}}, \bibinfo{pages}{717} (\bibinfo{year}{1981}).

\bibitem[{\citenamefont{{Ciomaga Hatnean} et~al.}(2013)\citenamefont{{Ciomaga
  Hatnean}, Lees, Paul, and Balakrishnan}}]{hat13}
\bibinfo{author}{\bibfnamefont{M.}~\bibnamefont{{Ciomaga Hatnean}}},
  \bibinfo{author}{\bibfnamefont{M.~R.} \bibnamefont{Lees}},
  \bibinfo{author}{\bibfnamefont{D.~M.} \bibnamefont{Paul}}, \bibnamefont{and}
  \bibinfo{author}{\bibfnamefont{G.}~\bibnamefont{Balakrishnan}},
  \bibinfo{journal}{Sci.\ Rep.} \textbf{\bibinfo{volume}{3}},
  \bibinfo{pages}{3071} (\bibinfo{year}{2013}).

\bibitem[{\citenamefont{Prokofiev}(2018)}]{pro18}
\bibinfo{author}{\bibfnamefont{A.}~\bibnamefont{Prokofiev}}, in
  \emph{\bibinfo{booktitle}{Crystal Growth of Intermetallics}}, edited by
  \bibinfo{editor}{\bibfnamefont{P.}~\bibnamefont{Gille}} \bibnamefont{and}
  \bibinfo{editor}{\bibfnamefont{Y.}~\bibnamefont{Grin}}
  (\bibinfo{publisher}{Berlin, Boston: De Gruyter}, \bibinfo{year}{2018}), pp.
  \bibinfo{pages}{91--116}.

\bibitem[{\citenamefont{Fuhrman et~al.}(2018)\citenamefont{Fuhrman, Chamorro,
  Alekseev, Mignot, Keller, Rodriguez-Rivera, Qiu, Nikoli\'{c}, McQueen, and
  Broholm}}]{fuh17}
\bibinfo{author}{\bibfnamefont{W.~T.} \bibnamefont{Fuhrman}},
  \bibinfo{author}{\bibfnamefont{J.~R.} \bibnamefont{Chamorro}},
  \bibinfo{author}{\bibfnamefont{P.}~\bibnamefont{Alekseev}},
  \bibinfo{author}{\bibfnamefont{J.-M.} \bibnamefont{Mignot}},
  \bibinfo{author}{\bibfnamefont{T.}~\bibnamefont{Keller}},
  \bibinfo{author}{\bibfnamefont{J.~A.} \bibnamefont{Rodriguez-Rivera}},
  \bibinfo{author}{\bibfnamefont{Y.}~\bibnamefont{Qiu}},
  \bibinfo{author}{\bibfnamefont{P.}~\bibnamefont{Nikoli\'{c}}},
  \bibinfo{author}{\bibfnamefont{T.~M.} \bibnamefont{McQueen}},
  \bibnamefont{and} \bibinfo{author}{\bibfnamefont{C.~L.}
  \bibnamefont{Broholm}}, \bibinfo{journal}{Nat.\ Commun.}
  \textbf{\bibinfo{volume}{9}}, \bibinfo{pages}{1539} (\bibinfo{year}{2018}).

\bibitem[{\citenamefont{Phelan et~al.}(2016)\citenamefont{Phelan, Koohpayeh,
  Cottingham, Tutmaher, Leiner, Lumsden, Lavelle, Wang, Hoffmann, Siegler
  et~al.}}]{phe16}
\bibinfo{author}{\bibfnamefont{W.~A.} \bibnamefont{Phelan}},
  \bibinfo{author}{\bibfnamefont{S.~M.} \bibnamefont{Koohpayeh}},
  \bibinfo{author}{\bibfnamefont{P.}~\bibnamefont{Cottingham}},
  \bibinfo{author}{\bibfnamefont{J.~A.} \bibnamefont{Tutmaher}},
  \bibinfo{author}{\bibfnamefont{J.~C.} \bibnamefont{Leiner}},
  \bibinfo{author}{\bibfnamefont{M.~D.} \bibnamefont{Lumsden}},
  \bibinfo{author}{\bibfnamefont{C.~M.} \bibnamefont{Lavelle}},
  \bibinfo{author}{\bibfnamefont{X.~P.} \bibnamefont{Wang}},
  \bibinfo{author}{\bibfnamefont{C.}~\bibnamefont{Hoffmann}},
  \bibinfo{author}{\bibfnamefont{M.~A.} \bibnamefont{Siegler}},
  \bibnamefont{et~al.}, \bibinfo{journal}{Sci.\ Rep.}
  \textbf{\bibinfo{volume}{6}}, \bibinfo{pages}{20860} (\bibinfo{year}{2016}).

\bibitem[{\citenamefont{Orend\'{a}\v{c}
  et~al.}(2017)\citenamefont{Orend\'{a}\v{c}, Gab\'{a}ni, Prist\'{a}\v{s},
  Ga\v{z}o, Diko, Farka\v{s}ovsk\'{y}, Levchenko, Shitsevalova, and
  Flachbart}}]{ore17}
\bibinfo{author}{\bibfnamefont{M.}~\bibnamefont{Orend\'{a}\v{c}}},
  \bibinfo{author}{\bibfnamefont{S.}~\bibnamefont{Gab\'{a}ni}},
  \bibinfo{author}{\bibfnamefont{G.}~\bibnamefont{Prist\'{a}\v{s}}},
  \bibinfo{author}{\bibfnamefont{E.}~\bibnamefont{Ga\v{z}o}},
  \bibinfo{author}{\bibfnamefont{P.}~\bibnamefont{Diko}},
  \bibinfo{author}{\bibfnamefont{P.}~\bibnamefont{Farka\v{s}ovsk\'{y}}},
  \bibinfo{author}{\bibfnamefont{A.}~\bibnamefont{Levchenko}},
  \bibinfo{author}{\bibfnamefont{N.}~\bibnamefont{Shitsevalova}},
  \bibnamefont{and}
  \bibinfo{author}{\bibfnamefont{K.}~\bibnamefont{Flachbart}},
  \bibinfo{journal}{Phys.\ Rev.\ B} \textbf{\bibinfo{volume}{96}},
  \bibinfo{pages}{115101} (\bibinfo{year}{2017}).

\bibitem[{\citenamefont{Eo et~al.}(2021)\citenamefont{Eo, Rakoski, Sinha,
  Mihaliov, Fuhrman, Saha, Rosa, Fisk, {Ciomaga Hatnean}, Balakrishnan
  et~al.}}]{eo20b}
\bibinfo{author}{\bibfnamefont{Y.~S.} \bibnamefont{Eo}},
  \bibinfo{author}{\bibfnamefont{A.}~\bibnamefont{Rakoski}},
  \bibinfo{author}{\bibfnamefont{S.}~\bibnamefont{Sinha}},
  \bibinfo{author}{\bibfnamefont{D.}~\bibnamefont{Mihaliov}},
  \bibinfo{author}{\bibfnamefont{W.~T.} \bibnamefont{Fuhrman}},
  \bibinfo{author}{\bibfnamefont{S.~R.} \bibnamefont{Saha}},
  \bibinfo{author}{\bibfnamefont{P.~F.~S.} \bibnamefont{Rosa}},
  \bibinfo{author}{\bibfnamefont{Z.}~\bibnamefont{Fisk}},
  \bibinfo{author}{\bibfnamefont{M.}~\bibnamefont{{Ciomaga Hatnean}}},
  \bibinfo{author}{\bibfnamefont{G.}~\bibnamefont{Balakrishnan}},
  \bibnamefont{et~al.}, \bibinfo{journal}{Phys.\ Rev.\ Materials}
  \textbf{\bibinfo{volume}{5}}, \bibinfo{pages}{055001} (\bibinfo{year}{2021}).

\bibitem[{\citenamefont{Thomas et~al.}(2019)\citenamefont{Thomas, Ding,
  Ronning, Zapf, Thompson, Fisk, Xia, and Rosa}}]{tho19}
\bibinfo{author}{\bibfnamefont{S.}~\bibnamefont{Thomas}},
  \bibinfo{author}{\bibfnamefont{X.}~\bibnamefont{Ding}},
  \bibinfo{author}{\bibfnamefont{F.}~\bibnamefont{Ronning}},
  \bibinfo{author}{\bibfnamefont{V.}~\bibnamefont{Zapf}},
  \bibinfo{author}{\bibfnamefont{J.}~\bibnamefont{Thompson}},
  \bibinfo{author}{\bibfnamefont{Z.}~\bibnamefont{Fisk}},
  \bibinfo{author}{\bibfnamefont{J.}~\bibnamefont{Xia}}, \bibnamefont{and}
  \bibinfo{author}{\bibfnamefont{P.}~\bibnamefont{Rosa}},
  \bibinfo{journal}{Phys.\ Rev.\ Lett.} \textbf{\bibinfo{volume}{122}},
  \bibinfo{pages}{166401} (\bibinfo{year}{2019}).

\bibitem[{\citenamefont{Grushko et~al.}(1985)\citenamefont{Grushko, Paderno,
  Mishin, Molkanov, Shadrina, Konovalova, and Dudnik}}]{gru85}
\bibinfo{author}{\bibfnamefont{Y.~S.} \bibnamefont{Grushko}},
  \bibinfo{author}{\bibfnamefont{Y.~B.} \bibnamefont{Paderno}},
  \bibinfo{author}{\bibfnamefont{K.~Y.} \bibnamefont{Mishin}},
  \bibinfo{author}{\bibfnamefont{L.~I.} \bibnamefont{Molkanov}},
  \bibinfo{author}{\bibfnamefont{G.~A.} \bibnamefont{Shadrina}},
  \bibinfo{author}{\bibfnamefont{E.~S.} \bibnamefont{Konovalova}},
  \bibnamefont{and} \bibinfo{author}{\bibfnamefont{E.~M.}
  \bibnamefont{Dudnik}}, \bibinfo{journal}{Phys. Stat. Sol. B}
  \textbf{\bibinfo{volume}{128}}, \bibinfo{pages}{591} (\bibinfo{year}{1985}).

\bibitem[{\citenamefont{{Ale Crivillero}
  et~al.}(2021{\natexlab{a}})\citenamefont{{Ale Crivillero}, R{\"o}{\ss}ler,
  Borrmann, Dawczak-D\c{e}bicki, Rosa, Fisk, and Wirth}}]{ale21}
\bibinfo{author}{\bibfnamefont{M.~V.} \bibnamefont{{Ale Crivillero}}},
  \bibinfo{author}{\bibfnamefont{S.}~\bibnamefont{R{\"o}{\ss}ler}},
  \bibinfo{author}{\bibfnamefont{H.}~\bibnamefont{Borrmann}},
  \bibinfo{author}{\bibfnamefont{H.}~\bibnamefont{Dawczak-D\c{e}bicki}},
  \bibinfo{author}{\bibfnamefont{P.~F.~S.} \bibnamefont{Rosa}},
  \bibinfo{author}{\bibfnamefont{Z.}~\bibnamefont{Fisk}}, \bibnamefont{and}
  \bibinfo{author}{\bibfnamefont{S.}~\bibnamefont{Wirth}},
  \bibinfo{journal}{Phys.\ Rev.\ Materials} \textbf{\bibinfo{volume}{5}},
  \bibinfo{pages}{044204} (\bibinfo{year}{2021}{\natexlab{a}}).

\bibitem[{\citenamefont{Hartstein et~al.}(2018)\citenamefont{Hartstein, Toews,
  Hsu, Zeng, Chen, {Ciomaga Hatnean}, Zhang, Nakamura, Padgett, {Rodway-Gant}
  et~al.}}]{har18}
\bibinfo{author}{\bibfnamefont{M.}~\bibnamefont{Hartstein}},
  \bibinfo{author}{\bibfnamefont{H.~W.} \bibnamefont{Toews}},
  \bibinfo{author}{\bibfnamefont{Y.-T.} \bibnamefont{Hsu}},
  \bibinfo{author}{\bibfnamefont{B.}~\bibnamefont{Zeng}},
  \bibinfo{author}{\bibfnamefont{X.}~\bibnamefont{Chen}},
  \bibinfo{author}{\bibfnamefont{M.}~\bibnamefont{{Ciomaga Hatnean}}},
  \bibinfo{author}{\bibfnamefont{Q.~R.} \bibnamefont{Zhang}},
  \bibinfo{author}{\bibfnamefont{S.}~\bibnamefont{Nakamura}},
  \bibinfo{author}{\bibfnamefont{A.~S.} \bibnamefont{Padgett}},
  \bibinfo{author}{\bibfnamefont{G.}~\bibnamefont{{Rodway-Gant}}},
  \bibnamefont{et~al.}, \bibinfo{journal}{Nat.\ Phys.}
  \textbf{\bibinfo{volume}{14}}, \bibinfo{pages}{166} (\bibinfo{year}{2018}).

\bibitem[{\citenamefont{Eo et~al.}(2020)\citenamefont{Eo, Wolgast, Rakoski,
  Mihaliov, Kang, Song, Cho, {Ciomaga Hatnean}, Balakrishnan, Fisk
  et~al.}}]{eo20}
\bibinfo{author}{\bibfnamefont{Y.~S.} \bibnamefont{Eo}},
  \bibinfo{author}{\bibfnamefont{S.}~\bibnamefont{Wolgast}},
  \bibinfo{author}{\bibfnamefont{A.}~\bibnamefont{Rakoski}},
  \bibinfo{author}{\bibfnamefont{D.}~\bibnamefont{Mihaliov}},
  \bibinfo{author}{\bibfnamefont{B.~Y.} \bibnamefont{Kang}},
  \bibinfo{author}{\bibfnamefont{M.~S.} \bibnamefont{Song}},
  \bibinfo{author}{\bibfnamefont{B.~K.} \bibnamefont{Cho}},
  \bibinfo{author}{\bibfnamefont{M.}~\bibnamefont{{Ciomaga Hatnean}}},
  \bibinfo{author}{\bibfnamefont{G.}~\bibnamefont{Balakrishnan}},
  \bibinfo{author}{\bibfnamefont{Z.}~\bibnamefont{Fisk}}, \bibnamefont{et~al.},
  \bibinfo{journal}{Phys.\ Rev.\ B} \textbf{\bibinfo{volume}{101}},
  \bibinfo{pages}{155109} (\bibinfo{year}{2020}).

\bibitem[{\citenamefont{{Ale Crivillero}
  et~al.}(2021{\natexlab{b}})\citenamefont{{Ale Crivillero}, K{\"o}nig, Souza,
  Pagliuso, Sichelschmidt, Rosa, Fisk, and Wirth}}]{ale21b}
\bibinfo{author}{\bibfnamefont{M.~V.} \bibnamefont{{Ale Crivillero}}},
  \bibinfo{author}{\bibfnamefont{M.}~\bibnamefont{K{\"o}nig}},
  \bibinfo{author}{\bibfnamefont{J.~C.} \bibnamefont{Souza}},
  \bibinfo{author}{\bibfnamefont{P.~G.} \bibnamefont{Pagliuso}},
  \bibinfo{author}{\bibfnamefont{J.}~\bibnamefont{Sichelschmidt}},
  \bibinfo{author}{\bibfnamefont{P.~F.~S.} \bibnamefont{Rosa}},
  \bibinfo{author}{\bibfnamefont{Z.}~\bibnamefont{Fisk}}, \bibnamefont{and}
  \bibinfo{author}{\bibfnamefont{S.}~\bibnamefont{Wirth}},
  \bibinfo{journal}{Phys. Rev. Research} \textbf{\bibinfo{volume}{3}},
  \bibinfo{pages}{023162} (\bibinfo{year}{2021}{\natexlab{b}}).

\bibitem[{\citenamefont{Lutz et~al.}(2016)\citenamefont{Lutz, Thees, Peixoto,
  Kang, Cho, Min, and Reinert}}]{lut16}
\bibinfo{author}{\bibfnamefont{P.}~\bibnamefont{Lutz}},
  \bibinfo{author}{\bibfnamefont{M.}~\bibnamefont{Thees}},
  \bibinfo{author}{\bibfnamefont{T.~R.~F.} \bibnamefont{Peixoto}},
  \bibinfo{author}{\bibfnamefont{B.~Y.} \bibnamefont{Kang}},
  \bibinfo{author}{\bibfnamefont{B.~K.} \bibnamefont{Cho}},
  \bibinfo{author}{\bibfnamefont{C.-H.} \bibnamefont{Min}}, \bibnamefont{and}
  \bibinfo{author}{\bibfnamefont{F.}~\bibnamefont{Reinert}},
  \bibinfo{journal}{Philos.\ Mag.} \textbf{\bibinfo{volume}{96}},
  \bibinfo{pages}{3307} (\bibinfo{year}{2016}).

\bibitem[{\citenamefont{Utsumi et~al.}(2017)\citenamefont{Utsumi, Kasinathan,
  Ko, Agrestini, Haverkort, Wirth, Wu, Tsuei, Kim, Fisk et~al.}}]{uts17}
\bibinfo{author}{\bibfnamefont{Y.}~\bibnamefont{Utsumi}},
  \bibinfo{author}{\bibfnamefont{D.}~\bibnamefont{Kasinathan}},
  \bibinfo{author}{\bibfnamefont{K.-T.} \bibnamefont{Ko}},
  \bibinfo{author}{\bibfnamefont{S.}~\bibnamefont{Agrestini}},
  \bibinfo{author}{\bibfnamefont{M.~W.} \bibnamefont{Haverkort}},
  \bibinfo{author}{\bibfnamefont{S.}~\bibnamefont{Wirth}},
  \bibinfo{author}{\bibfnamefont{Y.-H.} \bibnamefont{Wu}},
  \bibinfo{author}{\bibfnamefont{K.-D.} \bibnamefont{Tsuei}},
  \bibinfo{author}{\bibfnamefont{D.-J.} \bibnamefont{Kim}},
  \bibinfo{author}{\bibfnamefont{Z.}~\bibnamefont{Fisk}}, \bibnamefont{et~al.},
  \bibinfo{journal}{Phys. Rev. B} \textbf{\bibinfo{volume}{96}},
  \bibinfo{pages}{155130} (\bibinfo{year}{2017}).

\bibitem[{\citenamefont{Trenary}(2012)}]{tre12}
\bibinfo{author}{\bibfnamefont{M.}~\bibnamefont{Trenary}},
  \bibinfo{journal}{Sci.\ Technol.\ Adv.\ Mater.}
  \textbf{\bibinfo{volume}{13}}, \bibinfo{pages}{023002}
  (\bibinfo{year}{2012}).

\bibitem[{\citenamefont{Bellucci et~al.}(2019)\citenamefont{Bellucci,
  Mastellone, Orlando, Girolami, Generosi, Paci, Soltani, Mezzi, Kaciulis,
  Polini et~al.}}]{bel19}
\bibinfo{author}{\bibfnamefont{A.}~\bibnamefont{Bellucci}},
  \bibinfo{author}{\bibfnamefont{M.}~\bibnamefont{Mastellone}},
  \bibinfo{author}{\bibfnamefont{S.}~\bibnamefont{Orlando}},
  \bibinfo{author}{\bibfnamefont{M.}~\bibnamefont{Girolami}},
  \bibinfo{author}{\bibfnamefont{A.}~\bibnamefont{Generosi}},
  \bibinfo{author}{\bibfnamefont{B.}~\bibnamefont{Paci}},
  \bibinfo{author}{\bibfnamefont{P.}~\bibnamefont{Soltani}},
  \bibinfo{author}{\bibfnamefont{A.}~\bibnamefont{Mezzi}},
  \bibinfo{author}{\bibfnamefont{S.}~\bibnamefont{Kaciulis}},
  \bibinfo{author}{\bibfnamefont{R.}~\bibnamefont{Polini}},
  \bibnamefont{et~al.}, \bibinfo{journal}{Appl.\ Surf.\ Sci.}
  \textbf{\bibinfo{volume}{479}}, \bibinfo{pages}{296} (\bibinfo{year}{2019}).

\bibitem[{\citenamefont{Wolgast et~al.}(2015)\citenamefont{Wolgast, Eo,
  {\"O}zt{\"u}rk, Li, Xiang, Tinsman, Asaba, Lawson, Yu, Allen et~al.}}]{wol15}
\bibinfo{author}{\bibfnamefont{S.}~\bibnamefont{Wolgast}},
  \bibinfo{author}{\bibfnamefont{Y.~S.} \bibnamefont{Eo}},
  \bibinfo{author}{\bibfnamefont{T.}~\bibnamefont{{\"O}zt{\"u}rk}},
  \bibinfo{author}{\bibfnamefont{G.}~\bibnamefont{Li}},
  \bibinfo{author}{\bibfnamefont{Z.}~\bibnamefont{Xiang}},
  \bibinfo{author}{\bibfnamefont{C.}~\bibnamefont{Tinsman}},
  \bibinfo{author}{\bibfnamefont{T.}~\bibnamefont{Asaba}},
  \bibinfo{author}{\bibfnamefont{B.}~\bibnamefont{Lawson}},
  \bibinfo{author}{\bibfnamefont{F.}~\bibnamefont{Yu}},
  \bibinfo{author}{\bibfnamefont{J.~W.} \bibnamefont{Allen}},
  \bibnamefont{et~al.}, \bibinfo{journal}{Phys.\ Rev.\ B}
  \textbf{\bibinfo{volume}{92}}, \bibinfo{pages}{115110}
  (\bibinfo{year}{2015}).

\bibitem[{\citenamefont{Syers et~al.}(2015)\citenamefont{Syers, Kim, Fuhrer,
  and Paglione}}]{sye15}
\bibinfo{author}{\bibfnamefont{P.}~\bibnamefont{Syers}},
  \bibinfo{author}{\bibfnamefont{D.}~\bibnamefont{Kim}},
  \bibinfo{author}{\bibfnamefont{M.}~\bibnamefont{Fuhrer}}, \bibnamefont{and}
  \bibinfo{author}{\bibfnamefont{J.}~\bibnamefont{Paglione}},
  \bibinfo{journal}{Phys.\ Rev.\ Lett.} \textbf{\bibinfo{volume}{114}},
  \bibinfo{pages}{096601} (\bibinfo{year}{2015}).

\bibitem[{\citenamefont{Biswas et~al.}(2017)\citenamefont{Biswas, {Ciomaga
  Hatnean}, Balakrishnan, and Bid}}]{bis17}
\bibinfo{author}{\bibfnamefont{S.}~\bibnamefont{Biswas}},
  \bibinfo{author}{\bibfnamefont{M.}~\bibnamefont{{Ciomaga Hatnean}}},
  \bibinfo{author}{\bibfnamefont{G.}~\bibnamefont{Balakrishnan}},
  \bibnamefont{and} \bibinfo{author}{\bibfnamefont{A.}~\bibnamefont{Bid}},
  \bibinfo{journal}{Phys.\ Rev.\ B} \textbf{\bibinfo{volume}{95}},
  \bibinfo{pages}{205403} (\bibinfo{year}{2017}).

\bibitem[{\citenamefont{Fuhrman et~al.}(2019)\citenamefont{Fuhrman, Leiner,
  Freeland, {van Veenendaal}, Koohpayeh, Phelan, McQueen, and Broholm}}]{fuh19}
\bibinfo{author}{\bibfnamefont{W.~T.} \bibnamefont{Fuhrman}},
  \bibinfo{author}{\bibfnamefont{J.~C.} \bibnamefont{Leiner}},
  \bibinfo{author}{\bibfnamefont{J.~W.} \bibnamefont{Freeland}},
  \bibinfo{author}{\bibfnamefont{M.}~\bibnamefont{{van Veenendaal}}},
  \bibinfo{author}{\bibfnamefont{S.~M.} \bibnamefont{Koohpayeh}},
  \bibinfo{author}{\bibfnamefont{W.~A.} \bibnamefont{Phelan}},
  \bibinfo{author}{\bibfnamefont{T.~M.} \bibnamefont{McQueen}},
  \bibnamefont{and} \bibinfo{author}{\bibfnamefont{C.}~\bibnamefont{Broholm}},
  \bibinfo{journal}{Phys.\ Rev.\ B} \textbf{\bibinfo{volume}{99}},
  \bibinfo{pages}{020401(R)} (\bibinfo{year}{2019}).

\bibitem[{\citenamefont{Hess et~al.}(1989)\citenamefont{Hess, Robinson, Dynes,
  {Valles, Jr.}, and Waszczak}}]{hes89}
\bibinfo{author}{\bibfnamefont{H.~F.} \bibnamefont{Hess}},
  \bibinfo{author}{\bibfnamefont{R.~B.} \bibnamefont{Robinson}},
  \bibinfo{author}{\bibfnamefont{R.~C.} \bibnamefont{Dynes}},
  \bibinfo{author}{\bibfnamefont{J.~M.} \bibnamefont{{Valles, Jr.}}},
  \bibnamefont{and} \bibinfo{author}{\bibfnamefont{J.~V.}
  \bibnamefont{Waszczak}}, \bibinfo{journal}{Phys.\ Rev.\ Lett.}
  \textbf{\bibinfo{volume}{62}}, \bibinfo{pages}{214} (\bibinfo{year}{1989}).

\bibitem[{\citenamefont{R{\"o}{\ss}ler
  et~al.}(2016)\citenamefont{R{\"o}{\ss}ler, Jiao, Kim, Seiro, Rasim, Steglich,
  Tjeng, Fisk, and Wirth}}]{roe16}
\bibinfo{author}{\bibfnamefont{S.}~\bibnamefont{R{\"o}{\ss}ler}},
  \bibinfo{author}{\bibfnamefont{L.}~\bibnamefont{Jiao}},
  \bibinfo{author}{\bibfnamefont{D.~J.} \bibnamefont{Kim}},
  \bibinfo{author}{\bibfnamefont{S.}~\bibnamefont{Seiro}},
  \bibinfo{author}{\bibfnamefont{K.}~\bibnamefont{Rasim}},
  \bibinfo{author}{\bibfnamefont{F.}~\bibnamefont{Steglich}},
  \bibinfo{author}{\bibfnamefont{L.~H.} \bibnamefont{Tjeng}},
  \bibinfo{author}{\bibfnamefont{Z.}~\bibnamefont{Fisk}}, \bibnamefont{and}
  \bibinfo{author}{\bibfnamefont{S.}~\bibnamefont{Wirth}},
  \bibinfo{journal}{Philos.\ Mag.} \textbf{\bibinfo{volume}{96}},
  \bibinfo{pages}{3262} (\bibinfo{year}{2016}).

\bibitem[{\citenamefont{Sun et~al.}(2018)\citenamefont{Sun, Maldonado, Paz,
  Inosov, Schnyder, Palacios, Shitsevalova, Filipov, and Wahl}}]{sun18}
\bibinfo{author}{\bibfnamefont{Z.}~\bibnamefont{Sun}},
  \bibinfo{author}{\bibfnamefont{A.}~\bibnamefont{Maldonado}},
  \bibinfo{author}{\bibfnamefont{W.~S.} \bibnamefont{Paz}},
  \bibinfo{author}{\bibfnamefont{D.~S.} \bibnamefont{Inosov}},
  \bibinfo{author}{\bibfnamefont{A.~P.} \bibnamefont{Schnyder}},
  \bibinfo{author}{\bibfnamefont{J.~J.} \bibnamefont{Palacios}},
  \bibinfo{author}{\bibfnamefont{N.~Y.} \bibnamefont{Shitsevalova}},
  \bibinfo{author}{\bibfnamefont{V.~B.} \bibnamefont{Filipov}},
  \bibnamefont{and} \bibinfo{author}{\bibfnamefont{P.}~\bibnamefont{Wahl}},
  \bibinfo{journal}{Phys.\ Rev.\ B} \textbf{\bibinfo{volume}{97}},
  \bibinfo{pages}{235107} (\bibinfo{year}{2018}).

\bibitem[{\citenamefont{Ruan et~al.}(2014)\citenamefont{Ruan, Ye, Guo, Chen,
  Chen, Zhang, and Wang}}]{ruan14}
\bibinfo{author}{\bibfnamefont{W.}~\bibnamefont{Ruan}},
  \bibinfo{author}{\bibfnamefont{C.}~\bibnamefont{Ye}},
  \bibinfo{author}{\bibfnamefont{M.}~\bibnamefont{Guo}},
  \bibinfo{author}{\bibfnamefont{F.}~\bibnamefont{Chen}},
  \bibinfo{author}{\bibfnamefont{X.}~\bibnamefont{Chen}},
  \bibinfo{author}{\bibfnamefont{G.-M.} \bibnamefont{Zhang}}, \bibnamefont{and}
  \bibinfo{author}{\bibfnamefont{Y.}~\bibnamefont{Wang}},
  \bibinfo{journal}{Phys.\ Rev.\ Lett.} \textbf{\bibinfo{volume}{112}},
  \bibinfo{pages}{136401} (\bibinfo{year}{2014}).

\bibitem[{\citenamefont{Wirth et~al.}(2021)\citenamefont{Wirth, R{\"o}{\ss}ler,
  Jiao, {Ale Crivillero}, Rosa, and Fisk}}]{wir20}
\bibinfo{author}{\bibfnamefont{S.}~\bibnamefont{Wirth}},
  \bibinfo{author}{\bibfnamefont{S.}~\bibnamefont{R{\"o}{\ss}ler}},
  \bibinfo{author}{\bibfnamefont{L.}~\bibnamefont{Jiao}},
  \bibinfo{author}{\bibfnamefont{M.~V.} \bibnamefont{{Ale Crivillero}}},
  \bibinfo{author}{\bibfnamefont{P.~F.~S.} \bibnamefont{Rosa}},
  \bibnamefont{and} \bibinfo{author}{\bibfnamefont{Z.}~\bibnamefont{Fisk}},
  \bibinfo{journal}{Phys.\ Status Solidi B} \textbf{\bibinfo{volume}{258}},
  \bibinfo{pages}{2000022} (\bibinfo{year}{2021}).

\bibitem[{\citenamefont{Schubert et~al.}(2012)\citenamefont{Schubert, Fehske,
  Fritz, and Vojta}}]{schu12}
\bibinfo{author}{\bibfnamefont{G.}~\bibnamefont{Schubert}},
  \bibinfo{author}{\bibfnamefont{H.}~\bibnamefont{Fehske}},
  \bibinfo{author}{\bibfnamefont{L.}~\bibnamefont{Fritz}}, \bibnamefont{and}
  \bibinfo{author}{\bibfnamefont{M.}~\bibnamefont{Vojta}},
  \bibinfo{journal}{Phys.\ Rev.\ B} \textbf{\bibinfo{volume}{85}},
  \bibinfo{pages}{201105(R)} (\bibinfo{year}{2012}).

\bibitem[{\citenamefont{Yee et~al.}(2013)\citenamefont{Yee, He,
  Soumyanarayanan, Kim, Fisk, and Hoffman}}]{yee13}
\bibinfo{author}{\bibfnamefont{M.~M.} \bibnamefont{Yee}},
  \bibinfo{author}{\bibfnamefont{Y.}~\bibnamefont{He}},
  \bibinfo{author}{\bibfnamefont{A.}~\bibnamefont{Soumyanarayanan}},
  \bibinfo{author}{\bibfnamefont{D.-J.} \bibnamefont{Kim}},
  \bibinfo{author}{\bibfnamefont{Z.}~\bibnamefont{Fisk}}, \bibnamefont{and}
  \bibinfo{author}{\bibfnamefont{J.~E.} \bibnamefont{Hoffman}},
  \emph{\bibinfo{title}{Imaging the {K}ondo insulating gap on {S}m{B}$_6$}},
  \bibinfo{howpublished}{({P}reprint) {a}rXiv:1308.1085}
  (\bibinfo{year}{2013}).

\bibitem[{\citenamefont{R{\"o}{\ss}ler
  et~al.}(2014)\citenamefont{R{\"o}{\ss}ler, Jang, Kim, Tjeng, Fisk, Steglich,
  and Wirth}}]{roe14}
\bibinfo{author}{\bibfnamefont{S.}~\bibnamefont{R{\"o}{\ss}ler}},
  \bibinfo{author}{\bibfnamefont{T.-H.} \bibnamefont{Jang}},
  \bibinfo{author}{\bibfnamefont{D.~J.} \bibnamefont{Kim}},
  \bibinfo{author}{\bibfnamefont{L.~H.} \bibnamefont{Tjeng}},
  \bibinfo{author}{\bibfnamefont{Z.}~\bibnamefont{Fisk}},
  \bibinfo{author}{\bibfnamefont{F.}~\bibnamefont{Steglich}}, \bibnamefont{and}
  \bibinfo{author}{\bibfnamefont{S.}~\bibnamefont{Wirth}},
  \bibinfo{journal}{Proc.\ Natl.\ Acad.\ Sci.\ USA}
  \textbf{\bibinfo{volume}{111}}, \bibinfo{pages}{4798} (\bibinfo{year}{2014}).

\bibitem[{\citenamefont{Pirie et~al.}(2020)\citenamefont{Pirie, Liu,
  Soumyanarayanan, Chen, He, Yee, Rosa, Thompson, Kim, Fisk et~al.}}]{pir20}
\bibinfo{author}{\bibfnamefont{H.}~\bibnamefont{Pirie}},
  \bibinfo{author}{\bibfnamefont{Y.}~\bibnamefont{Liu}},
  \bibinfo{author}{\bibfnamefont{A.}~\bibnamefont{Soumyanarayanan}},
  \bibinfo{author}{\bibfnamefont{P.}~\bibnamefont{Chen}},
  \bibinfo{author}{\bibfnamefont{Y.}~\bibnamefont{He}},
  \bibinfo{author}{\bibfnamefont{M.~M.} \bibnamefont{Yee}},
  \bibinfo{author}{\bibfnamefont{P.~F.~S.} \bibnamefont{Rosa}},
  \bibinfo{author}{\bibfnamefont{J.~D.} \bibnamefont{Thompson}},
  \bibinfo{author}{\bibfnamefont{D.-J.} \bibnamefont{Kim}},
  \bibinfo{author}{\bibfnamefont{Z.}~\bibnamefont{Fisk}}, \bibnamefont{et~al.},
  \bibinfo{journal}{Nat.\ Phys.} \textbf{\bibinfo{volume}{16}},
  \bibinfo{pages}{52} (\bibinfo{year}{2020}).

\bibitem[{\citenamefont{Herrmann et~al.}(2020)\citenamefont{Herrmann, Hlawenka,
  Siemensmeyer, Weschke, S\'{a}nchez-Barriga, Varykhalov, Shitsevalova,
  Dukhnenko, Filipov, Gab\'{a}ni et~al.}}]{her20}
\bibinfo{author}{\bibfnamefont{H.}~\bibnamefont{Herrmann}},
  \bibinfo{author}{\bibfnamefont{P.}~\bibnamefont{Hlawenka}},
  \bibinfo{author}{\bibfnamefont{K.}~\bibnamefont{Siemensmeyer}},
  \bibinfo{author}{\bibfnamefont{E.}~\bibnamefont{Weschke}},
  \bibinfo{author}{\bibfnamefont{J.}~\bibnamefont{S\'{a}nchez-Barriga}},
  \bibinfo{author}{\bibfnamefont{A.}~\bibnamefont{Varykhalov}},
  \bibinfo{author}{\bibfnamefont{N.~Y.} \bibnamefont{Shitsevalova}},
  \bibinfo{author}{\bibfnamefont{A.~V.} \bibnamefont{Dukhnenko}},
  \bibinfo{author}{\bibfnamefont{V.~B.} \bibnamefont{Filipov}},
  \bibinfo{author}{\bibfnamefont{S.}~\bibnamefont{Gab\'{a}ni}},
  \bibnamefont{et~al.}, \bibinfo{journal}{Adv.\ Mater.}
  \textbf{\bibinfo{volume}{32}}, \bibinfo{pages}{1906725}
  (\bibinfo{year}{2020}).

\bibitem[{\citenamefont{Aono et~al.}(1978)\citenamefont{Aono, Nishitani,
  Tanaka, Bannai, and Kawai}}]{aon78}
\bibinfo{author}{\bibfnamefont{M.}~\bibnamefont{Aono}},
  \bibinfo{author}{\bibfnamefont{R.}~\bibnamefont{Nishitani}},
  \bibinfo{author}{\bibfnamefont{T.}~\bibnamefont{Tanaka}},
  \bibinfo{author}{\bibfnamefont{E.}~\bibnamefont{Bannai}}, \bibnamefont{and}
  \bibinfo{author}{\bibfnamefont{S.}~\bibnamefont{Kawai}},
  \bibinfo{journal}{Solid State Commun.} \textbf{\bibinfo{volume}{28}},
  \bibinfo{pages}{409} (\bibinfo{year}{1978}).

\bibitem[{\citenamefont{Miyazaki et~al.}(2012)\citenamefont{Miyazaki, Hajiri,
  Ito, Kunii, and Kimura}}]{miy12}
\bibinfo{author}{\bibfnamefont{H.}~\bibnamefont{Miyazaki}},
  \bibinfo{author}{\bibfnamefont{T.}~\bibnamefont{Hajiri}},
  \bibinfo{author}{\bibfnamefont{T.}~\bibnamefont{Ito}},
  \bibinfo{author}{\bibfnamefont{S.}~\bibnamefont{Kunii}}, \bibnamefont{and}
  \bibinfo{author}{\bibfnamefont{S.~I.} \bibnamefont{Kimura}},
  \bibinfo{journal}{Phys.\ Rev.\ B} \textbf{\bibinfo{volume}{86}},
  \bibinfo{pages}{075105} (\bibinfo{year}{2012}).

\bibitem[{\citenamefont{Ramankutty et~al.}(2016)\citenamefont{Ramankutty,
  de~Jong, Huang, Zwartsenberg, Massee, Bay, Golden, and
  Frantzeskakis}}]{ram16}
\bibinfo{author}{\bibfnamefont{S.~V.} \bibnamefont{Ramankutty}},
  \bibinfo{author}{\bibfnamefont{N.}~\bibnamefont{de~Jong}},
  \bibinfo{author}{\bibfnamefont{Y.~K.} \bibnamefont{Huang}},
  \bibinfo{author}{\bibfnamefont{B.}~\bibnamefont{Zwartsenberg}},
  \bibinfo{author}{\bibfnamefont{F.}~\bibnamefont{Massee}},
  \bibinfo{author}{\bibfnamefont{T.~V.} \bibnamefont{Bay}},
  \bibinfo{author}{\bibfnamefont{M.~S.} \bibnamefont{Golden}},
  \bibnamefont{and}
  \bibinfo{author}{\bibfnamefont{E.}~\bibnamefont{Frantzeskakis}},
  \bibinfo{journal}{J. Electron Spectrosc.\ Relat.\ Phenom.}
  \textbf{\bibinfo{volume}{208}}, \bibinfo{pages}{43} (\bibinfo{year}{2016}).

\bibitem[{\citenamefont{Miyamachi et~al.}(2017)\citenamefont{Miyamachi, Suga,
  Ellguth, Tusche, Schneider, Iga, and Komori}}]{miy17}
\bibinfo{author}{\bibfnamefont{T.}~\bibnamefont{Miyamachi}},
  \bibinfo{author}{\bibfnamefont{S.}~\bibnamefont{Suga}},
  \bibinfo{author}{\bibfnamefont{M.}~\bibnamefont{Ellguth}},
  \bibinfo{author}{\bibfnamefont{C.}~\bibnamefont{Tusche}},
  \bibinfo{author}{\bibfnamefont{C.~M.} \bibnamefont{Schneider}},
  \bibinfo{author}{\bibfnamefont{F.}~\bibnamefont{Iga}}, \bibnamefont{and}
  \bibinfo{author}{\bibfnamefont{F.}~\bibnamefont{Komori}},
  \bibinfo{journal}{Sci.\ Rep.} \textbf{\bibinfo{volume}{7}},
  \bibinfo{pages}{12837} (\bibinfo{year}{2017}).

\bibitem[{\citenamefont{Frantzeskakis et~al.}(2013)\citenamefont{Frantzeskakis,
  {de Jong}, Zwartsenberg, Huang, Pan, Zhang, Zhang, Zhang, Bao, Tegus
  et~al.}}]{fra13}
\bibinfo{author}{\bibfnamefont{E.}~\bibnamefont{Frantzeskakis}},
  \bibinfo{author}{\bibfnamefont{N.}~\bibnamefont{{de Jong}}},
  \bibinfo{author}{\bibfnamefont{B.}~\bibnamefont{Zwartsenberg}},
  \bibinfo{author}{\bibfnamefont{Y.~K.} \bibnamefont{Huang}},
  \bibinfo{author}{\bibfnamefont{Y.}~\bibnamefont{Pan}},
  \bibinfo{author}{\bibfnamefont{X.}~\bibnamefont{Zhang}},
  \bibinfo{author}{\bibfnamefont{J.~X.} \bibnamefont{Zhang}},
  \bibinfo{author}{\bibfnamefont{F.~X.} \bibnamefont{Zhang}},
  \bibinfo{author}{\bibfnamefont{L.~H.} \bibnamefont{Bao}},
  \bibinfo{author}{\bibfnamefont{O.}~\bibnamefont{Tegus}},
  \bibnamefont{et~al.}, \bibinfo{journal}{Phys.\ Rev.\ X}
  \textbf{\bibinfo{volume}{3}}, \bibinfo{pages}{041024} (\bibinfo{year}{2013}).

\bibitem[{\citenamefont{Zabolotnyy et~al.}(2018)\citenamefont{Zabolotnyy,
  F{\"u}rsich, Green, Lutz, Treiber, Min, Dukhnenko, Shitsevalova, Filipov,
  Kang et~al.}}]{zab18}
\bibinfo{author}{\bibfnamefont{V.~B.} \bibnamefont{Zabolotnyy}},
  \bibinfo{author}{\bibfnamefont{K.}~\bibnamefont{F{\"u}rsich}},
  \bibinfo{author}{\bibfnamefont{R.~J.} \bibnamefont{Green}},
  \bibinfo{author}{\bibfnamefont{P.}~\bibnamefont{Lutz}},
  \bibinfo{author}{\bibfnamefont{K.}~\bibnamefont{Treiber}},
  \bibinfo{author}{\bibfnamefont{C.-H.} \bibnamefont{Min}},
  \bibinfo{author}{\bibfnamefont{A.~V.} \bibnamefont{Dukhnenko}},
  \bibinfo{author}{\bibfnamefont{N.~Y.} \bibnamefont{Shitsevalova}},
  \bibinfo{author}{\bibfnamefont{V.~B.} \bibnamefont{Filipov}},
  \bibinfo{author}{\bibfnamefont{B.~Y.} \bibnamefont{Kang}},
  \bibnamefont{et~al.}, \bibinfo{journal}{Phys.\ Rev.\ B}
  \textbf{\bibinfo{volume}{97}}, \bibinfo{pages}{205416}
  (\bibinfo{year}{2018}).

\bibitem[{\citenamefont{Yoo and Weitering}(2002)}]{yoo02}
\bibinfo{author}{\bibfnamefont{K.}~\bibnamefont{Yoo}} \bibnamefont{and}
  \bibinfo{author}{\bibfnamefont{H.~H.} \bibnamefont{Weitering}},
  \bibinfo{journal}{Phys.\ Rev.\ B} \textbf{\bibinfo{volume}{65}},
  \bibinfo{pages}{115424} (\bibinfo{year}{2002}).

\bibitem[{\citenamefont{Matt et~al.}(2020)\citenamefont{Matt, Pirie,
  Soumyanarayanan, He, Yee, Chen, Liu, Larson, Paz, Palacios et~al.}}]{mat20}
\bibinfo{author}{\bibfnamefont{C.~E.} \bibnamefont{Matt}},
  \bibinfo{author}{\bibfnamefont{H.}~\bibnamefont{Pirie}},
  \bibinfo{author}{\bibfnamefont{A.}~\bibnamefont{Soumyanarayanan}},
  \bibinfo{author}{\bibfnamefont{Y.}~\bibnamefont{He}},
  \bibinfo{author}{\bibfnamefont{M.~M.} \bibnamefont{Yee}},
  \bibinfo{author}{\bibfnamefont{P.}~\bibnamefont{Chen}},
  \bibinfo{author}{\bibfnamefont{Y.}~\bibnamefont{Liu}},
  \bibinfo{author}{\bibfnamefont{D.~T.} \bibnamefont{Larson}},
  \bibinfo{author}{\bibfnamefont{W.~S.} \bibnamefont{Paz}},
  \bibinfo{author}{\bibfnamefont{J.~J.} \bibnamefont{Palacios}},
  \bibnamefont{et~al.}, \bibinfo{journal}{Phys.\ Rev.\ B}
  \textbf{\bibinfo{volume}{101}}, \bibinfo{pages}{085142}
  (\bibinfo{year}{2020}).

\bibitem[{\citenamefont{Schmidt et~al.}(2018)\citenamefont{Schmidt, Jaime,
  Cahill, Edwards, Misture, Graeve, and Vasquez}}]{sch18}
\bibinfo{author}{\bibfnamefont{K.~M.} \bibnamefont{Schmidt}},
  \bibinfo{author}{\bibfnamefont{O.}~\bibnamefont{Jaime}},
  \bibinfo{author}{\bibfnamefont{J.~T.} \bibnamefont{Cahill}},
  \bibinfo{author}{\bibfnamefont{D.}~\bibnamefont{Edwards}},
  \bibinfo{author}{\bibfnamefont{S.~T.} \bibnamefont{Misture}},
  \bibinfo{author}{\bibfnamefont{O.~A.} \bibnamefont{Graeve}},
  \bibnamefont{and} \bibinfo{author}{\bibfnamefont{V.~R.}
  \bibnamefont{Vasquez}}, \bibinfo{journal}{Acta Mater.}
  \textbf{\bibinfo{volume}{144}}, \bibinfo{pages}{187} (\bibinfo{year}{2018}).

\bibitem[{\citenamefont{Allen}(2016)}]{all16}
\bibinfo{author}{\bibfnamefont{J.~W.} \bibnamefont{Allen}},
  \bibinfo{journal}{Philos.\ Mag.} \textbf{\bibinfo{volume}{96}},
  \bibinfo{pages}{3227} (\bibinfo{year}{2016}).

\bibitem[{\citenamefont{Jiao et~al.}(2016)\citenamefont{Jiao, R{\"o}{\ss}ler,
  Kim, Tjeng, Fisk, Steglich, and Wirth}}]{jiao16}
\bibinfo{author}{\bibfnamefont{L.}~\bibnamefont{Jiao}},
  \bibinfo{author}{\bibfnamefont{S.}~\bibnamefont{R{\"o}{\ss}ler}},
  \bibinfo{author}{\bibfnamefont{D.~J.} \bibnamefont{Kim}},
  \bibinfo{author}{\bibfnamefont{L.~H.} \bibnamefont{Tjeng}},
  \bibinfo{author}{\bibfnamefont{Z.}~\bibnamefont{Fisk}},
  \bibinfo{author}{\bibfnamefont{F.}~\bibnamefont{Steglich}}, \bibnamefont{and}
  \bibinfo{author}{\bibfnamefont{S.}~\bibnamefont{Wirth}},
  \bibinfo{journal}{Nat.\ Commun.} \textbf{\bibinfo{volume}{7}},
  \bibinfo{pages}{13762} (\bibinfo{year}{2016}).

\bibitem[{\citenamefont{Erten et~al.}(2016)\citenamefont{Erten, Ghaemi, and
  Coleman}}]{ert16}
\bibinfo{author}{\bibfnamefont{O.}~\bibnamefont{Erten}},
  \bibinfo{author}{\bibfnamefont{P.}~\bibnamefont{Ghaemi}}, \bibnamefont{and}
  \bibinfo{author}{\bibfnamefont{P.}~\bibnamefont{Coleman}},
  \bibinfo{journal}{Phys.\ Rev.\ Lett.} \textbf{\bibinfo{volume}{116}},
  \bibinfo{pages}{046403} (\bibinfo{year}{2016}).

\bibitem[{\citenamefont{Peters et~al.}(2016)\citenamefont{Peters, Yoshida,
  Sakakibara, and Kawakami}}]{pet16}
\bibinfo{author}{\bibfnamefont{R.}~\bibnamefont{Peters}},
  \bibinfo{author}{\bibfnamefont{T.}~\bibnamefont{Yoshida}},
  \bibinfo{author}{\bibfnamefont{H.}~\bibnamefont{Sakakibara}},
  \bibnamefont{and} \bibinfo{author}{\bibfnamefont{N.}~\bibnamefont{Kawakami}},
  \bibinfo{journal}{Phys.\ Rev.\ B} \textbf{\bibinfo{volume}{93}},
  \bibinfo{pages}{235159} (\bibinfo{year}{2016}).

\bibitem[{\citenamefont{Denlinger et~al.}(2016)\citenamefont{Denlinger, Jang,
  Li, Chen, Lawson, Asaba, Tinsman, Yu, Sun, Allen et~al.}}]{den16}
\bibinfo{author}{\bibfnamefont{J.~D.} \bibnamefont{Denlinger}},
  \bibinfo{author}{\bibfnamefont{S.}~\bibnamefont{Jang}},
  \bibinfo{author}{\bibfnamefont{G.}~\bibnamefont{Li}},
  \bibinfo{author}{\bibfnamefont{L.}~\bibnamefont{Chen}},
  \bibinfo{author}{\bibfnamefont{B.~J.} \bibnamefont{Lawson}},
  \bibinfo{author}{\bibfnamefont{T.}~\bibnamefont{Asaba}},
  \bibinfo{author}{\bibfnamefont{C.}~\bibnamefont{Tinsman}},
  \bibinfo{author}{\bibfnamefont{F.}~\bibnamefont{Yu}},
  \bibinfo{author}{\bibfnamefont{K.}~\bibnamefont{Sun}},
  \bibinfo{author}{\bibfnamefont{J.~W.} \bibnamefont{Allen}},
  \bibnamefont{et~al.}, \emph{\bibinfo{title}{Consistency of photoemission and
  quantum oscillations for surface states of {S}m{B}$_6$}},
  \bibinfo{howpublished}{({P}reprint) {a}rXiv:1601.07408}
  (\bibinfo{year}{2016}).

\bibitem[{\citenamefont{Inosov}(2021)}]{ino21}
\bibinfo{editor}{\bibfnamefont{D.~S.} \bibnamefont{Inosov}}, ed.,
  \emph{\bibinfo{title}{Rare-earth borides}} (\bibinfo{publisher}{Jenny
  Stanford Publishing, Singapore}, \bibinfo{year}{2021}).

\bibitem[{\citenamefont{Nikitin et~al.}(2018)\citenamefont{Nikitin,
  Portnichenko, Dukhnenko, Shitsevalova, Filipov, Qiu, Rodriguez-Rivera,
  Ollivier, and Inosov}}]{nik18}
\bibinfo{author}{\bibfnamefont{S.~E.} \bibnamefont{Nikitin}},
  \bibinfo{author}{\bibfnamefont{P.~Y.} \bibnamefont{Portnichenko}},
  \bibinfo{author}{\bibfnamefont{A.~V.} \bibnamefont{Dukhnenko}},
  \bibinfo{author}{\bibfnamefont{N.~Y.} \bibnamefont{Shitsevalova}},
  \bibinfo{author}{\bibfnamefont{V.~B.} \bibnamefont{Filipov}},
  \bibinfo{author}{\bibfnamefont{Y.}~\bibnamefont{Qiu}},
  \bibinfo{author}{\bibfnamefont{J.~A.} \bibnamefont{Rodriguez-Rivera}},
  \bibinfo{author}{\bibfnamefont{J.}~\bibnamefont{Ollivier}}, \bibnamefont{and}
  \bibinfo{author}{\bibfnamefont{D.~S.} \bibnamefont{Inosov}},
  \bibinfo{journal}{Phys.\ Rev.\ B} \textbf{\bibinfo{volume}{97}},
  \bibinfo{pages}{075116} (\bibinfo{year}{2018}).

\bibitem[{\citenamefont{Nie et~al.}(2020)\citenamefont{Nie, Sun, Prinz, Wang,
  Weng, Fang, and Dai}}]{nie20}
\bibinfo{author}{\bibfnamefont{S.}~\bibnamefont{Nie}},
  \bibinfo{author}{\bibfnamefont{Y.}~\bibnamefont{Sun}},
  \bibinfo{author}{\bibfnamefont{F.~B.} \bibnamefont{Prinz}},
  \bibinfo{author}{\bibfnamefont{Z.}~\bibnamefont{Wang}},
  \bibinfo{author}{\bibfnamefont{H.}~\bibnamefont{Weng}},
  \bibinfo{author}{\bibfnamefont{Z.}~\bibnamefont{Fang}}, \bibnamefont{and}
  \bibinfo{author}{\bibfnamefont{X.}~\bibnamefont{Dai}},
  \bibinfo{journal}{Phys.\ Rev.\ Lett.} \textbf{\bibinfo{volume}{124}},
  \bibinfo{pages}{076403} (\bibinfo{year}{2020}).

\bibitem[{\citenamefont{Geballe}(1970)}]{geb70}
\bibinfo{author}{\bibfnamefont{T.~H.} \bibnamefont{Geballe}},
  \bibinfo{journal}{J.\ Appl.\ Phys.} \textbf{\bibinfo{volume}{41}},
  \bibinfo{pages}{904} (\bibinfo{year}{1970}).

\bibitem[{\citenamefont{Yeo et~al.}(2012)\citenamefont{Yeo, Song, Hur, Fisk,
  and Schlottmann}}]{yeo12}
\bibinfo{author}{\bibfnamefont{S.}~\bibnamefont{Yeo}},
  \bibinfo{author}{\bibfnamefont{K.}~\bibnamefont{Song}},
  \bibinfo{author}{\bibfnamefont{N.}~\bibnamefont{Hur}},
  \bibinfo{author}{\bibfnamefont{Z.}~\bibnamefont{Fisk}}, \bibnamefont{and}
  \bibinfo{author}{\bibfnamefont{P.}~\bibnamefont{Schlottmann}},
  \bibinfo{journal}{Phys.\ Rev.\ B} \textbf{\bibinfo{volume}{85}},
  \bibinfo{pages}{115125} (\bibinfo{year}{2012}).

\bibitem[{\citenamefont{Miao et~al.}(2021)\citenamefont{Miao, Min, Xu, Huang,
  Kotta, Basak, Song, Kang, Cho, Ki{\ss}ner et~al.}}]{mia21}
\bibinfo{author}{\bibfnamefont{L.}~\bibnamefont{Miao}},
  \bibinfo{author}{\bibfnamefont{C.-H.} \bibnamefont{Min}},
  \bibinfo{author}{\bibfnamefont{Y.}~\bibnamefont{Xu}},
  \bibinfo{author}{\bibfnamefont{Z.}~\bibnamefont{Huang}},
  \bibinfo{author}{\bibfnamefont{E.~C.} \bibnamefont{Kotta}},
  \bibinfo{author}{\bibfnamefont{R.}~\bibnamefont{Basak}},
  \bibinfo{author}{\bibfnamefont{M.}~\bibnamefont{Song}},
  \bibinfo{author}{\bibfnamefont{B.}~\bibnamefont{Kang}},
  \bibinfo{author}{\bibfnamefont{B.}~\bibnamefont{Cho}},
  \bibinfo{author}{\bibfnamefont{K.}~\bibnamefont{Ki{\ss}ner}},
  \bibnamefont{et~al.}, \bibinfo{journal}{Phys.\ Rev.\ Lett.}
  \textbf{\bibinfo{volume}{126}}, \bibinfo{pages}{136401}
  (\bibinfo{year}{2021}).

\bibitem[{\citenamefont{Kim et~al.}(2014{\natexlab{b}})\citenamefont{Kim, Xia,
  and Fisk}}]{kim14}
\bibinfo{author}{\bibfnamefont{D.~J.} \bibnamefont{Kim}},
  \bibinfo{author}{\bibfnamefont{J.}~\bibnamefont{Xia}}, \bibnamefont{and}
  \bibinfo{author}{\bibfnamefont{Z.}~\bibnamefont{Fisk}},
  \bibinfo{journal}{Nat.\ Mater.} \textbf{\bibinfo{volume}{13}},
  \bibinfo{pages}{466} (\bibinfo{year}{2014}{\natexlab{b}}).

\bibitem[{\citenamefont{{Ciomaga Hatnean} et~al.}(2020)\citenamefont{{Ciomaga
  Hatnean}, Ahmad, Walker, Lees, and Balakrishnan}}]{hat20}
\bibinfo{author}{\bibfnamefont{M.}~\bibnamefont{{Ciomaga Hatnean}}},
  \bibinfo{author}{\bibfnamefont{T.}~\bibnamefont{Ahmad}},
  \bibinfo{author}{\bibfnamefont{M.}~\bibnamefont{Walker}},
  \bibinfo{author}{\bibfnamefont{M.~R.} \bibnamefont{Lees}}, \bibnamefont{and}
  \bibinfo{author}{\bibfnamefont{G.}~\bibnamefont{Balakrishnan}},
  \bibinfo{journal}{Crystals} \textbf{\bibinfo{volume}{10}},
  \bibinfo{pages}{827} (\bibinfo{year}{2020}).

\bibitem[{\citenamefont{Altshuler et~al.}(2005)\citenamefont{Altshuler,
  Bresler, and Goryunov}}]{alt05}
\bibinfo{author}{\bibfnamefont{T.~S.} \bibnamefont{Altshuler}},
  \bibinfo{author}{\bibfnamefont{M.~S.} \bibnamefont{Bresler}},
  \bibnamefont{and} \bibinfo{author}{\bibfnamefont{Y.~V.}
  \bibnamefont{Goryunov}}, \bibinfo{journal}{JETP Lett.}
  \textbf{\bibinfo{volume}{81}}, \bibinfo{pages}{475} (\bibinfo{year}{2005}).

\bibitem[{\citenamefont{Akintola et~al.}(2017)\citenamefont{Akintola, Pal,
  Potma, Saha, Wang, Paglione, and Sonier}}]{aki17}
\bibinfo{author}{\bibfnamefont{K.}~\bibnamefont{Akintola}},
  \bibinfo{author}{\bibfnamefont{A.}~\bibnamefont{Pal}},
  \bibinfo{author}{\bibfnamefont{M.}~\bibnamefont{Potma}},
  \bibinfo{author}{\bibfnamefont{S.~R.} \bibnamefont{Saha}},
  \bibinfo{author}{\bibfnamefont{X.~F.} \bibnamefont{Wang}},
  \bibinfo{author}{\bibfnamefont{J.}~\bibnamefont{Paglione}}, \bibnamefont{and}
  \bibinfo{author}{\bibfnamefont{J.~E.} \bibnamefont{Sonier}},
  \bibinfo{journal}{Phys.\ Rev.\ B} \textbf{\bibinfo{volume}{95}},
  \bibinfo{pages}{245107} (\bibinfo{year}{2017}).

\bibitem[{\citenamefont{Gab\'{a}ni et~al.}(2016)\citenamefont{Gab\'{a}ni,
  Orend\'{a}\v{c}, Prist\'{a}\v{s}, Ga\v{z}o, Diko, and Piovar\v{c}i}}]{gab16}
\bibinfo{author}{\bibfnamefont{S.}~\bibnamefont{Gab\'{a}ni}},
  \bibinfo{author}{\bibfnamefont{M.}~\bibnamefont{Orend\'{a}\v{c}}},
  \bibinfo{author}{\bibfnamefont{G.}~\bibnamefont{Prist\'{a}\v{s}}},
  \bibinfo{author}{\bibfnamefont{E.}~\bibnamefont{Ga\v{z}o}},
  \bibinfo{author}{\bibfnamefont{P.}~\bibnamefont{Diko}}, \bibnamefont{and}
  \bibinfo{author}{\bibfnamefont{S.}~\bibnamefont{Piovar\v{c}i}},
  \bibinfo{journal}{Philos.\ Mag.} \textbf{\bibinfo{volume}{96}},
  \bibinfo{pages}{3274} (\bibinfo{year}{2016}).

\bibitem[{\citenamefont{Valentine et~al.}(2016)\citenamefont{Valentine,
  Koohpayeh, Phelan, McQueen, Rosa, Fisk, and Drichko}}]{val16}
\bibinfo{author}{\bibfnamefont{M.~E.} \bibnamefont{Valentine}},
  \bibinfo{author}{\bibfnamefont{S.}~\bibnamefont{Koohpayeh}},
  \bibinfo{author}{\bibfnamefont{W.~A.} \bibnamefont{Phelan}},
  \bibinfo{author}{\bibfnamefont{T.~M.} \bibnamefont{McQueen}},
  \bibinfo{author}{\bibfnamefont{P.~F.~S.} \bibnamefont{Rosa}},
  \bibinfo{author}{\bibfnamefont{Z.}~\bibnamefont{Fisk}}, \bibnamefont{and}
  \bibinfo{author}{\bibfnamefont{N.}~\bibnamefont{Drichko}},
  \bibinfo{journal}{Phys.\ Rev.\ B} \textbf{\bibinfo{volume}{94}},
  \bibinfo{pages}{075102} (\bibinfo{year}{2016}).

\bibitem[{\citenamefont{Li et~al.}(2020)\citenamefont{Li, Sun, Kurdak, and
  Allen}}]{li20}
\bibinfo{author}{\bibfnamefont{L.}~\bibnamefont{Li}},
  \bibinfo{author}{\bibfnamefont{K.}~\bibnamefont{Sun}},
  \bibinfo{author}{\bibfnamefont{C.}~\bibnamefont{Kurdak}}, \bibnamefont{and}
  \bibinfo{author}{\bibfnamefont{J.~W.} \bibnamefont{Allen}},
  \bibinfo{journal}{Nat.\ Rev.\ Phys.} \textbf{\bibinfo{volume}{2}},
  \bibinfo{pages}{463} (\bibinfo{year}{2020}).

\bibitem[{\citenamefont{Ozcomert and Trenary}(1992)}]{ozc92}
\bibinfo{author}{\bibfnamefont{J.~S.} \bibnamefont{Ozcomert}} \bibnamefont{and}
  \bibinfo{author}{\bibfnamefont{M.}~\bibnamefont{Trenary}},
  \bibinfo{journal}{Surf.\ Sci.\ Lett.} \textbf{\bibinfo{volume}{265}},
  \bibinfo{pages}{L227} (\bibinfo{year}{1992}).

\bibitem[{\citenamefont{Buchsteiner et~al.}(2019)\citenamefont{Buchsteiner,
  Sohn, Horstmann, Voigt, {Ciomaga Hatnean}, Balakrishnan, Ropers, Bl{\"o}chl,
  and Wenderoth}}]{buc19}
\bibinfo{author}{\bibfnamefont{P.}~\bibnamefont{Buchsteiner}},
  \bibinfo{author}{\bibfnamefont{F.}~\bibnamefont{Sohn}},
  \bibinfo{author}{\bibfnamefont{J.~G.} \bibnamefont{Horstmann}},
  \bibinfo{author}{\bibfnamefont{J.}~\bibnamefont{Voigt}},
  \bibinfo{author}{\bibfnamefont{M.}~\bibnamefont{{Ciomaga Hatnean}}},
  \bibinfo{author}{\bibfnamefont{G.}~\bibnamefont{Balakrishnan}},
  \bibinfo{author}{\bibfnamefont{C.}~\bibnamefont{Ropers}},
  \bibinfo{author}{\bibfnamefont{P.~E.} \bibnamefont{Bl{\"o}chl}},
  \bibnamefont{and}
  \bibinfo{author}{\bibfnamefont{M.}~\bibnamefont{Wenderoth}},
  \bibinfo{journal}{Phys.\ Rev.\ B} \textbf{\bibinfo{volume}{100}},
  \bibinfo{pages}{205407} (\bibinfo{year}{2019}).

\bibitem[{\citenamefont{Buchsteiner et~al.}(2020)\citenamefont{Buchsteiner,
  Harmsen, {Ciomaga Hatnean}, Balakrishnan, and Wenderoth}}]{buc20}
\bibinfo{author}{\bibfnamefont{P.}~\bibnamefont{Buchsteiner}},
  \bibinfo{author}{\bibfnamefont{L.}~\bibnamefont{Harmsen}},
  \bibinfo{author}{\bibfnamefont{M.}~\bibnamefont{{Ciomaga Hatnean}}},
  \bibinfo{author}{\bibfnamefont{G.}~\bibnamefont{Balakrishnan}},
  \bibnamefont{and}
  \bibinfo{author}{\bibfnamefont{M.}~\bibnamefont{Wenderoth}},
  \bibinfo{journal}{Phys.\ Rev.\ B} \textbf{\bibinfo{volume}{102}},
  \bibinfo{pages}{205403} (\bibinfo{year}{2020}).

\bibitem[{\citenamefont{Enayat}(2014)}]{ena-the}
\bibinfo{author}{\bibfnamefont{M.}~\bibnamefont{Enayat}}, Ph.D. thesis,
  \bibinfo{school}{{\'E}cole Polytechnique F{\'e}d{\'e}rale de Lausanne,
  Switzerland} (\bibinfo{year}{2014}).

\bibitem[{\citenamefont{Pohlit et~al.}(2018)\citenamefont{Pohlit,
  R{\"o}{\ss}ler, Ohno, Ohno, {von Moln{\'a}r}, Fisk, M{\"u}ller, and
  Wirth}}]{poh18}
\bibinfo{author}{\bibfnamefont{M.}~\bibnamefont{Pohlit}},
  \bibinfo{author}{\bibfnamefont{S.}~\bibnamefont{R{\"o}{\ss}ler}},
  \bibinfo{author}{\bibfnamefont{Y.}~\bibnamefont{Ohno}},
  \bibinfo{author}{\bibfnamefont{H.}~\bibnamefont{Ohno}},
  \bibinfo{author}{\bibfnamefont{S.}~\bibnamefont{{von Moln{\'a}r}}},
  \bibinfo{author}{\bibfnamefont{Z.}~\bibnamefont{Fisk}},
  \bibinfo{author}{\bibfnamefont{J.}~\bibnamefont{M{\"u}ller}},
  \bibnamefont{and} \bibinfo{author}{\bibfnamefont{S.}~\bibnamefont{Wirth}},
  \bibinfo{journal}{Phys.\ Rev.\ Lett.} \textbf{\bibinfo{volume}{120}},
  \bibinfo{pages}{257201} (\bibinfo{year}{2018}).

\bibitem[{\citenamefont{R{\"o}{\ss}ler
  et~al.}(2020)\citenamefont{R{\"o}{\ss}ler, Jiao, Seiro, Rosa, Fisk,
  R{\"o}{\ss}ler, and Wirth}}]{roe20}
\bibinfo{author}{\bibfnamefont{S.}~\bibnamefont{R{\"o}{\ss}ler}},
  \bibinfo{author}{\bibfnamefont{L.}~\bibnamefont{Jiao}},
  \bibinfo{author}{\bibfnamefont{S.}~\bibnamefont{Seiro}},
  \bibinfo{author}{\bibfnamefont{P.~F.~S.} \bibnamefont{Rosa}},
  \bibinfo{author}{\bibfnamefont{Z.}~\bibnamefont{Fisk}},
  \bibinfo{author}{\bibfnamefont{U.~K.} \bibnamefont{R{\"o}{\ss}ler}},
  \bibnamefont{and} \bibinfo{author}{\bibfnamefont{S.}~\bibnamefont{Wirth}},
  \bibinfo{journal}{Phys.\ Rev.\ B} \textbf{\bibinfo{volume}{101}},
  \bibinfo{pages}{235421} (\bibinfo{year}{2020}).

\bibitem[{\citenamefont{G{\"u}ntherodt
  et~al.}(1982)\citenamefont{G{\"u}ntherodt, Thompson, Holtzberg, and
  Fisk}}]{gue82}
\bibinfo{author}{\bibfnamefont{G.}~\bibnamefont{G{\"u}ntherodt}},
  \bibinfo{author}{\bibfnamefont{W.~A.} \bibnamefont{Thompson}},
  \bibinfo{author}{\bibfnamefont{F.}~\bibnamefont{Holtzberg}},
  \bibnamefont{and} \bibinfo{author}{\bibfnamefont{Z.}~\bibnamefont{Fisk}},
  \bibinfo{journal}{Phys.\ Rev.\ Lett.} \textbf{\bibinfo{volume}{49}},
  \bibinfo{pages}{1030} (\bibinfo{year}{1982}).

\bibitem[{\citenamefont{Ba\v{t}kov\'{a}
  et~al.}(2008)\citenamefont{Ba\v{t}kov\'{a}, Ba\v{t}ko, Konovalova, and
  Shitsevalova}}]{bat07}
\bibinfo{author}{\bibfnamefont{M.}~\bibnamefont{Ba\v{t}kov\'{a}}},
  \bibinfo{author}{\bibfnamefont{I.}~\bibnamefont{Ba\v{t}ko}},
  \bibinfo{author}{\bibfnamefont{E.~S.} \bibnamefont{Konovalova}},
  \bibnamefont{and}
  \bibinfo{author}{\bibfnamefont{N.}~\bibnamefont{Shitsevalova}},
  \bibinfo{journal}{Acta Phys.\ Pol.\ A} \textbf{\bibinfo{volume}{113}},
  \bibinfo{pages}{255} (\bibinfo{year}{2008}).

\bibitem[{\citenamefont{Amsler et~al.}(1998)\citenamefont{Amsler, Fisk, Sarrao,
  {von Moln{\'a}r}, Meisel, and Sharifi}}]{ams98}
\bibinfo{author}{\bibfnamefont{B.}~\bibnamefont{Amsler}},
  \bibinfo{author}{\bibfnamefont{Z.}~\bibnamefont{Fisk}},
  \bibinfo{author}{\bibfnamefont{J.~L.} \bibnamefont{Sarrao}},
  \bibinfo{author}{\bibfnamefont{S.}~\bibnamefont{{von Moln{\'a}r}}},
  \bibinfo{author}{\bibfnamefont{M.~W.} \bibnamefont{Meisel}},
  \bibnamefont{and} \bibinfo{author}{\bibfnamefont{F.}~\bibnamefont{Sharifi}},
  \bibinfo{journal}{Phys.\ Rev.\ B} \textbf{\bibinfo{volume}{57}},
  \bibinfo{pages}{8747} (\bibinfo{year}{1998}).

\bibitem[{\citenamefont{Flachbart et~al.}(2001)\citenamefont{Flachbart, Gloos,
  Konovalova, Paderno, Reiffers, Samuely, and \v{S}vec}}]{fla01}
\bibinfo{author}{\bibfnamefont{K.}~\bibnamefont{Flachbart}},
  \bibinfo{author}{\bibfnamefont{K.}~\bibnamefont{Gloos}},
  \bibinfo{author}{\bibfnamefont{E.}~\bibnamefont{Konovalova}},
  \bibinfo{author}{\bibfnamefont{Y.}~\bibnamefont{Paderno}},
  \bibinfo{author}{\bibfnamefont{M.}~\bibnamefont{Reiffers}},
  \bibinfo{author}{\bibfnamefont{P.}~\bibnamefont{Samuely}}, \bibnamefont{and}
  \bibinfo{author}{\bibfnamefont{P.}~\bibnamefont{\v{S}vec}},
  \bibinfo{journal}{Phys.\ Rev.\ B} \textbf{\bibinfo{volume}{64}},
  \bibinfo{pages}{085104} (\bibinfo{year}{2001}).

\bibitem[{\citenamefont{Zhang et~al.}(2013)\citenamefont{Zhang, Butch, Syers,
  Ziemak, Greene, and Paglione}}]{zha13}
\bibinfo{author}{\bibfnamefont{X.}~\bibnamefont{Zhang}},
  \bibinfo{author}{\bibfnamefont{N.~P.} \bibnamefont{Butch}},
  \bibinfo{author}{\bibfnamefont{P.}~\bibnamefont{Syers}},
  \bibinfo{author}{\bibfnamefont{S.}~\bibnamefont{Ziemak}},
  \bibinfo{author}{\bibfnamefont{R.~L.} \bibnamefont{Greene}},
  \bibnamefont{and} \bibinfo{author}{\bibfnamefont{J.}~\bibnamefont{Paglione}},
  \bibinfo{journal}{Phys.\ Rev.\ X} \textbf{\bibinfo{volume}{3}},
  \bibinfo{pages}{011011} (\bibinfo{year}{2013}).

\bibitem[{\citenamefont{Costi}(2000)}]{cos00}
\bibinfo{author}{\bibfnamefont{T.~A.} \bibnamefont{Costi}},
  \bibinfo{journal}{Phys.\ Rev.\ Lett.} \textbf{\bibinfo{volume}{85}},
  \bibinfo{pages}{1504} (\bibinfo{year}{2000}).

\bibitem[{\citenamefont{Wirth and Steglich}(2016)}]{wir16}
\bibinfo{author}{\bibfnamefont{S.}~\bibnamefont{Wirth}} \bibnamefont{and}
  \bibinfo{author}{\bibfnamefont{F.}~\bibnamefont{Steglich}},
  \bibinfo{journal}{Nat.\ Rev.\ Mat.} \textbf{\bibinfo{volume}{1}},
  \bibinfo{pages}{16051} (\bibinfo{year}{2016}).

\bibitem[{\citenamefont{Fano}(1961)}]{fan61}
\bibinfo{author}{\bibfnamefont{U.}~\bibnamefont{Fano}},
  \bibinfo{journal}{Phys.\ Rev.} \textbf{\bibinfo{volume}{124}},
  \bibinfo{pages}{1866} (\bibinfo{year}{1961}).

\bibitem[{\citenamefont{Maltseva et~al.}(2009)\citenamefont{Maltseva, Dzero,
  and Coleman}}]{mal09}
\bibinfo{author}{\bibfnamefont{M.}~\bibnamefont{Maltseva}},
  \bibinfo{author}{\bibfnamefont{M.}~\bibnamefont{Dzero}}, \bibnamefont{and}
  \bibinfo{author}{\bibfnamefont{P.}~\bibnamefont{Coleman}},
  \bibinfo{journal}{Phys.\ Rev.\ Lett.} \textbf{\bibinfo{volume}{103}},
  \bibinfo{pages}{206402} (\bibinfo{year}{2009}).

\bibitem[{\citenamefont{Figgins and Morr}(2010)}]{fig10}
\bibinfo{author}{\bibfnamefont{J.}~\bibnamefont{Figgins}} \bibnamefont{and}
  \bibinfo{author}{\bibfnamefont{D.~K.} \bibnamefont{Morr}},
  \bibinfo{journal}{Phys.\ Rev.\ Lett.} \textbf{\bibinfo{volume}{104}},
  \bibinfo{pages}{187202} (\bibinfo{year}{2010}).

\bibitem[{\citenamefont{W{\"o}lfle et~al.}(2010)\citenamefont{W{\"o}lfle, Dubi,
  and Balatsky}}]{woe10}
\bibinfo{author}{\bibfnamefont{P.}~\bibnamefont{W{\"o}lfle}},
  \bibinfo{author}{\bibfnamefont{Y.}~\bibnamefont{Dubi}}, \bibnamefont{and}
  \bibinfo{author}{\bibfnamefont{A.~V.} \bibnamefont{Balatsky}},
  \bibinfo{journal}{Phys.\ Rev.\ Lett.} \textbf{\bibinfo{volume}{105}},
  \bibinfo{pages}{246401} (\bibinfo{year}{2010}).

\bibitem[{\citenamefont{Schiller and Hershfield}(2000)}]{sch00}
\bibinfo{author}{\bibfnamefont{A.}~\bibnamefont{Schiller}} \bibnamefont{and}
  \bibinfo{author}{\bibfnamefont{S.}~\bibnamefont{Hershfield}},
  \bibinfo{journal}{Phys.\ Rev.\ B} \textbf{\bibinfo{volume}{61}},
  \bibinfo{pages}{9036} (\bibinfo{year}{2000}).

\bibitem[{\citenamefont{Ernst et~al.}(2011)\citenamefont{Ernst, Kirchner,
  Krellner, Geibel, Zwicknagl, Steglich, and Wirth}}]{ern11}
\bibinfo{author}{\bibfnamefont{S.}~\bibnamefont{Ernst}},
  \bibinfo{author}{\bibfnamefont{S.}~\bibnamefont{Kirchner}},
  \bibinfo{author}{\bibfnamefont{C.}~\bibnamefont{Krellner}},
  \bibinfo{author}{\bibfnamefont{C.}~\bibnamefont{Geibel}},
  \bibinfo{author}{\bibfnamefont{G.}~\bibnamefont{Zwicknagl}},
  \bibinfo{author}{\bibfnamefont{F.}~\bibnamefont{Steglich}}, \bibnamefont{and}
  \bibinfo{author}{\bibfnamefont{S.}~\bibnamefont{Wirth}},
  \bibinfo{journal}{Nature} \textbf{\bibinfo{volume}{474}},
  \bibinfo{pages}{362} (\bibinfo{year}{2011}).

\bibitem[{\citenamefont{Aynajian et~al.}(2012)\citenamefont{Aynajian, {da Silva
  Neto}, Gyenis, Baumbach, Thompson, Fisk, Bauer, and Yazdani}}]{ayn12}
\bibinfo{author}{\bibfnamefont{P.}~\bibnamefont{Aynajian}},
  \bibinfo{author}{\bibfnamefont{E.~H.} \bibnamefont{{da Silva Neto}}},
  \bibinfo{author}{\bibfnamefont{A.}~\bibnamefont{Gyenis}},
  \bibinfo{author}{\bibfnamefont{R.~E.} \bibnamefont{Baumbach}},
  \bibinfo{author}{\bibfnamefont{J.~D.} \bibnamefont{Thompson}},
  \bibinfo{author}{\bibfnamefont{Z.}~\bibnamefont{Fisk}},
  \bibinfo{author}{\bibfnamefont{E.~D.} \bibnamefont{Bauer}}, \bibnamefont{and}
  \bibinfo{author}{\bibfnamefont{A.}~\bibnamefont{Yazdani}},
  \bibinfo{journal}{Nature} \textbf{\bibinfo{volume}{486}},
  \bibinfo{pages}{201} (\bibinfo{year}{2012}).

\bibitem[{\citenamefont{Arab et~al.}(2016)\citenamefont{Arab, Gray,
  Nem\v{s}\'{a}k, Evtushinsky, Schneider, Kim, Fisk, Rosa, Durakiewicz, and
  Riseborough}}]{ara16}
\bibinfo{author}{\bibfnamefont{A.}~\bibnamefont{Arab}},
  \bibinfo{author}{\bibfnamefont{A.~X.} \bibnamefont{Gray}},
  \bibinfo{author}{\bibfnamefont{S.}~\bibnamefont{Nem\v{s}\'{a}k}},
  \bibinfo{author}{\bibfnamefont{D.~V.} \bibnamefont{Evtushinsky}},
  \bibinfo{author}{\bibfnamefont{C.~M.} \bibnamefont{Schneider}},
  \bibinfo{author}{\bibfnamefont{D.-J.} \bibnamefont{Kim}},
  \bibinfo{author}{\bibfnamefont{Z.}~\bibnamefont{Fisk}},
  \bibinfo{author}{\bibfnamefont{P.~F.~S.} \bibnamefont{Rosa}},
  \bibinfo{author}{\bibfnamefont{T.}~\bibnamefont{Durakiewicz}},
  \bibnamefont{and} \bibinfo{author}{\bibfnamefont{P.~S.}
  \bibnamefont{Riseborough}}, \bibinfo{journal}{Phys.\ Rev.\ B}
  \textbf{\bibinfo{volume}{94}}, \bibinfo{pages}{235125}
  (\bibinfo{year}{2016}).

\bibitem[{\citenamefont{Sun et~al.}(2017)\citenamefont{Sun, Kim, Fisk, and
  Park}}]{sun17}
\bibinfo{author}{\bibfnamefont{L.}~\bibnamefont{Sun}},
  \bibinfo{author}{\bibfnamefont{D.-J.} \bibnamefont{Kim}},
  \bibinfo{author}{\bibfnamefont{Z.}~\bibnamefont{Fisk}}, \bibnamefont{and}
  \bibinfo{author}{\bibfnamefont{W.~K.} \bibnamefont{Park}},
  \bibinfo{journal}{Phys.\ Rev.\ B} \textbf{\bibinfo{volume}{95}},
  \bibinfo{pages}{195129} (\bibinfo{year}{2017}).

\bibitem[{\citenamefont{Zhang et~al.}(2018)\citenamefont{Zhang, Yong, Takeuchi,
  Greene, and Averitt}}]{zha18}
\bibinfo{author}{\bibfnamefont{J.}~\bibnamefont{Zhang}},
  \bibinfo{author}{\bibfnamefont{J.}~\bibnamefont{Yong}},
  \bibinfo{author}{\bibfnamefont{I.}~\bibnamefont{Takeuchi}},
  \bibinfo{author}{\bibfnamefont{R.~L.} \bibnamefont{Greene}},
  \bibnamefont{and} \bibinfo{author}{\bibfnamefont{R.~D.}
  \bibnamefont{Averitt}}, \bibinfo{journal}{Phys.\ Rev.\ B}
  \textbf{\bibinfo{volume}{97}}, \bibinfo{pages}{155119}
  (\bibinfo{year}{2018}).

\bibitem[{\citenamefont{Alekseev
  et~al.}(1993{\natexlab{b}})\citenamefont{Alekseev, Lazukov, Osborn, Rainford,
  Sadikov, Konovalova, and Paderno}}]{ale93}
\bibinfo{author}{\bibfnamefont{P.~A.} \bibnamefont{Alekseev}},
  \bibinfo{author}{\bibfnamefont{V.~N.} \bibnamefont{Lazukov}},
  \bibinfo{author}{\bibfnamefont{R.}~\bibnamefont{Osborn}},
  \bibinfo{author}{\bibfnamefont{B.~D.} \bibnamefont{Rainford}},
  \bibinfo{author}{\bibfnamefont{I.~P.} \bibnamefont{Sadikov}},
  \bibinfo{author}{\bibfnamefont{E.~S.} \bibnamefont{Konovalova}},
  \bibnamefont{and} \bibinfo{author}{\bibfnamefont{Y.~B.}
  \bibnamefont{Paderno}}, \bibinfo{journal}{Europhys.\ Lett.}
  \textbf{\bibinfo{volume}{23}}, \bibinfo{pages}{347}
  (\bibinfo{year}{1993}{\natexlab{b}}).

\bibitem[{\citenamefont{Kim et~al.}(2019)\citenamefont{Kim, Jang, Wang,
  Paglione, Hong, Lee, Choi, and Kim}}]{kim19}
\bibinfo{author}{\bibfnamefont{J.}~\bibnamefont{Kim}},
  \bibinfo{author}{\bibfnamefont{C.}~\bibnamefont{Jang}},
  \bibinfo{author}{\bibfnamefont{X.}~\bibnamefont{Wang}},
  \bibinfo{author}{\bibfnamefont{J.}~\bibnamefont{Paglione}},
  \bibinfo{author}{\bibfnamefont{S.}~\bibnamefont{Hong}},
  \bibinfo{author}{\bibfnamefont{J.}~\bibnamefont{Lee}},
  \bibinfo{author}{\bibfnamefont{H.}~\bibnamefont{Choi}}, \bibnamefont{and}
  \bibinfo{author}{\bibfnamefont{D.}~\bibnamefont{Kim}},
  \bibinfo{journal}{Phys. Rev. B} \textbf{\bibinfo{volume}{99}},
  \bibinfo{pages}{245148} (\bibinfo{year}{2019}).

\bibitem[{\citenamefont{Gab\'{a}ni et~al.}(1999)\citenamefont{Gab\'{a}ni,
  Flachbart, Farka\v{s}ovsk\'{y}, Pavl\'{\i}k, Bat'ko, Herrmannsd{\"o}rfer,
  Konovalova, and Paderno}}]{gab99}
\bibinfo{author}{\bibfnamefont{S.}~\bibnamefont{Gab\'{a}ni}},
  \bibinfo{author}{\bibfnamefont{K.}~\bibnamefont{Flachbart}},
  \bibinfo{author}{\bibfnamefont{P.}~\bibnamefont{Farka\v{s}ovsk\'{y}}},
  \bibinfo{author}{\bibfnamefont{V.}~\bibnamefont{Pavl\'{\i}k}},
  \bibinfo{author}{\bibfnamefont{I.}~\bibnamefont{Bat'ko}},
  \bibinfo{author}{\bibfnamefont{T.}~\bibnamefont{Herrmannsd{\"o}rfer}},
  \bibinfo{author}{\bibfnamefont{E.}~\bibnamefont{Konovalova}},
  \bibnamefont{and} \bibinfo{author}{\bibfnamefont{Y.}~\bibnamefont{Paderno}},
  \bibinfo{journal}{Physica B} \textbf{\bibinfo{volume}{259}},
  \bibinfo{pages}{345} (\bibinfo{year}{1999}).

\bibitem[{\citenamefont{Sluchanko et~al.}(2000)\citenamefont{Sluchanko,
  Glushkov, Gorshunov, Demishev, Kondrin, Pronin, Volkov, Savchenko,
  Gr{\"u}ner, Bruynseraede et~al.}}]{slu00}
\bibinfo{author}{\bibfnamefont{N.~E.} \bibnamefont{Sluchanko}},
  \bibinfo{author}{\bibfnamefont{V.~V.} \bibnamefont{Glushkov}},
  \bibinfo{author}{\bibfnamefont{B.~P.} \bibnamefont{Gorshunov}},
  \bibinfo{author}{\bibfnamefont{S.~V.} \bibnamefont{Demishev}},
  \bibinfo{author}{\bibfnamefont{M.~V.} \bibnamefont{Kondrin}},
  \bibinfo{author}{\bibfnamefont{A.~A.} \bibnamefont{Pronin}},
  \bibinfo{author}{\bibfnamefont{A.~A.} \bibnamefont{Volkov}},
  \bibinfo{author}{\bibfnamefont{A.~K.} \bibnamefont{Savchenko}},
  \bibinfo{author}{\bibfnamefont{G.}~\bibnamefont{Gr{\"u}ner}},
  \bibinfo{author}{\bibfnamefont{Y.}~\bibnamefont{Bruynseraede}},
  \bibnamefont{et~al.}, \bibinfo{journal}{Phys.\ Rev.\ B}
  \textbf{\bibinfo{volume}{61}}, \bibinfo{pages}{9906} (\bibinfo{year}{2000}).

\bibitem[{\citenamefont{Gab\'{a}ni et~al.}(2001)\citenamefont{Gab\'{a}ni,
  Flachbart, Konovalova, Orend\'{a}\v{c}, Paderno, Pavl\'{\i}k, and
  \v{S}ebek}}]{gab01}
\bibinfo{author}{\bibfnamefont{S.}~\bibnamefont{Gab\'{a}ni}},
  \bibinfo{author}{\bibfnamefont{K.}~\bibnamefont{Flachbart}},
  \bibinfo{author}{\bibfnamefont{E.}~\bibnamefont{Konovalova}},
  \bibinfo{author}{\bibfnamefont{M.}~\bibnamefont{Orend\'{a}\v{c}}},
  \bibinfo{author}{\bibfnamefont{Y.}~\bibnamefont{Paderno}},
  \bibinfo{author}{\bibfnamefont{V.}~\bibnamefont{Pavl\'{\i}k}},
  \bibnamefont{and}
  \bibinfo{author}{\bibfnamefont{J.}~\bibnamefont{\v{S}ebek}},
  \bibinfo{journal}{Solid State Comm.} \textbf{\bibinfo{volume}{117}},
  \bibinfo{pages}{641} (\bibinfo{year}{2001}).

\bibitem[{\citenamefont{Xu et~al.}(2016)\citenamefont{Xu, Ding, and
  Shi}}]{nxu16}
\bibinfo{author}{\bibfnamefont{N.}~\bibnamefont{Xu}},
  \bibinfo{author}{\bibfnamefont{H.}~\bibnamefont{Ding}}, \bibnamefont{and}
  \bibinfo{author}{\bibfnamefont{M.}~\bibnamefont{Shi}}, \bibinfo{journal}{J.\
  Phys.: Condens.\ Matter} \textbf{\bibinfo{volume}{28}},
  \bibinfo{pages}{363001} (\bibinfo{year}{2016}).

\bibitem[{\citenamefont{Allen and Martin}(1980)}]{all80}
\bibinfo{author}{\bibfnamefont{J.~W.} \bibnamefont{Allen}} \bibnamefont{and}
  \bibinfo{author}{\bibfnamefont{R.~M.} \bibnamefont{Martin}},
  \bibinfo{journal}{J. Phys.\ Colloq.} \textbf{\bibinfo{volume}{C5}},
  \bibinfo{pages}{171} (\bibinfo{year}{1980}).

\bibitem[{\citenamefont{Chen et~al.}(2015)\citenamefont{Chen, Shang, Jin, Zhao,
  Wu, Xiang, Xia, Wang, Luo, Wu et~al.}}]{chen15}
\bibinfo{author}{\bibfnamefont{F.}~\bibnamefont{Chen}},
  \bibinfo{author}{\bibfnamefont{C.}~\bibnamefont{Shang}},
  \bibinfo{author}{\bibfnamefont{Z.}~\bibnamefont{Jin}},
  \bibinfo{author}{\bibfnamefont{D.}~\bibnamefont{Zhao}},
  \bibinfo{author}{\bibfnamefont{Y.~P.} \bibnamefont{Wu}},
  \bibinfo{author}{\bibfnamefont{Z.~J.} \bibnamefont{Xiang}},
  \bibinfo{author}{\bibfnamefont{Z.~C.} \bibnamefont{Xia}},
  \bibinfo{author}{\bibfnamefont{A.~F.} \bibnamefont{Wang}},
  \bibinfo{author}{\bibfnamefont{X.~G.} \bibnamefont{Luo}},
  \bibinfo{author}{\bibfnamefont{T.}~\bibnamefont{Wu}}, \bibnamefont{et~al.},
  \bibinfo{journal}{Phys.\ Rev.\ B} \textbf{\bibinfo{volume}{91}},
  \bibinfo{pages}{205133} (\bibinfo{year}{2015}).

\bibitem[{\citenamefont{Amorese et~al.}(2019)\citenamefont{Amorese, Stockert,
  Kummer, Brookes, Kim, Fisk, Haverkort, Thalmeier, Tjeng, and
  Severing}}]{amo19}
\bibinfo{author}{\bibfnamefont{A.}~\bibnamefont{Amorese}},
  \bibinfo{author}{\bibfnamefont{O.}~\bibnamefont{Stockert}},
  \bibinfo{author}{\bibfnamefont{K.}~\bibnamefont{Kummer}},
  \bibinfo{author}{\bibfnamefont{N.~B.} \bibnamefont{Brookes}},
  \bibinfo{author}{\bibfnamefont{D.-J.} \bibnamefont{Kim}},
  \bibinfo{author}{\bibfnamefont{Z.}~\bibnamefont{Fisk}},
  \bibinfo{author}{\bibfnamefont{M.~W.} \bibnamefont{Haverkort}},
  \bibinfo{author}{\bibfnamefont{P.}~\bibnamefont{Thalmeier}},
  \bibinfo{author}{\bibfnamefont{L.~H.} \bibnamefont{Tjeng}}, \bibnamefont{and}
  \bibinfo{author}{\bibfnamefont{A.}~\bibnamefont{Severing}},
  \bibinfo{journal}{Phys.\ Rev.\ B} \textbf{\bibinfo{volume}{100}},
  \bibinfo{pages}{241107(R)} (\bibinfo{year}{2019}).

\bibitem[{\citenamefont{Kapilevich et~al.}(2015)\citenamefont{Kapilevich,
  Riseborough, Gray, Gulacsi, Durakiewicz, and Smith}}]{kap15}
\bibinfo{author}{\bibfnamefont{G.~A.} \bibnamefont{Kapilevich}},
  \bibinfo{author}{\bibfnamefont{P.~S.} \bibnamefont{Riseborough}},
  \bibinfo{author}{\bibfnamefont{A.~X.} \bibnamefont{Gray}},
  \bibinfo{author}{\bibfnamefont{M.}~\bibnamefont{Gulacsi}},
  \bibinfo{author}{\bibfnamefont{T.}~\bibnamefont{Durakiewicz}},
  \bibnamefont{and} \bibinfo{author}{\bibfnamefont{J.~L.} \bibnamefont{Smith}},
  \bibinfo{journal}{Phys.\ Rev.\ B} \textbf{\bibinfo{volume}{92}},
  \bibinfo{pages}{085133} (\bibinfo{year}{2015}).

\bibitem[{\citenamefont{Fuhrman et~al.}(2015)\citenamefont{Fuhrman, Leiner,
  Nikoli\'{c}, Granroth, Stone, Lumsden, {DeBeer-Schmitt}, Alekseev, Mignot,
  Koohpayeh et~al.}}]{fuh15}
\bibinfo{author}{\bibfnamefont{W.~T.} \bibnamefont{Fuhrman}},
  \bibinfo{author}{\bibfnamefont{J.}~\bibnamefont{Leiner}},
  \bibinfo{author}{\bibfnamefont{P.}~\bibnamefont{Nikoli\'{c}}},
  \bibinfo{author}{\bibfnamefont{G.~E.} \bibnamefont{Granroth}},
  \bibinfo{author}{\bibfnamefont{M.~B.} \bibnamefont{Stone}},
  \bibinfo{author}{\bibfnamefont{M.~D.} \bibnamefont{Lumsden}},
  \bibinfo{author}{\bibfnamefont{L.}~\bibnamefont{{DeBeer-Schmitt}}},
  \bibinfo{author}{\bibfnamefont{P.~A.} \bibnamefont{Alekseev}},
  \bibinfo{author}{\bibfnamefont{J.-M.} \bibnamefont{Mignot}},
  \bibinfo{author}{\bibfnamefont{S.~M.} \bibnamefont{Koohpayeh}},
  \bibnamefont{et~al.}, \bibinfo{journal}{Phys.\ Rev.\ Lett.}
  \textbf{\bibinfo{volume}{114}}, \bibinfo{pages}{036401}
  (\bibinfo{year}{2015}).

\bibitem[{\citenamefont{Caldwell et~al.}(2007)\citenamefont{Caldwell, Reyes,
  Moulton, Kuhns, Hoch, Schlottmann, and Fisk}}]{cal07}
\bibinfo{author}{\bibfnamefont{T.}~\bibnamefont{Caldwell}},
  \bibinfo{author}{\bibfnamefont{A.~P.} \bibnamefont{Reyes}},
  \bibinfo{author}{\bibfnamefont{W.~G.} \bibnamefont{Moulton}},
  \bibinfo{author}{\bibfnamefont{P.~L.} \bibnamefont{Kuhns}},
  \bibinfo{author}{\bibfnamefont{M.~J.~R.} \bibnamefont{Hoch}},
  \bibinfo{author}{\bibfnamefont{P.}~\bibnamefont{Schlottmann}},
  \bibnamefont{and} \bibinfo{author}{\bibfnamefont{Z.}~\bibnamefont{Fisk}},
  \bibinfo{journal}{Phys.\ Rev.\ B} \textbf{\bibinfo{volume}{75}},
  \bibinfo{pages}{075106} (\bibinfo{year}{2007}).

\bibitem[{\citenamefont{Park et~al.}(2016)\citenamefont{Park, Sun, Noddings,
  Kim, Fisk, and Greene}}]{par16}
\bibinfo{author}{\bibfnamefont{W.~K.} \bibnamefont{Park}},
  \bibinfo{author}{\bibfnamefont{L.}~\bibnamefont{Sun}},
  \bibinfo{author}{\bibfnamefont{A.}~\bibnamefont{Noddings}},
  \bibinfo{author}{\bibfnamefont{D.-J.} \bibnamefont{Kim}},
  \bibinfo{author}{\bibfnamefont{Z.}~\bibnamefont{Fisk}}, \bibnamefont{and}
  \bibinfo{author}{\bibfnamefont{L.~H.} \bibnamefont{Greene}},
  \bibinfo{journal}{Proc.\ Natl.\ Acad.\ Sci.\ USA}
  \textbf{\bibinfo{volume}{113}}, \bibinfo{pages}{6599} (\bibinfo{year}{2016}).

\bibitem[{\citenamefont{Min et~al.}(2017)\citenamefont{Min, Goth, Lutz,
  Bentmann, Kang, Cho, Werner, Chen, Assaad, and Reinert}}]{min17}
\bibinfo{author}{\bibfnamefont{C.-H.} \bibnamefont{Min}},
  \bibinfo{author}{\bibfnamefont{F.}~\bibnamefont{Goth}},
  \bibinfo{author}{\bibfnamefont{P.}~\bibnamefont{Lutz}},
  \bibinfo{author}{\bibfnamefont{H.}~\bibnamefont{Bentmann}},
  \bibinfo{author}{\bibfnamefont{B.~Y.} \bibnamefont{Kang}},
  \bibinfo{author}{\bibfnamefont{B.~K.} \bibnamefont{Cho}},
  \bibinfo{author}{\bibfnamefont{J.}~\bibnamefont{Werner}},
  \bibinfo{author}{\bibfnamefont{K.-S.} \bibnamefont{Chen}},
  \bibinfo{author}{\bibfnamefont{F.}~\bibnamefont{Assaad}}, \bibnamefont{and}
  \bibinfo{author}{\bibfnamefont{F.}~\bibnamefont{Reinert}},
  \bibinfo{journal}{Sci. Rep.} \textbf{\bibinfo{volume}{7}},
  \bibinfo{pages}{11980} (\bibinfo{year}{2017}).

\bibitem[{\citenamefont{Alexandrov et~al.}(2015)\citenamefont{Alexandrov,
  Coleman, and Erten}}]{ale15}
\bibinfo{author}{\bibfnamefont{V.}~\bibnamefont{Alexandrov}},
  \bibinfo{author}{\bibfnamefont{P.}~\bibnamefont{Coleman}}, \bibnamefont{and}
  \bibinfo{author}{\bibfnamefont{O.}~\bibnamefont{Erten}},
  \bibinfo{journal}{Phys.\ Rev.\ Lett.} \textbf{\bibinfo{volume}{114}},
  \bibinfo{pages}{177202} (\bibinfo{year}{2015}).

\bibitem[{\citenamefont{Luo et~al.}(2015)\citenamefont{Luo, Chen, Dai, Xu, and
  Thompson}}]{luo15}
\bibinfo{author}{\bibfnamefont{Y.}~\bibnamefont{Luo}},
  \bibinfo{author}{\bibfnamefont{H.}~\bibnamefont{Chen}},
  \bibinfo{author}{\bibfnamefont{J.}~\bibnamefont{Dai}},
  \bibinfo{author}{\bibfnamefont{Z.}~\bibnamefont{Xu}}, \bibnamefont{and}
  \bibinfo{author}{\bibfnamefont{J.~D.} \bibnamefont{Thompson}},
  \bibinfo{journal}{Phys.\ Rev.\ B} \textbf{\bibinfo{volume}{91}},
  \bibinfo{pages}{075130} (\bibinfo{year}{2015}).

\bibitem[{\citenamefont{Petersen et~al.}(1998)\citenamefont{Petersen, Sprunger,
  Hofmann, L{\ae}gsgaard, Briner, Doering, Rust, Bradshaw, Besenbacher, and
  Plummer}}]{pet98}
\bibinfo{author}{\bibfnamefont{L.}~\bibnamefont{Petersen}},
  \bibinfo{author}{\bibfnamefont{P.~T.} \bibnamefont{Sprunger}},
  \bibinfo{author}{\bibfnamefont{P.}~\bibnamefont{Hofmann}},
  \bibinfo{author}{\bibfnamefont{E.}~\bibnamefont{L{\ae}gsgaard}},
  \bibinfo{author}{\bibfnamefont{B.~G.} \bibnamefont{Briner}},
  \bibinfo{author}{\bibfnamefont{M.}~\bibnamefont{Doering}},
  \bibinfo{author}{\bibfnamefont{H.-P.} \bibnamefont{Rust}},
  \bibinfo{author}{\bibfnamefont{A.~M.} \bibnamefont{Bradshaw}},
  \bibinfo{author}{\bibfnamefont{F.}~\bibnamefont{Besenbacher}},
  \bibnamefont{and} \bibinfo{author}{\bibfnamefont{E.~W.}
  \bibnamefont{Plummer}}, \bibinfo{journal}{Phys.\ Rev.\ B}
  \textbf{\bibinfo{volume}{57}}, \bibinfo{pages}{R6858} (\bibinfo{year}{1998}).

\bibitem[{\citenamefont{Hoffman}(2011)}]{hof11}
\bibinfo{author}{\bibfnamefont{J.~E.} \bibnamefont{Hoffman}},
  \bibinfo{journal}{Rep.\ Prog.\ Phys.} \textbf{\bibinfo{volume}{74}},
  \bibinfo{pages}{124513} (\bibinfo{year}{2011}).

\bibitem[{\citenamefont{Oka et~al.}(2014)\citenamefont{Oka, Brovko, Corbetta,
  Stepanyuk, Sander, and Kirschner}}]{oka14}
\bibinfo{author}{\bibfnamefont{H.}~\bibnamefont{Oka}},
  \bibinfo{author}{\bibfnamefont{O.~O.} \bibnamefont{Brovko}},
  \bibinfo{author}{\bibfnamefont{M.}~\bibnamefont{Corbetta}},
  \bibinfo{author}{\bibfnamefont{V.~S.} \bibnamefont{Stepanyuk}},
  \bibinfo{author}{\bibfnamefont{D.}~\bibnamefont{Sander}}, \bibnamefont{and}
  \bibinfo{author}{\bibfnamefont{J.}~\bibnamefont{Kirschner}},
  \bibinfo{journal}{Rev.\ Mod.\ Phys.} \textbf{\bibinfo{volume}{86}},
  \bibinfo{pages}{1127} (\bibinfo{year}{2014}).

\bibitem[{\citenamefont{Zhang et~al.}(2009)\citenamefont{Zhang, Cheng, Chen,
  Jia, Ma, He, Wang, Zhang, Dai, Fang et~al.}}]{zha09}
\bibinfo{author}{\bibfnamefont{T.}~\bibnamefont{Zhang}},
  \bibinfo{author}{\bibfnamefont{P.}~\bibnamefont{Cheng}},
  \bibinfo{author}{\bibfnamefont{X.}~\bibnamefont{Chen}},
  \bibinfo{author}{\bibfnamefont{J.-F.} \bibnamefont{Jia}},
  \bibinfo{author}{\bibfnamefont{X.}~\bibnamefont{Ma}},
  \bibinfo{author}{\bibfnamefont{K.}~\bibnamefont{He}},
  \bibinfo{author}{\bibfnamefont{L.}~\bibnamefont{Wang}},
  \bibinfo{author}{\bibfnamefont{H.}~\bibnamefont{Zhang}},
  \bibinfo{author}{\bibfnamefont{X.}~\bibnamefont{Dai}},
  \bibinfo{author}{\bibfnamefont{Z.}~\bibnamefont{Fang}}, \bibnamefont{et~al.},
  \bibinfo{journal}{Phys.\ Rev.\ Lett.} \textbf{\bibinfo{volume}{103}},
  \bibinfo{pages}{266803} (\bibinfo{year}{2009}).

\bibitem[{\citenamefont{Guo and Franz}(2010)}]{guo10}
\bibinfo{author}{\bibfnamefont{H.~M.} \bibnamefont{Guo}} \bibnamefont{and}
  \bibinfo{author}{\bibfnamefont{M.}~\bibnamefont{Franz}},
  \bibinfo{journal}{Phys.\ Rev.\ B} \textbf{\bibinfo{volume}{81}},
  \bibinfo{pages}{041102} (\bibinfo{year}{2010}).

\bibitem[{\citenamefont{Thompson}(2011)}]{jdt11}
\bibinfo{author}{\bibfnamefont{J.~D.} \bibnamefont{Thompson}},
  \bibinfo{journal}{Proc.\ Natl.\ Acad.\ Sci.\ USA}
  \textbf{\bibinfo{volume}{108}}, \bibinfo{pages}{18191}
  (\bibinfo{year}{2011}).

\bibitem[{\citenamefont{Figgins and Morr}(2011)}]{fig11}
\bibinfo{author}{\bibfnamefont{J.}~\bibnamefont{Figgins}} \bibnamefont{and}
  \bibinfo{author}{\bibfnamefont{D.~K.} \bibnamefont{Morr}},
  \bibinfo{journal}{Phys.\ Rev.\ Lett.} \textbf{\bibinfo{volume}{107}},
  \bibinfo{pages}{066401} (\bibinfo{year}{2011}).

\bibitem[{\citenamefont{Baum and Stern}(2012)}]{bau12}
\bibinfo{author}{\bibfnamefont{Y.}~\bibnamefont{Baum}} \bibnamefont{and}
  \bibinfo{author}{\bibfnamefont{A.}~\bibnamefont{Stern}},
  \bibinfo{journal}{Phys.\ Rev.\ B} \textbf{\bibinfo{volume}{85}},
  \bibinfo{pages}{121105} (\bibinfo{year}{2012}).

\bibitem[{\citenamefont{Nakajima et~al.}(2015)\citenamefont{Nakajima, Syers,
  Wang, Wang, and Paglione}}]{nak15}
\bibinfo{author}{\bibfnamefont{Y.}~\bibnamefont{Nakajima}},
  \bibinfo{author}{\bibfnamefont{P.}~\bibnamefont{Syers}},
  \bibinfo{author}{\bibfnamefont{X.}~\bibnamefont{Wang}},
  \bibinfo{author}{\bibfnamefont{R.}~\bibnamefont{Wang}}, \bibnamefont{and}
  \bibinfo{author}{\bibfnamefont{J.}~\bibnamefont{Paglione}},
  \bibinfo{journal}{Nature Phys.} \textbf{\bibinfo{volume}{12}},
  \bibinfo{pages}{213} (\bibinfo{year}{2015}).

\bibitem[{\citenamefont{Okada et~al.}(2011)\citenamefont{Okada, Dhital, Zhou,
  Huemiller, Lin, Basak, Bansil, Huang, Ding, Wang et~al.}}]{oka11}
\bibinfo{author}{\bibfnamefont{Y.}~\bibnamefont{Okada}},
  \bibinfo{author}{\bibfnamefont{C.}~\bibnamefont{Dhital}},
  \bibinfo{author}{\bibfnamefont{W.}~\bibnamefont{Zhou}},
  \bibinfo{author}{\bibfnamefont{E.~D.} \bibnamefont{Huemiller}},
  \bibinfo{author}{\bibfnamefont{H.}~\bibnamefont{Lin}},
  \bibinfo{author}{\bibfnamefont{S.}~\bibnamefont{Basak}},
  \bibinfo{author}{\bibfnamefont{A.}~\bibnamefont{Bansil}},
  \bibinfo{author}{\bibfnamefont{Y.-B.} \bibnamefont{Huang}},
  \bibinfo{author}{\bibfnamefont{H.}~\bibnamefont{Ding}},
  \bibinfo{author}{\bibfnamefont{Z.}~\bibnamefont{Wang}}, \bibnamefont{et~al.},
  \bibinfo{journal}{Phys.\ Rev.\ Lett.} \textbf{\bibinfo{volume}{106}},
  \bibinfo{pages}{206805} (\bibinfo{year}{2011}).

\bibitem[{\citenamefont{Teng et~al.}(2019)\citenamefont{Teng, Liu, and
  Li}}]{ten19}
\bibinfo{author}{\bibfnamefont{J.}~\bibnamefont{Teng}},
  \bibinfo{author}{\bibfnamefont{N.}~\bibnamefont{Liu}}, \bibnamefont{and}
  \bibinfo{author}{\bibfnamefont{Y.}~\bibnamefont{Li}}, \bibinfo{journal}{J.\
  Semicond.} \textbf{\bibinfo{volume}{40}}, \bibinfo{pages}{081507}
  (\bibinfo{year}{2019}).

\bibitem[{\citenamefont{Kang et~al.}(2016)\citenamefont{Kang, Min, Lee, Song,
  Cho, and Cho}}]{kan16}
\bibinfo{author}{\bibfnamefont{B.~Y.} \bibnamefont{Kang}},
  \bibinfo{author}{\bibfnamefont{C.-H.} \bibnamefont{Min}},
  \bibinfo{author}{\bibfnamefont{S.~S.} \bibnamefont{Lee}},
  \bibinfo{author}{\bibfnamefont{M.~S.} \bibnamefont{Song}},
  \bibinfo{author}{\bibfnamefont{K.~K.} \bibnamefont{Cho}}, \bibnamefont{and}
  \bibinfo{author}{\bibfnamefont{B.~K.} \bibnamefont{Cho}},
  \bibinfo{journal}{Phys.\ Rev.\ B} \textbf{\bibinfo{volume}{94}},
  \bibinfo{pages}{165102} (\bibinfo{year}{2016}).

\bibitem[{\citenamefont{Wakeham
  et~al.}(2016{\natexlab{b}})\citenamefont{Wakeham, Wen, Wang, Fisk, Ronning,
  and Thompson}}]{wak16}
\bibinfo{author}{\bibfnamefont{N.}~\bibnamefont{Wakeham}},
  \bibinfo{author}{\bibfnamefont{J.}~\bibnamefont{Wen}},
  \bibinfo{author}{\bibfnamefont{Y.~Q.} \bibnamefont{Wang}},
  \bibinfo{author}{\bibfnamefont{Z.}~\bibnamefont{Fisk}},
  \bibinfo{author}{\bibfnamefont{F.}~\bibnamefont{Ronning}}, \bibnamefont{and}
  \bibinfo{author}{\bibfnamefont{J.~D.} \bibnamefont{Thompson}},
  \bibinfo{journal}{J.\ Magn.\ Magn.\ Mater.} \textbf{\bibinfo{volume}{400}},
  \bibinfo{pages}{62} (\bibinfo{year}{2016}{\natexlab{b}}).

\bibitem[{\citenamefont{Lesseux et~al.}(2017)\citenamefont{Lesseux, Rosa, Fisk,
  Schlottmann, Pagliuso, Urbano, and Rettori}}]{les17}
\bibinfo{author}{\bibfnamefont{G.~G.} \bibnamefont{Lesseux}},
  \bibinfo{author}{\bibfnamefont{P.~F.~S.} \bibnamefont{Rosa}},
  \bibinfo{author}{\bibfnamefont{Z.}~\bibnamefont{Fisk}},
  \bibinfo{author}{\bibfnamefont{P.}~\bibnamefont{Schlottmann}},
  \bibinfo{author}{\bibfnamefont{P.~G.} \bibnamefont{Pagliuso}},
  \bibinfo{author}{\bibfnamefont{R.~R.} \bibnamefont{Urbano}},
  \bibnamefont{and} \bibinfo{author}{\bibfnamefont{C.}~\bibnamefont{Rettori}},
  \bibinfo{journal}{AIP Adv.} \textbf{\bibinfo{volume}{7}},
  \bibinfo{pages}{055709} (\bibinfo{year}{2017}).

\bibitem[{\citenamefont{Demishev et~al.}(2018)\citenamefont{Demishev, Gilmanov,
  Samarin, Semeno, Sluchanko, Samarin, Bogach, Shitsevalova, Filipov, Karasev
  et~al.}}]{dem18}
\bibinfo{author}{\bibfnamefont{S.~V.} \bibnamefont{Demishev}},
  \bibinfo{author}{\bibfnamefont{M.~I.} \bibnamefont{Gilmanov}},
  \bibinfo{author}{\bibfnamefont{A.~N.} \bibnamefont{Samarin}},
  \bibinfo{author}{\bibfnamefont{A.~V.} \bibnamefont{Semeno}},
  \bibinfo{author}{\bibfnamefont{N.~E.} \bibnamefont{Sluchanko}},
  \bibinfo{author}{\bibfnamefont{N.~A.} \bibnamefont{Samarin}},
  \bibinfo{author}{\bibfnamefont{A.~V.} \bibnamefont{Bogach}},
  \bibinfo{author}{\bibfnamefont{N.~Y.} \bibnamefont{Shitsevalova}},
  \bibinfo{author}{\bibfnamefont{V.~B.} \bibnamefont{Filipov}},
  \bibinfo{author}{\bibfnamefont{M.~S.} \bibnamefont{Karasev}},
  \bibnamefont{et~al.}, \bibinfo{journal}{Sci.\ Rep.}
  \textbf{\bibinfo{volume}{8}}, \bibinfo{pages}{7125} (\bibinfo{year}{2018}).

\bibitem[{\citenamefont{Liu et~al.}(2009)\citenamefont{Liu, Liu, Xu, Qi, and
  Zhang}}]{liu09}
\bibinfo{author}{\bibfnamefont{Q.}~\bibnamefont{Liu}},
  \bibinfo{author}{\bibfnamefont{C.-X.} \bibnamefont{Liu}},
  \bibinfo{author}{\bibfnamefont{C.}~\bibnamefont{Xu}},
  \bibinfo{author}{\bibfnamefont{X.-L.} \bibnamefont{Qi}}, \bibnamefont{and}
  \bibinfo{author}{\bibfnamefont{S.-C.} \bibnamefont{Zhang}},
  \bibinfo{journal}{Phys.\ Rev.\ Lett.} \textbf{\bibinfo{volume}{102}},
  \bibinfo{pages}{156603} (\bibinfo{year}{2009}).

\bibitem[{\citenamefont{Baruselli and Vojta}(2016)}]{bar16}
\bibinfo{author}{\bibfnamefont{P.~P.} \bibnamefont{Baruselli}}
  \bibnamefont{and} \bibinfo{author}{\bibfnamefont{M.}~\bibnamefont{Vojta}},
  \bibinfo{journal}{Phys.\ Rev.\ B} \textbf{\bibinfo{volume}{93}},
  \bibinfo{pages}{195117} (\bibinfo{year}{2016}).

\bibitem[{\citenamefont{Enayat et~al.}(2014)\citenamefont{Enayat, Sun, Singh,
  Aluru, Schmaus, Yaresko, Liu, Lin, Tsurkan, Loidl et~al.}}]{ena14}
\bibinfo{author}{\bibfnamefont{M.}~\bibnamefont{Enayat}},
  \bibinfo{author}{\bibfnamefont{Z.}~\bibnamefont{Sun}},
  \bibinfo{author}{\bibfnamefont{U.~R.} \bibnamefont{Singh}},
  \bibinfo{author}{\bibfnamefont{R.}~\bibnamefont{Aluru}},
  \bibinfo{author}{\bibfnamefont{S.}~\bibnamefont{Schmaus}},
  \bibinfo{author}{\bibfnamefont{A.}~\bibnamefont{Yaresko}},
  \bibinfo{author}{\bibfnamefont{Y.}~\bibnamefont{Liu}},
  \bibinfo{author}{\bibfnamefont{C.}~\bibnamefont{Lin}},
  \bibinfo{author}{\bibfnamefont{V.}~\bibnamefont{Tsurkan}},
  \bibinfo{author}{\bibfnamefont{A.}~\bibnamefont{Loidl}},
  \bibnamefont{et~al.}, \bibinfo{journal}{Science}
  \textbf{\bibinfo{volume}{345}}, \bibinfo{pages}{653} (\bibinfo{year}{2014}).

\bibitem[{\citenamefont{Fisk et~al.}(1979)\citenamefont{Fisk, Johnston, Cornut,
  {von Moln{\'a}r}, Oseroff, and Calvo}}]{fis79}
\bibinfo{author}{\bibfnamefont{Z.}~\bibnamefont{Fisk}},
  \bibinfo{author}{\bibfnamefont{D.~C.} \bibnamefont{Johnston}},
  \bibinfo{author}{\bibfnamefont{B.}~\bibnamefont{Cornut}},
  \bibinfo{author}{\bibfnamefont{S.}~\bibnamefont{{von Moln{\'a}r}}},
  \bibinfo{author}{\bibfnamefont{S.}~\bibnamefont{Oseroff}}, \bibnamefont{and}
  \bibinfo{author}{\bibfnamefont{R.}~\bibnamefont{Calvo}},
  \bibinfo{journal}{J.\ Appl.\ Phys.} \textbf{\bibinfo{volume}{50}},
  \bibinfo{pages}{1911} (\bibinfo{year}{1979}).

\bibitem[{\citenamefont{S{\"u}llow et~al.}(2000)\citenamefont{S{\"u}llow,
  Prasad, Aronson, Bogdanovich, Sarrao, and Fisk}}]{sue00}
\bibinfo{author}{\bibfnamefont{S.}~\bibnamefont{S{\"u}llow}},
  \bibinfo{author}{\bibfnamefont{I.}~\bibnamefont{Prasad}},
  \bibinfo{author}{\bibfnamefont{M.~C.} \bibnamefont{Aronson}},
  \bibinfo{author}{\bibfnamefont{S.}~\bibnamefont{Bogdanovich}},
  \bibinfo{author}{\bibfnamefont{J.~L.} \bibnamefont{Sarrao}},
  \bibnamefont{and} \bibinfo{author}{\bibfnamefont{Z.}~\bibnamefont{Fisk}},
  \bibinfo{journal}{Phys.\ Rev.\ B} \textbf{\bibinfo{volume}{62}},
  \bibinfo{pages}{11626} (\bibinfo{year}{2000}).

\bibitem[{\citenamefont{{von Moln{\'a}r}}(2018)}]{mol01}
\bibinfo{author}{\bibfnamefont{S.}~\bibnamefont{{von Moln{\'a}r}}},
  \bibinfo{journal}{Sens.\ Actuators A} \textbf{\bibinfo{volume}{91}},
  \bibinfo{pages}{161} (\bibinfo{year}{2018}).

\bibitem[{\citenamefont{Rao and Raveau}(1998)}]{rao98}
\bibinfo{author}{\bibfnamefont{C.~N.~R.} \bibnamefont{Rao}} \bibnamefont{and}
  \bibinfo{author}{\bibfnamefont{B.}~\bibnamefont{Raveau}},
  \emph{\bibinfo{title}{Colossal {M}agnetoresistance, {C}harge {O}rdering and
  {R}elated {P}roperties of {M}anganites}} (\bibinfo{publisher}{World
  Scientific, Singapore}, \bibinfo{year}{1998}).

\bibitem[{\citenamefont{Das et~al.}(2012)\citenamefont{Das, Amyan, Brandenburg,
  M{\"u}ller, Xiong, {von Moln{\'a}r}, and Fisk}}]{das12}
\bibinfo{author}{\bibfnamefont{P.}~\bibnamefont{Das}},
  \bibinfo{author}{\bibfnamefont{A.}~\bibnamefont{Amyan}},
  \bibinfo{author}{\bibfnamefont{J.}~\bibnamefont{Brandenburg}},
  \bibinfo{author}{\bibfnamefont{J.}~\bibnamefont{M{\"u}ller}},
  \bibinfo{author}{\bibfnamefont{P.}~\bibnamefont{Xiong}},
  \bibinfo{author}{\bibfnamefont{S.}~\bibnamefont{{von Moln{\'a}r}}},
  \bibnamefont{and} \bibinfo{author}{\bibfnamefont{Z.}~\bibnamefont{Fisk}},
  \bibinfo{journal}{Phys.\ Rev.\ B} \textbf{\bibinfo{volume}{86}},
  \bibinfo{pages}{184425} (\bibinfo{year}{2012}).

\bibitem[{\citenamefont{Manna et~al.}(2014)\citenamefont{Manna, Das, {de
  Souza}, Schnelle, Lang, M{\"u}ller, {von Moln{\'a}r}, and Fisk}}]{man14}
\bibinfo{author}{\bibfnamefont{R.~S.} \bibnamefont{Manna}},
  \bibinfo{author}{\bibfnamefont{P.}~\bibnamefont{Das}},
  \bibinfo{author}{\bibfnamefont{M.}~\bibnamefont{{de Souza}}},
  \bibinfo{author}{\bibfnamefont{F.}~\bibnamefont{Schnelle}},
  \bibinfo{author}{\bibfnamefont{M.}~\bibnamefont{Lang}},
  \bibinfo{author}{\bibfnamefont{J.}~\bibnamefont{M{\"u}ller}},
  \bibinfo{author}{\bibfnamefont{S.}~\bibnamefont{{von Moln{\'a}r}}},
  \bibnamefont{and} \bibinfo{author}{\bibfnamefont{Z.}~\bibnamefont{Fisk}},
  \bibinfo{journal}{Phys.\ Rev.\ Lett.} \textbf{\bibinfo{volume}{113}},
  \bibinfo{pages}{067202} (\bibinfo{year}{2014}).

\bibitem[{\citenamefont{R{\o}nnow et~al.}(2006)\citenamefont{R{\o}nnow, Renner,
  Aeppli, Kimura, and Tokura}}]{ron06}
\bibinfo{author}{\bibfnamefont{H.~M.} \bibnamefont{R{\o}nnow}},
  \bibinfo{author}{\bibfnamefont{C.}~\bibnamefont{Renner}},
  \bibinfo{author}{\bibfnamefont{G.}~\bibnamefont{Aeppli}},
  \bibinfo{author}{\bibfnamefont{T.}~\bibnamefont{Kimura}}, \bibnamefont{and}
  \bibinfo{author}{\bibfnamefont{Y.}~\bibnamefont{Tokura}},
  \bibinfo{journal}{Nature} \textbf{\bibinfo{volume}{440}},
  \bibinfo{pages}{1025} (\bibinfo{year}{2006}).

\bibitem[{\citenamefont{Buchsteiner}(2020)}]{buc-the}
\bibinfo{author}{\bibfnamefont{P.}~\bibnamefont{Buchsteiner}}, Ph.D. thesis,
  \bibinfo{school}{Georg-August-University G\"{o}ttingen, Germany}
  (\bibinfo{year}{2020}).

\bibitem[{\citenamefont{Etourneau and Hagenmuller}(1985)}]{eto85}
\bibinfo{author}{\bibfnamefont{J.}~\bibnamefont{Etourneau}} \bibnamefont{and}
  \bibinfo{author}{\bibfnamefont{P.}~\bibnamefont{Hagenmuller}},
  \bibinfo{journal}{Philos. Mag.} \textbf{\bibinfo{volume}{52}},
  \bibinfo{pages}{589} (\bibinfo{year}{1985}).

\bibitem[{\citenamefont{Marchenko et~al.}(1998)\citenamefont{Marchenko,
  Cherepanov, Tarashchenko, Kazantseva, and Naumovets}}]{mar98}
\bibinfo{author}{\bibfnamefont{A.~A.} \bibnamefont{Marchenko}},
  \bibinfo{author}{\bibfnamefont{V.~V.} \bibnamefont{Cherepanov}},
  \bibinfo{author}{\bibfnamefont{D.~T.} \bibnamefont{Tarashchenko}},
  \bibinfo{author}{\bibfnamefont{Z.~I.} \bibnamefont{Kazantseva}},
  \bibnamefont{and} \bibinfo{author}{\bibfnamefont{A.~G.}
  \bibnamefont{Naumovets}}, \bibinfo{journal}{Surf.\ Sci.}
  \textbf{\bibinfo{volume}{416}}, \bibinfo{pages}{460} (\bibinfo{year}{1998}).

\bibitem[{\citenamefont{Nagaoka and Ohmi}(2020)}]{nag20}
\bibinfo{author}{\bibfnamefont{K.}~\bibnamefont{Nagaoka}} \bibnamefont{and}
  \bibinfo{author}{\bibfnamefont{S.-i.} \bibnamefont{Ohmi}},
  \bibinfo{journal}{J.\ Vac.\ Sci.\ Technol.\ B} \textbf{\bibinfo{volume}{38}},
  \bibinfo{pages}{062801} (\bibinfo{year}{2020}).

\bibitem[{\citenamefont{Aono et~al.}(1979)\citenamefont{Aono, Nishitani,
  Oshima, Tanaka, Bannai, and Kawai}}]{aon79}
\bibinfo{author}{\bibfnamefont{M.}~\bibnamefont{Aono}},
  \bibinfo{author}{\bibfnamefont{R.}~\bibnamefont{Nishitani}},
  \bibinfo{author}{\bibfnamefont{C.}~\bibnamefont{Oshima}},
  \bibinfo{author}{\bibfnamefont{T.}~\bibnamefont{Tanaka}},
  \bibinfo{author}{\bibfnamefont{E.}~\bibnamefont{Bannai}}, \bibnamefont{and}
  \bibinfo{author}{\bibfnamefont{S.}~\bibnamefont{Kawai}},
  \bibinfo{journal}{Surf.\ Sci.} \textbf{\bibinfo{volume}{86}},
  \bibinfo{pages}{631} (\bibinfo{year}{1979}).

\bibitem[{\citenamefont{Michaelson}(1977)}]{mic77}
\bibinfo{author}{\bibfnamefont{H.~B.} \bibnamefont{Michaelson}},
  \bibinfo{journal}{J.\ Appl.\ Phys.} \textbf{\bibinfo{volume}{48}},
  \bibinfo{pages}{4729} (\bibinfo{year}{1977}).

\bibitem[{\citenamefont{Mitterer}(1997)}]{mit97}
\bibinfo{author}{\bibfnamefont{C.}~\bibnamefont{Mitterer}},
  \bibinfo{journal}{J.\ Solid State Chem.} \textbf{\bibinfo{volume}{133}},
  \bibinfo{pages}{279} (\bibinfo{year}{1997}).

\bibitem[{\citenamefont{Shishido et~al.}(2014)\citenamefont{Shishido, Kawai,
  Futagami, Noguchi, and Ishida}}]{shi14}
\bibinfo{author}{\bibfnamefont{H.}~\bibnamefont{Shishido}},
  \bibinfo{author}{\bibfnamefont{K.}~\bibnamefont{Kawai}},
  \bibinfo{author}{\bibfnamefont{A.}~\bibnamefont{Futagami}},
  \bibinfo{author}{\bibfnamefont{S.}~\bibnamefont{Noguchi}}, \bibnamefont{and}
  \bibinfo{author}{\bibfnamefont{T.}~\bibnamefont{Ishida}},
  \bibinfo{journal}{JPS Conf.\ Proc.} \textbf{\bibinfo{volume}{3}},
  \bibinfo{pages}{011045} (\bibinfo{year}{2014}).

\bibitem[{\citenamefont{Yong et~al.}(2014)\citenamefont{Yong, Jiang, Usanmaz,
  Curtarolo, Zhang, Li, Pan, Shin, Takeuchi, and Greene}}]{yon14}
\bibinfo{author}{\bibfnamefont{J.}~\bibnamefont{Yong}},
  \bibinfo{author}{\bibfnamefont{Y.}~\bibnamefont{Jiang}},
  \bibinfo{author}{\bibfnamefont{D.}~\bibnamefont{Usanmaz}},
  \bibinfo{author}{\bibfnamefont{S.}~\bibnamefont{Curtarolo}},
  \bibinfo{author}{\bibfnamefont{X.}~\bibnamefont{Zhang}},
  \bibinfo{author}{\bibfnamefont{L.}~\bibnamefont{Li}},
  \bibinfo{author}{\bibfnamefont{X.}~\bibnamefont{Pan}},
  \bibinfo{author}{\bibfnamefont{J.}~\bibnamefont{Shin}},
  \bibinfo{author}{\bibfnamefont{I.}~\bibnamefont{Takeuchi}}, \bibnamefont{and}
  \bibinfo{author}{\bibfnamefont{R.~L.} \bibnamefont{Greene}},
  \bibinfo{journal}{Appl.\ Phys.\ Lett.} \textbf{\bibinfo{volume}{105}},
  \bibinfo{pages}{222403} (\bibinfo{year}{2014}).

\bibitem[{\citenamefont{Li et~al.}(2018)\citenamefont{Li, Ma, Huang, and
  Chien}}]{li18}
\bibinfo{author}{\bibfnamefont{Y.}~\bibnamefont{Li}},
  \bibinfo{author}{\bibfnamefont{Q.}~\bibnamefont{Ma}},
  \bibinfo{author}{\bibfnamefont{S.~X.} \bibnamefont{Huang}}, \bibnamefont{and}
  \bibinfo{author}{\bibfnamefont{C.~L.} \bibnamefont{Chien}},
  \bibinfo{journal}{Sci.\ Adv.} \textbf{\bibinfo{volume}{4}},
  \bibinfo{pages}{eaap8294} (\bibinfo{year}{2018}).

\bibitem[{\citenamefont{Ba\v{t}kov\'{a}
  et~al.}(2018)\citenamefont{Ba\v{t}kov\'{a}, Ba\v{t}ko, Stobiecki,
  Szyma\'{n}ski, Ku\'{s}wik, Mackov\'{a}, and Malinsk\'{y}}}]{bat18}
\bibinfo{author}{\bibfnamefont{M.}~\bibnamefont{Ba\v{t}kov\'{a}}},
  \bibinfo{author}{\bibfnamefont{I.}~\bibnamefont{Ba\v{t}ko}},
  \bibinfo{author}{\bibfnamefont{F.}~\bibnamefont{Stobiecki}},
  \bibinfo{author}{\bibfnamefont{B.}~\bibnamefont{Szyma\'{n}ski}},
  \bibinfo{author}{\bibfnamefont{P.}~\bibnamefont{Ku\'{s}wik}},
  \bibinfo{author}{\bibfnamefont{A.}~\bibnamefont{Mackov\'{a}}},
  \bibnamefont{and}
  \bibinfo{author}{\bibfnamefont{P.}~\bibnamefont{Malinsk\'{y}}},
  \bibinfo{journal}{J.\ Alloys Comp.} \textbf{\bibinfo{volume}{744}},
  \bibinfo{pages}{821} (\bibinfo{year}{2018}).

\bibitem[{\citenamefont{Liu et~al.}(2018)\citenamefont{Liu, Li, Gu, Ding,
  Chang, Janantha, Kalinikos, Novosad, Hoffmann, Wu et~al.}}]{liu18}
\bibinfo{author}{\bibfnamefont{T.}~\bibnamefont{Liu}},
  \bibinfo{author}{\bibfnamefont{Y.}~\bibnamefont{Li}},
  \bibinfo{author}{\bibfnamefont{L.}~\bibnamefont{Gu}},
  \bibinfo{author}{\bibfnamefont{J.}~\bibnamefont{Ding}},
  \bibinfo{author}{\bibfnamefont{H.}~\bibnamefont{Chang}},
  \bibinfo{author}{\bibfnamefont{P.~A.~P.} \bibnamefont{Janantha}},
  \bibinfo{author}{\bibfnamefont{B.}~\bibnamefont{Kalinikos}},
  \bibinfo{author}{\bibfnamefont{V.}~\bibnamefont{Novosad}},
  \bibinfo{author}{\bibfnamefont{A.}~\bibnamefont{Hoffmann}},
  \bibinfo{author}{\bibfnamefont{R.}~\bibnamefont{Wu}}, \bibnamefont{et~al.},
  \bibinfo{journal}{Phys.\ Rev.\ Lett.} \textbf{\bibinfo{volume}{120}},
  \bibinfo{pages}{207206} (\bibinfo{year}{2018}).

\bibitem[{\citenamefont{Lee et~al.}(2016)\citenamefont{Lee, Zhang, Liang,
  Fackler, Yong, Wang, Paglione, Greene, and Takeuchi}}]{lee16}
\bibinfo{author}{\bibfnamefont{S.}~\bibnamefont{Lee}},
  \bibinfo{author}{\bibfnamefont{X.}~\bibnamefont{Zhang}},
  \bibinfo{author}{\bibfnamefont{Y.}~\bibnamefont{Liang}},
  \bibinfo{author}{\bibfnamefont{S.~W.} \bibnamefont{Fackler}},
  \bibinfo{author}{\bibfnamefont{J.}~\bibnamefont{Yong}},
  \bibinfo{author}{\bibfnamefont{X.}~\bibnamefont{Wang}},
  \bibinfo{author}{\bibfnamefont{J.}~\bibnamefont{Paglione}},
  \bibinfo{author}{\bibfnamefont{R.~L.} \bibnamefont{Greene}},
  \bibnamefont{and} \bibinfo{author}{\bibfnamefont{I.}~\bibnamefont{Takeuchi}},
  \bibinfo{journal}{Phys.\ Rev.\ X} \textbf{\bibinfo{volume}{6}},
  \bibinfo{pages}{031031} (\bibinfo{year}{2016}).

\bibitem[{\citenamefont{Wakeham et~al.}(2015)\citenamefont{Wakeham, Wang, Fisk,
  Ronning, and Thompson}}]{wak15}
\bibinfo{author}{\bibfnamefont{N.}~\bibnamefont{Wakeham}},
  \bibinfo{author}{\bibfnamefont{Y.}~\bibnamefont{Wang}},
  \bibinfo{author}{\bibfnamefont{Z.}~\bibnamefont{Fisk}},
  \bibinfo{author}{\bibfnamefont{F.}~\bibnamefont{Ronning}}, \bibnamefont{and}
  \bibinfo{author}{\bibfnamefont{J.~D.} \bibnamefont{Thompson}},
  \bibinfo{journal}{Phys.\ Rev.\ B} \textbf{\bibinfo{volume}{91}},
  \bibinfo{pages}{085107} (\bibinfo{year}{2015}).

\bibitem[{\citenamefont{Sen et~al.}(2020)\citenamefont{Sen, Vidhyadhiraja,
  Miranda, Dobrosavljevi{\'c}, and Ku}}]{sen20}
\bibinfo{author}{\bibfnamefont{S.}~\bibnamefont{Sen}},
  \bibinfo{author}{\bibfnamefont{N.~S.} \bibnamefont{Vidhyadhiraja}},
  \bibinfo{author}{\bibfnamefont{E.}~\bibnamefont{Miranda}},
  \bibinfo{author}{\bibfnamefont{V.}~\bibnamefont{Dobrosavljevi{\'c}}},
  \bibnamefont{and} \bibinfo{author}{\bibfnamefont{W.}~\bibnamefont{Ku}},
  \bibinfo{journal}{Phys.\ Rev.\ Research} \textbf{\bibinfo{volume}{2}},
  \bibinfo{pages}{033370} (\bibinfo{year}{2020}).

\bibitem[{\citenamefont{Abele et~al.}(2020)\citenamefont{Abele, Yuan, and
  Riseborough}}]{abe20}
\bibinfo{author}{\bibfnamefont{M.}~\bibnamefont{Abele}},
  \bibinfo{author}{\bibfnamefont{X.}~\bibnamefont{Yuan}}, \bibnamefont{and}
  \bibinfo{author}{\bibfnamefont{P.~S.} \bibnamefont{Riseborough}},
  \bibinfo{journal}{Phys.\ Rev.\ B} \textbf{\bibinfo{volume}{101}},
  \bibinfo{pages}{094101} (\bibinfo{year}{2020}).

\bibitem[{\citenamefont{Stern et~al.}(2017)\citenamefont{Stern, Dzero,
  Galitski, Fisk, and Xia}}]{ste17}
\bibinfo{author}{\bibfnamefont{A.}~\bibnamefont{Stern}},
  \bibinfo{author}{\bibfnamefont{M.}~\bibnamefont{Dzero}},
  \bibinfo{author}{\bibfnamefont{V.~M.} \bibnamefont{Galitski}},
  \bibinfo{author}{\bibfnamefont{Z.}~\bibnamefont{Fisk}}, \bibnamefont{and}
  \bibinfo{author}{\bibfnamefont{J.}~\bibnamefont{Xia}},
  \bibinfo{journal}{Nat.\ Mater.} \textbf{\bibinfo{volume}{16}},
  \bibinfo{pages}{708} (\bibinfo{year}{2017}).

\bibitem[{\citenamefont{Nahas et~al.}(2013)\citenamefont{Nahas, Berneau,
  Bonneville, Coupeau, Drouet, Lamongie, Marteau, Michel, Tanguy, and
  Tromas}}]{nah13}
\bibinfo{author}{\bibfnamefont{Y.}~\bibnamefont{Nahas}},
  \bibinfo{author}{\bibfnamefont{F.}~\bibnamefont{Berneau}},
  \bibinfo{author}{\bibfnamefont{J.}~\bibnamefont{Bonneville}},
  \bibinfo{author}{\bibfnamefont{C.}~\bibnamefont{Coupeau}},
  \bibinfo{author}{\bibfnamefont{M.}~\bibnamefont{Drouet}},
  \bibinfo{author}{\bibfnamefont{B.}~\bibnamefont{Lamongie}},
  \bibinfo{author}{\bibfnamefont{M.}~\bibnamefont{Marteau}},
  \bibinfo{author}{\bibfnamefont{J.}~\bibnamefont{Michel}},
  \bibinfo{author}{\bibfnamefont{P.}~\bibnamefont{Tanguy}}, \bibnamefont{and}
  \bibinfo{author}{\bibfnamefont{C.}~\bibnamefont{Tromas}},
  \bibinfo{journal}{Rev.\ Sci.\ Instrum.} \textbf{\bibinfo{volume}{84}},
  \bibinfo{pages}{105117} (\bibinfo{year}{2013}).

\bibitem[{\citenamefont{Lee et~al.}(2019)\citenamefont{Lee, Stanev, Zhang,
  Stasak, Flowers, Higgins, Dai, Blum, Pan, Yakovenko et~al.}}]{lee19}
\bibinfo{author}{\bibfnamefont{S.}~\bibnamefont{Lee}},
  \bibinfo{author}{\bibfnamefont{V.}~\bibnamefont{Stanev}},
  \bibinfo{author}{\bibfnamefont{X.}~\bibnamefont{Zhang}},
  \bibinfo{author}{\bibfnamefont{D.}~\bibnamefont{Stasak}},
  \bibinfo{author}{\bibfnamefont{J.}~\bibnamefont{Flowers}},
  \bibinfo{author}{\bibfnamefont{J.~S.} \bibnamefont{Higgins}},
  \bibinfo{author}{\bibfnamefont{S.}~\bibnamefont{Dai}},
  \bibinfo{author}{\bibfnamefont{T.}~\bibnamefont{Blum}},
  \bibinfo{author}{\bibfnamefont{X.}~\bibnamefont{Pan}},
  \bibinfo{author}{\bibfnamefont{V.~M.} \bibnamefont{Yakovenko}},
  \bibnamefont{et~al.}, \bibinfo{journal}{Nature}
  \textbf{\bibinfo{volume}{570}}, \bibinfo{pages}{344} (\bibinfo{year}{2019}).

\bibitem[{\citenamefont{Yim et~al.}(2018)\citenamefont{Yim, Trainer, Aluru,
  Chi, Hardy, Liang, Bonn, and Wahl}}]{yim18}
\bibinfo{author}{\bibfnamefont{C.~M.} \bibnamefont{Yim}},
  \bibinfo{author}{\bibfnamefont{C.}~\bibnamefont{Trainer}},
  \bibinfo{author}{\bibfnamefont{R.}~\bibnamefont{Aluru}},
  \bibinfo{author}{\bibfnamefont{S.}~\bibnamefont{Chi}},
  \bibinfo{author}{\bibfnamefont{W.~N.} \bibnamefont{Hardy}},
  \bibinfo{author}{\bibfnamefont{R.}~\bibnamefont{Liang}},
  \bibinfo{author}{\bibfnamefont{D.}~\bibnamefont{Bonn}}, \bibnamefont{and}
  \bibinfo{author}{\bibfnamefont{P.}~\bibnamefont{Wahl}},
  \bibinfo{journal}{Nat.\ Commun.} \textbf{\bibinfo{volume}{9}},
  \bibinfo{pages}{2602} (\bibinfo{year}{2018}).

\end{thebibliography}
\end{document}